\documentclass[prd,twocolumn,superscriptaddress,floatfix,amsmath,amssymb,floatfix]{revtex4-1}

\usepackage{float} 

\usepackage{graphicx}

\usepackage[normalem]{ulem}
\usepackage[english]{babel}
\usepackage{graphicx}
\usepackage{dcolumn}
\usepackage{bm}
\usepackage{verbatim}
\usepackage{mathrsfs}
\usepackage{cancel}
\usepackage{epstopdf}
\usepackage{mathtools}
\usepackage{color}
\usepackage{tensor}

\newcommand{\ket}[1]{\left| {#1} \right\rangle}
\newcommand{\bra}[1]{\left\langle {#1} \right|}

\newcommand{\dd}{\mathrm{d}}
\newcommand{\ii}{\mathrm{i}}
\newcommand{\II}{\textsc{i}}
\newcommand{\RR}{\textsc{r}}

\newcommand{\Tr}[1]{\text{tr}\left(#1\right)}
\newcommand{\Abs}[1]{\vert #1\vert}

\newcommand{\mean}[1]{\big\langle #1\big\rangle}
\newcommand{\abs}[1]{\vert #1\vert}
\newcommand{\Det}[1]{\left\vert #1\right\vert}

\newcommand{\Imag}[1]{\mathbb{I}\text{m}\left[ #1 \right]}
\newcommand{\Real}[1]{\mathbb{R}\text{e}\left[ #1 \right]}
\newcommand{\Eq}[1]{Eq.~\eqref{#1}}

\usepackage[usenames,dvipsnames]{pstricks}
\usepackage{epsfig}
\usepackage{pst-grad} 
\usepackage{pst-plot} 

\usepackage{hyperref}

\usepackage{verbatim}

\DeclareMathOperator*{\sumint}{%
\mathchoice%
  {\ooalign{$\displaystyle\sum$\cr\hidewidth$\displaystyle\int$\hidewidth\cr}}
  {\ooalign{\raisebox{.14\height}{\scalebox{.7}{$\textstyle\sum$}}\cr\hidewidth$\textstyle\int$\hidewidth\cr}}
  {\ooalign{\raisebox{.2\height}{\scalebox{.6}{$\scriptstyle\sum$}}\cr$\scriptstyle\int$\cr}}
  {\ooalign{\raisebox{.2\height}{\scalebox{.6}{$\scriptstyle\sum$}}\cr$\scriptstyle\int$\cr}}
}

\begin{document}

\title{Casimir forces and quantum friction of finite-size atoms in relativistic trajectories}

\author{Pablo Rodriguez-Lopez}
\affiliation{Materials Science Factory, Instituto de Ciencia de Materiales de Madrid (ICMM),
Consejo Superior de Investigaciones Científicas (CSIC),
Sor Juana Inés de la Cruz 3, 28049 Madrid Spain.}

\author{Eduardo Mart\'{i}n-Mart\'{i}nez}
\email{emartinm@uwaterloo.ca}
\affiliation{Institute for Quantum Computing, University of Waterloo, Waterloo, Ontario, N2L 3G1, Canada}
\affiliation{Department of Applied Mathematics, University of Waterloo, Waterloo, Ontario, N2L 3G1, Canada}
\affiliation{Perimeter Institute for Theoretical Physics, Waterloo, Ontario, N2L 2Y5, Canada}

\begin{abstract}
We study quantum friction and Casimir forces with a full-relativistic formalism for atoms modelled as Unruh-DeWitt detectors in the presence of arbitrary macroscopic objects. We consider the general case of atoms with arbitrary relativistic trajectories in arbitrary quantum states (including  coherent superpositions) close to objects that impose arbitrary boundary conditions. Particularizing for conducting plates, we show that, for relative velocities close to the speed of light, the quantum friction diverges while the Casimir force is almost independent of the velocity. Since we include the effect of the finite size of the detector and the finite interaction time, we also obtain quantum friction when the detector is isolated but follows a non-inertial trajectory.
\end{abstract}

\maketitle



\section{Introduction}

Quantum friction is the appearance of a reaction force to the movement of a neutral object in the presence of a quantum field \cite{Milton2016}. Quantum friction is strongly related with the dynamical Casimir effect, which is the emission of real particles by moving objects in the presence of a quantum field \cite{Wilson2009,Wilson2010,Wilson2011,Souza2018}. 

The study of quantum friction between atoms, and between atoms and dielectric (and metallic) plates has been studied in great detail, since it is the simplest non-trivial model for the interaction of microscopic objects (e.g., atoms, molecules, etc.) and macroscopic objects (e.g., mirrors, dielectric spheres, etc.). It is impossible to cite all works on the topic due to their sheer abundance, but as a token, the quantum friction between atoms has been studied in, e.g., \cite{Donaire2016B,Jentschura2017,Scheel2009}. The friction between atoms and plates has been studied in  \cite{Dalvitian,Donaire2016,Intravaia-Non-Markovianity2016,Reiche-Spatial-dispersion-plate2017,Klatt2017,BartonII2010,Farias-Topological-quantum-friction2017,Parsegian1974,Reiche2017,Hoye2014}, among others. The case of rotation friction has been analyzed in, e.g., \cite{Manjavacas2010,Manjavacas2010B,Manjavacas2017,Volokitin2018}. The effect of acceleration in quantum friction in the context of the Unruh effect has been studied in, e.g., \cite{Parentani1995,Reznik1998,Hu2004,Lin2007,Marino2014}. Analyses from the perspective of the dynamical Casimir effect for non-uniform trajectories and in curved spaces have been performed in e.g., \cite{Corona-Ugalde2016,Lock2017}. The friction forces between macroscopic objects has also been studied in, e.g., \cite{Scattering_formalism_Dynamical_em2013,Maghrebi2013,Brevik2015,Sherkunov2005,Volokitin2016,Volokitin2011}. The relation of quantum friction with the fluctuation-dissipation theorem has been considered in \cite{Intravaia2014,Golyk2013,Intravaia-Vacuum-Incalescence2016,NefedovRubi2017}, etc. 

The Casimir forces \cite{Libro_Dalvit} or Van der Waals forces in the non-relativistic regime, were studied between atoms in the seminal work \cite{CasimirPolderada}, and between parallel perfect metal plates in \cite{CasimirPlacas1948}. Lifshitz included the material properties of the plates into the formalism in \cite{Lifshitz1957}. Several approaches to the calculation of those forces, as the pairwise summation approximation \cite{Milton2008} and the proximity force approximation \cite{BLOCKI1977}, have been widely employed, although their range of validity have been justified only recently \cite{Rodriguez-Lopez2009,FoscoPFA2011}. A formal complete solution for multiple macroscopic arbitrary shaped linear optical materials has been obtained in \cite{Casimir_Scattering2007,Scattering_formalism_em,Lambrecht2006}. Since then, this solution inspired new numerical methods \cite{Reid2013} and theoretical research, with the development of the trace formalism \cite{Trace_formalism_em} for non-equilibrium and non-stationary set-ups \cite{Scattering_formalism_Dynamical_em2013}. See, for example, \cite{Woods2016} for a review.

Usually, in the calculations of the Casimir and Quantum Friction forces with atoms, the atoms are modelled as pointlike objects. One may justify this arguing that, at the end of the day, atoms are small and usually their characteristic length is the smallest one of all scales involved in any experiment. However, it is not unusual that the distance between interacting atoms would be of the order of their respective atomic radii, this is the case, for example, of the Van der Waals interaction that appears in the Lienard-Jones potential. It would be natural that, at those scales, the shape of the atom will be relevant. Additionally, the pointlike nature of the atom has been responsible for the appearance of divergences in the calculation of these forces in the past (e.g., \cite{Armata2016}). In the study of the relativistic aspects of the light matter interaction it is well known, however, that considering the finite-size of the atoms cures the models from these divergences \cite{Louko2006,Louko2008}.



In this paper, we perform a time-dependent  relativistic analysis of the Casimir force and the quantum friction on atoms (taking into account their internal dynamics) in the presence of extended objects. We also analyze the dependence of these forces on the internal state of the atoms (the full density matrix characterizing the quantum state of the atoms). Furthermore, we do not approximate the atoms as a pointlike objects, instead we consider the full spatial extension of the atom, which removes the divergences present in other pointlike calculations.

To do a covariant study of the Casimir and quantum friction forces, we model the interaction of the atoms and the quantum field with the Unruh-DeWitt model. While this model simplifies the nature of the field and the atom and considers a scalar coupling, it has been shown that Unruh-DeWitt detectors capture the relevant features of the light-matter interaction when the exchange of angular momentum between the field and the detector is not dominant for the phenomenology studied \cite{Pozas-Kerstjens2016,Nuestro1}. In any case, the formalism developed here is easily generalizable to the electromagnetic field case as seen e.g., in \cite{Pozas-Kerstjens2016,Lopp2018}. 

Concretely, we develop the formalism of finite-time Casimir and quantum friction forces for relativistic atoms undergoing arbitrary trajectories, in the presence of macroscopic objects modelled by their T scattering matrix via the Lippmann-Schwinger equation. 




By performing a fully relativistic finite-time study of the Casimir force and quantum friction, we find that there is an additional term of the  quantum friction force for free atoms moving in the presence of the field vacuum in non-equilibrium regimes. As far as the author's knowledge goes, this has not been the focus of  previous literature. Namely, the fact that there is a four-force always present when the detector interacts with a field regardless whether there is any other object present other than the detector. This force will depend on the shape of the detector and on its trajectory, and it is present even when the field and the detector are in their ground state, in contrast to the case studied in \cite{Volokitin2008,Volokitin2015}, where a thermal bath was present.



\section{Setup}

Let us first introduce the atom-field interaction model that we will use throughout the paper. For simplicity, we will model the atom as a two-level quantum system (we denote $\ket{g}$ the ground state and $\ket{e}$ the excited state), monopolarly coupled to a scalar field. This is the well-known Unruh-DeWitt (UDW) detector model \cite{DeWittBook}, which has been shown to capture the features of the light-matter interaction when there is no exchange of angular momentum \cite{Martin-Martinez2013,Alhambra2014} (see \cite{Pozas-Kerstjens2016} for a more in-depth discussion).

The Hamiltonian of the atom-field system in the interaction picture is given by
\begin{align}
\hat H=\hat H_\textsc{a}+\hat H_\phi+\hat H_I.
\end{align}

As we will see later on, only the interaction part of the Hamiltonian will be relevant to the Casimir and quantum friction forces. In any case, a full discussion of the motivation and the form of the different terms in the Hamiltonian can be found, e.g., in \cite{Nuestro1}.

Let us consider a detector, modelling an atom, that moves with an arbitrary trajectory coupled to the scalar field. There are two relevant reference frames in the problem: the laboratory frame $(t,\bm{x})$, assumed to be inertial, and the detector's center of mass reference frame $(\tau,\bm{\xi})$. In this work we will not consider the detector to be a pointlike object. Rather, we study the more general case of a finite-size detector smeared in its own frame moving in an arbitrary trajectory (for which a pointlike detector is a particular case). The reasonable hypothesis for a physical detector (for example, an atom)  is that it has to keep internal coherence as it moves. We will make the assumption that the detector is rigid (keeps its shape) in its center of mass reference frame  \cite{Schlicht2004,Louko2006,Martin-Martinez2013}. That means that the internal forces that keep the detector together  will prevent it from being further smeared due to its (possibly) non-inertial motion, up to accelerations where atomic coherence is compromised (which can be surprisingly large, see \cite{Nuestro1} and \cite{Kazantsev1974Original}). 

The interaction Hamiltonian that generates translations with respect to the atom's proper time $\tau$, in the interaction picture (notated with a prefix subindex $D$), can be written in a very compact way following the formalism of \cite{Schlicht2004,Louko2006,Nuestro1}. Namely,
\begin{equation}\label{Hamiltoniano_Unruh_DeWitt}
_{D}\hat{H}_{\II}^{\tau} = \lambda\hbar c\,\chi(\tau(t))\int \dd\bm{\xi}f(\bm{\xi})\hat{\mu}(\tau)\hat{\phi}(t(\tau,\bm{\xi}), \bm{x}(\tau,\bm{\xi})),
\end{equation}
where:
\begin{enumerate}
\item $\chi(\tau(t))$ is the switching function which we have written as an implicit function of $t$ since it will be assumed to be set in the lab's reference frame (the experimenter controls the switching). 
\item $f(\bm{\xi})$ is the spatial smearing of the detector (can be thought of as the density of the detector in its centre of mass reference frame). This generically shaped detector includes the pointlike case as the particular choice $f(\bm{\xi})=\delta(\xi)$.

\item $\hat{\mu}(\tau) = \hat{\sigma}^{+}e^{\ii\Omega\tau} + \hat{\sigma}^{-}e^{-\ii\Omega\tau}$ is the detector's mono-pole moment. $\hbar\Omega$ is the difference between the energy of the ground state and the excited state (energy gap).

\item $\hat{\phi}(t(\tau,\bm{\xi}), \bm{x}(\tau,\bm{\xi}))$ is a massless scalar field. For simplicity, the field quantization frame $(t,\bm x)$ is assumed to be inertial (for example the lab frame).
\end{enumerate}

For further detail on how the prescription of the Hamiltonian \eqref{Hamiltoniano_Unruh_DeWitt} comes from first principles in the relativistic approach to the light-matter interaction one can check \cite{Nuestro1}.

\section{Quantum friction forces for a relativistic inertial detector}
\subsection{Relativistic detector in an arbitrary trajectory}
For any point of the detector's trajectory, even if the trajectory is arbitrary, at any given point in time $t$ there is a Lorentz transformation between the inertial lab's frame ($t,\bm{x}$) and the comoving frame with the detector's centre of mass \cite{Nuestro1}  $(\tau,\bm{\xi})$ at that instant. Let us notate that instantaneous Lorentz transformation as $\Lambda_{\mu}^{\phantom{\mu}\nu}(t)$:
\begin{eqnarray}\label{Relacion_local_instantanea_referenciales}
\left(\begin{array}{c}
c\dd t\\
\dd\bm{x}
\end{array}\right)^{\nu} = \tensor{\Lambda}{_{\mu}^{\nu}}\left(\tau\right)\left(\begin{array}{c}
c\dd\tau\\
\dd\bm{\xi}
\end{array}\right)^{\mu}.
\end{eqnarray}
Where the general infinitesimal Lorentz transformation is defined as
\begin{eqnarray}\label{Transformacion_Lorentz_local_instantanea}
\Lambda(t) & = & e^{\ii\left[\bm{K}\cdot\bm{\zeta}(t) + \bm{J}\cdot\bm{\theta}(t) \right]}.
\end{eqnarray}
Each possible trajectory is defined locally by the vectors $\bm{\zeta}(t)$ and $\bm{\theta}(t)$ (instant rapidity and instantaneous rotation vector respectively) and by an initial condition. The rapidity is defined as
\begin{eqnarray}
\bm{\zeta}(t)=\bm{e_{\bm v}}(t) \operatorname{arctanh}\left(\frac{|\bm{v}(t)|}{c}\right),
\end{eqnarray}
where the unitary vector in the direction of $\bm v$ is $\bm{e_{\bm v}}(t)\coloneqq \bm{v}(t)/|\bm{v}(t)|$ and \mbox{$\bm{v}(t)\coloneqq\frac{\dd\bm{x}(t)}{\dd t}$}. 

The instantaneous rotation vector $\bm{\theta}$ is a vector whose direction is the instantaneous rotation axis and whose modulus is the instantaneous rotation angle.

The infinitesimal generators of Lorentz transformations can be represented as
\begin{eqnarray}
\tensor{{(K_{j})}}{_{\mu}^{\nu}} & = & - \ii\left( \tensor{\delta}{_{j}^{\nu}} + \tensor{\delta}{_{\mu}^{j}}\right),\\
\tensor{{(J_{j})}}{_{\mu}^{\nu}} & = &   \phantom{+}\ii\,\delta_{\mu\alpha}\epsilon^{0j\alpha\nu},
\end{eqnarray}
where the Einstein summation convention is assumed, latin indices go from 1 to 3, and Greek indices go from 0 to 3.

\subsection{General expression for arbitrary trajectories}
Substituting \Eq{Transformacion_Lorentz_local_instantanea} into \Eq{Relacion_local_instantanea_referenciales}, we get the full coordinate transformation between the inertial and non-inertial frames as
\begin{eqnarray}\label{De_General_a_Inercial_b}
x^{\mu} & = & \int_{\tau_{0}}^{\tau}\dd\tau'\tensor{\Lambda}{_{0}^{\mu}}(\tau') + \tensor{\Lambda}{_{j}^{\mu}}(\tau)\xi^{j}.
\end{eqnarray}
The trajectory of the detector's centre of mass in its own reference frame is
\begin{eqnarray}\label{Trayectoria_propia_detector}
\xi_{\textsc{cm}}^{\mu}(\tau) & = & \left(\begin{array}{c}
c\tau\\
\textbf{0}
\end{array}\right).
\end{eqnarray}
This trajectory, seen in the lab's reference frame (i.e., the trajectory of the centre of mass of the detector ---which is the origin of the coordinates $(\tau,\bm\xi)$--- as seen from the lab) is
\begin{equation}
x^{\mu}_{\textsc{cm}}(\tau) =(ct(\tau,\bm 0), \bm{x}(\tau,\bm 0)),
\end{equation}
therefore, using \Eq{Trayectoria_propia_detector}
in \Eq{De_General_a_Inercial_b}, we get the trajectory in the lab's frame as
\begin{eqnarray}
x^{\mu}_{\textsc{cm}}(\tau) & = & \int_{\tau_{0}}^{\tau}\dd\tau'\tensor{\Lambda}{_{0}^{\mu}}(\tau').
\end{eqnarray}

We need to substitute this trajectory in the interaction picture Hamiltonian \eqref{Hamiltoniano_Unruh_DeWitt}. The scalar field operator in the interaction picture can be expanded in an arbitrary basis $u_{\bm k}(t,\bm x)$ as
\begin{eqnarray}\label{Campo_Escalar}
\hat{\phi}(t,  \bm{x}) & = & \int \dd^{d} \bm{k}\left[ 
   \hat{a}_{ \bm{k}}^{\dagger}u_{ \bm{k}}(t,  \bm{x})
 + \hat{a}_{ \bm{k}}u_{ \bm{k}}^{*}(t,  \bm{x})
\right].
\end{eqnarray}
If we now express the field in terms of the proper coordinates of the detector $\xi^{\mu} = (\tau,\bm \xi)$ we obtain
\begin{eqnarray}\label{Expansion_campo}
& & \hat{\phi}(t(\tau,\bm{\xi}), \bm{x}(\tau,\bm{\xi}))\\
& &=  \int\dd^{d} \bm{k}\left[ \begin{array}{l}
   \hat{a}_{\bm{k}}^{\dagger}u_{ \bm{k}}(x^{\mu}(\tau,\bm{\xi}))
 + \hat{a}_{\bm{k}}u_{ \bm{k}}^{*}(x^{\mu}(\tau,\bm{\xi}))
\end{array}\right]\nonumber\\
& &=  \int\dd^{d} \bm{k}\left[ \begin{array}{l}
   \hat{a}_{ \bm{k}}^{\dagger}u_{ \bm{k}}\left(\displaystyle{\int_{\tau_{0}}^{\tau}}\dd\tau'\tensor{\Lambda}{_{0}^{\mu}}(\tau') + \tensor{\Lambda}{_{j}^{\mu}}(\tau)\xi^{j}\right) + \text{H.c}
\end{array}\right].\nonumber
\end{eqnarray}
In this fashion, we can now write the interaction Hamiltonian as an explicit function of the detector's centre of mass trajectory $x^{\mu}_{\textsc{cm}} (\tau)$ as
\begin{align}\label{consolega}
\nonumber & _{D}\hat{H}_{\II}^{\tau}( x_\textsc{cm}^{\mu}(\tau))=  \lambda\,\hbar\, c\chi(\tau(t))\\
&\qquad\;\times
 \int \dd\bm{\xi}f(\bm{\xi})\hat{\mu}(\tau)\hat{\phi}\left(\int_{\tau_{0}}^{\tau}\dd\tau'\tensor{\Lambda}{_{0}^{\mu}}(\tau') + \tensor{\Lambda}{_{j}^{\mu}}(\tau)\xi^{j}\right)\nonumber\\
&=  \lambda\,\hbar\, c\chi(\tau(t))\!\!
 \int \!\!\dd\bm{\xi}f(\bm{\xi})\hat{\mu}(\tau)\hat{\phi}\left(x^{\mu}_{\textsc{cm}}(\tau) + \tensor{\Lambda}{_{j}^{\mu}}(\tau)\xi^{j}\right)\!.
\end{align}
We recall that $x^{\mu}_{\textsc{cm}}(\tau) = (t(\tau),\bm{x}(\tau))$ is the trajectory of the centre of mass of the detector. The interaction Hamiltonian is the only component of the system's full Hamiltonian that depends on the trajectory, and will be the relevant part to compute the reaction force on the detector.

\subsection{Reaction force on the particle detector}
The Hamiltonian of the detector-field system, generating translations with respect to the proper time of the detector is given by
\begin{eqnarray}
_{D}\hat H^{\tau}(\tau) & = & _{D}\hat H_{d}^{\tau} + _{D\!}\hat H_{\phi}^{\tau} + _{D\!}\hat H_{\II}^{\tau}( x^{\mu}_{\textsc{cm}}(\tau)).
\end{eqnarray}
We can compute the four-force operator on the detector from Newton's second law as the derivative of the four-momentum with respect to the detector's proper time. The fastest way to compute this is through  Hamilton's equations~\cite{goldstein2002classical}, since the derivative of the four-momentum with respect to proper time is equal to the derivative of the Hamiltonian with respect to the trajectory of the detector:   
\begin{align}\label{Def_Operador_4Fuerza_v3}
\hat{F}_{\mu}(\tau) & = - \partial_{{x}^{\mu}_{\textsc{cm}}(\tau)}\hat H^{\tau}(\tau),
\end{align}
where $\partial_{{x}^{\mu}_{\textsc{cm}}(\tau)}$ is the derivative with respect to the trajectory of the detector's centre of mass. 
Since the only Hamiltonian component that depends on the trajectory of the detector is the interaction Hamiltonian, and the explicit dependence of the interaction Hamiltonian on the trajectory of the centre of mass is given in \eqref{consolega}, this yields
\begin{align}
\hat{F}_{\mu}(\tau)  &= - \partial_{{x}^{\mu}_{\textsc{cm}}(\tau)}{_{D}\hat H_{\II}^{\tau}}( x^{\mu}(\xi^{\nu})) \\
&=  - \partial_{{x}^{\mu}_{\textsc{cm}}(\tau)}\Big[ \lambda\hbar c\,\chi(\tau)\nonumber
\int \dd\bm{\xi}f(\bm{\xi})\hat{\mu}(\tau)\hat{\phi}\left(x^{\mu}(\tau,\bm{\xi})\right)\Big]. 
\end{align}
Notice that the derivative commutes with the integral over the smearing function since we are taking derivatives with respect to the trajectory of the detector ${x}^{\mu}_{\textsc{cm}}(\tau)$. We can use the mode expansion \eqref{Expansion_campo} to compute the derivative with respect to the trajectory explicitly. For simplicity let us choose, without loss of generality, a plane-wave basis:
\begin{equation}\label{Plane-Wave_basis}
u_{\bm{k}}(t,\bm{x}) = \frac{1}{\sqrt{2(2\pi)^{d}\abs{\bm{k}}}}e^{\ii (c\abs{\bm{k}} t -\bm{k} \cdot \bm{x})}.
\end{equation}
Working in this basis it is useful to realize that
\begin{align}
\partial_{{x}^{\mu}_{\textsc{cm}}(\tau)}&e^{ - \ii\,k_{\mu} \left({x}^{\mu}_{\textsc{cm}}(\tau) + \tensor{\Lambda}{_{j}^{\mu}}(\tau)\xi^{j}\right)}\nonumber
\\&= - \ii\,k_{\mu}e^{ - \ii\,k_{\mu}\left({x}^{\mu}_{\textsc{cm}}(\tau) + \tensor{\Lambda}{_{j}^{\mu}}(\tau)\xi^{j}\right)},
\end{align}
where we recall that both  $\tensor{\Lambda}{_{\nu}^{\mu}}$ and $x^\mu$ are dependent on the proper time parameter. In turn, this implies that
\begin{eqnarray}\label{Trajectory_Derived_Field}
- \partial_{{x}^{\mu}_{\textsc{cm}}(\tau)}\hat{\phi}
& = & \ii\int\dd^{d}\bm{k}\,k_{\mu}\\
& & \times\Big[ 
   \hat{a}_{\bm{k}}^{\dagger}u_{\bm{k}}(x^{\mu}(\tau,\bm{\xi})) - \hat{a}_{\bm{k}}u_{\bm{k}}^{*}(x^{\mu}(\tau,\bm{\xi}))
\Big].\nonumber
\end{eqnarray}
With all these ingredients we can now write the expression of the four-force as
\begin{align}\label{Def_Operador_4Fuerza_v2}
\nonumber \hat{F}_{\mu}(\tau)  &= - \ii\lambda\hbar c\,\chi(\tau(t))\int\dd\bm{\xi}f(\bm{\xi})\hat{\mu}(\tau)
\int\dd^{d}\bm{k}\,k_{\mu}\\
&\times\left[
\hat{a}_{ \bm{k}}^{\dagger}u_{ \bm{k}}(x^{\mu}(\tau,\bm{\xi}))
 - \hat{a}_{ \bm{k}}u_{ \bm{k}}^{*}(x^{\mu}(\tau,\bm{\xi}))
\right].
\end{align}
We would like to compute the time evolution of the expectation value of the four-force. For this, we can use leading-order perturbation theory. 

In summary, if the initial density operator of the system is an uncorrelated state of detector and field, i.e., $\hat \rho_{0} = \hat \rho_{0,\phi}\otimes\hat \rho_{0,d}$, the system will evolve to a density operator $\hat \rho_{f} = \hat{U}\hat \rho_{0}\hat{U}^{\dagger}$, where $\hat U$ is the time evolution operator in the interaction picture (and $\mathcal{T}$ represents the time ordering operation)
\begin{equation}
\hat{U} = \mathcal{T}\,\,\,\text{exp}\left( - \frac{\ii}{\hbar}\int_{-\infty}^{\infty}\dd\tau \hat H_{\II}(\tau) \right).
\end{equation}
For small enough coupling strength $\lambda$ we could consider the perturbative corrections to the initial state:
\begin{equation}
\hat \rho_{f} = \hat \rho_{0} + \hat \rho_{f}^{(1)} + \hat \rho_{f}^{(2)} + \mathcal{O}\left(\lambda^{3}\right),
\end{equation}
where the different order corrections are given by
\begin{align}
\hat \rho_{f}^{(1)} &= \hat U^{(1)}\hat{\rho}_{0} + \hat{\rho}_{0}{\hat {U}^{(1)}}{}^{\dagger},\\
\hat{\rho}_{f}^{(2)} &= \hat{U}^{(1)}\hat{\rho}_{0}{\hat{U}^{(1)}}{}^{\dagger} + \hat{U}^{(2)}\hat{\rho}_{0} + \hat{\rho}_{0}{\hat{U}^{(2)}}{}^{\dagger},\\
\nonumber\vdots
\end{align}
and where the time evolution operator has been expanded in Dyson series as
\begin{equation}
\hat U = \hat U^{(0)} + \hat U^{(1)} + \hat U^{(2)} + \mathcal{O}\left(\lambda^{3}\right),
\end{equation}
where
\begin{align}
\hat U^{(0)} &= \openone,\\
\label{Def_U1}\hat U^{(1)} &= - \frac{\ii}{\hbar} \int_{-\infty}^{\infty}\dd\tau \hat H_{\II}(\tau),\\
\hat U^{(2)} &= \frac{-1}{\hbar^{2}} \int_{-\infty}^{\infty}\dd\tau_{1}\int_{-\infty}^{\tau_{1}}\dd\tau_{2} \hat{H}_{\II}(\tau_{1})\hat{H}_{\II}(\tau_{2}),\\
&\nonumber\vdots \nonumber
\end{align}
Performing this perturbative analysis, the expectation value of the four-force, at leading order in $\lambda$, is obtained as
\begin{align}\label{Def_Operador_4Fuerza}
\mean{\hat{F}_{\mu}} & = \Tr{\hat{F}_{\mu}\hat U^{(1)}\hat \rho_{0}} + \Tr{\hat U^{(1)}\hat\rho_{0}\hat{F}_{\mu} }\nonumber\\
 & = 2\,\Real{\Tr{\hat{F}_{\mu}\hat U^{(1)}\hat \rho_{0}}}.
\end{align}
We can rewrite $\hat U^{(1)}$ and $\hat{F}_\mu$ in terms of the field mode expansion. Namely, we can substitute the mode expansion \eqref{Campo_Escalar} and \eqref{Trajectory_Derived_Field} to compute the expression for the operator $\hat{F}_{\mu} \hat{U}^{(1)}$
\begin{align}\label{Fuerza_1er_orden}
 \nonumber  \hat{F}_{\mu}& \hat{U}^{(1)}\hat{\rho}_{0} = \frac{\ii}{\hbar}\, \partial_{{x}^{\mu}(\tau)}\hat{H}_{\II}(\tau) \int_{-\infty}^{\infty}\dd \tau'\hat{H}_{\II}(\tau')\\
&= - \hbar c^{2}\lambda^{2}\chi(\tau)\int \dd^{d}\bm{\xi}f(\bm{\xi}) \int \dd^{d}\bm{\xi'}f(\bm{\xi}')
\int_{-\infty}^{\tau}\!\!\!\dd \tau'\chi(\tau')\nonumber\\
&    \times\hat{\mathcal{F}}_{\mu}(\tau,\bm{\xi},\tau',\bm{\xi}')
\hat{\mu}(\tau)\hat{\mu}(\tau')\hat{\rho}_{0,d},
\end{align}
where 
\begin{align}\label{Definicion_mathcalF}
&\hat{\mathcal{F}}_{\mu}(\tau,\bm{\xi},\tau',\bm{\xi}') \coloneqq 
\int\dd^{d}\bm{k}\,k_{\mu}\int \dd^{d} \bm{k}'\nonumber\\
 &\times\!\left[
   \hat{a}_{ \bm{k}}^{\dagger}u_{\bm{k}}(x^{\mu}(\tau,\bm{\xi}))
 - \hat{a}_{ \bm{k}}u_{\bm{k}}^{*}(x^{\mu}(\tau,\bm{\xi}))
\right]\nonumber\\
 &\times\!\Big[
  \hat{a}_{ \bm{k}'}^{\dagger}u_{\bm{k}'}(x^{\mu}(\tau',\bm{\xi}'))
+ \hat{a}_{ \bm{k}'}u_{ \bm{k}'}^{*}(x^{\mu}(\tau',\bm{\xi}'))
\Big]\hat{\rho}_{0,\phi},
\end{align}
where we have used that the initial state of the detector and the field is uncorrelated: $\hat{\rho}_{0} = \hat{\rho}_{0,\phi}\otimes\hat{ \rho}_{0,d}$. Notice that, because of causality, the expression above would not make sense for switching functions that were supported for $\tau'>\tau$. Under this constraint, \Eq{Fuerza_1er_orden} will allow us to compute the force at an instant time $\tau$ taking into account the time evolution of the atomic state from some initial preparation time (origin of the support of $\chi(\tau)$) to the time $\tau$ where the force is evaluated.

We will consider the most general possible initial detector state:
\begin{align}\label{Definicion_Matriz_Densidad_Atomo}
\hat{\rho}_{0,d} & = \begin{pmatrix}
a & b \\
b^{*} & 1-a
\end{pmatrix},
\end{align}
in the basis $\{\ket{e},\ket{g}\}$. In this basis, the monopole moment takes the form
\begin{align}
\hat{\mu}(\tau) & = \begin{pmatrix}
0 & e^{\ii \Omega \tau} \\
e^{-\ii \Omega \tau} & 0
\end{pmatrix},
\end{align}
and therefore
\begin{equation}\label{Pabla34}
\hat{\mu}(\tau)\hat{\mu}(\tau')\hat{\rho}_{0,d}=\begin{pmatrix}
a\, e^{\ii \Omega(\tau - \tau')} & b\, e^{\ii \Omega  (\tau -\tau')} \\
b^{*}e^{-\ii \Omega  (\tau -\tau')} & (1-a) e^{-\ii \Omega  (\tau -\tau')}
\end{pmatrix}.
\end{equation}

Given the form of \eqref{Pabla34} we can already conclude that, at leading order, the coherences of the quantum state of the detector, $b$, do not have any influence on the reaction force on the detector. The force however, will be influenced by the diagonal elements of the density matrix of the detector.

For the particular case where the state of the field is the vacuum $\rho_{0,\phi}=\ket{0}\!\bra{0}$, we can re-express the expectation value of $\hat{\mathcal{F}}_{\mu}$ in terms of one of the field two-point correlators. In fact, it is easy to see that (see Appendix \ref{Apendice_Calculo_Traza})
\begin{align}\label{Traza_MathcalF}
\nonumber  & \text{tr}\Big[ \hat{\mathcal{F}}_{\mu}(\tau,\bm{\xi},\tau',\bm{\xi}')\Big]\\
&=-\int\dd^{d}\bm{k}\,k_{\mu}
u_{ \bm{k}}^{*}(x^{\mu}(\tau,\bm{\xi}))
u_{ \bm{k}}(x^{\mu}(\tau',\bm{\xi}'))\nonumber\\
&=-\int\dd^{d}\bm{k}\,k_{\mu}
\mathbb{G}_{\bm{k}}(\tau, \bm{\xi}, \tau', \bm{\xi}'),
\end{align}
where $\mathbb{G}_{\bm{k}}(\mathsf{x}, \mathsf{x}')\coloneqq\bra{0} \hat{\phi}(\mathsf{x})\hat{\phi}(\mathsf{x'})\ket{0}$ is the two point correlator of the field, and $\mathsf{x}$ is a four-vector.

Putting all together, we obtain an expression for the expectation value of the force for the field vacuum and a general state of the detector as
\begin{align}\label{Fuerza_1er_ordenb}
\nonumber \mean{\hat{F}_{\mu}} \!=&2\hbar c^{2}\lambda^{2}\,\text{Re}\Bigg[\chi(\tau)\!\!\! \int_{-\infty}^{\tau}\!\!\!\!\!\!\!\dd\tau'\chi(\tau')\!
\int\dd^{d}\bm{\xi}f(\bm{\xi})\!
\int\dd^{d}\bm{\xi}'f(\bm{\xi}')\nonumber\\
&\times\int\dd^{d}\bm{k}\,k_{\mu}
\mathbb{G}_{\bm{k}}(\tau, \bm{\xi}, \tau', \bm{\xi}')\nonumber\\
&\times\left( a\, e^{\ii \Omega(\tau - \tau')}  + (1-a) e^{-\ii \Omega  (\tau -\tau')} \right)
\Bigg].
\end{align}
This is a convenient way, for computational purposes, of writing the expectation value of the four-force. Notice that this expression is general for arbitrary trajectories and for any kind of linear boundary conditions on the field. The way in which different boundary conditions are implemented in practice is through finding the specific form of $\mathbb{G}_{\bm{k}}$ through Lippmann-Schwinger, see appendix \ref{Appendix_Lippman_Schwinger} for the exact details).

Except for the fact that we couple the atom to the field as opposed to the gradient of the field, this result is a relativistic generalization of previous literature in the low speed limit \cite{Dalvitian}. In comparison with \cite{Dalvitian}, we also consider any general atomic shape and size (being the pointlike atom a particular case) and the possibility of considering the atom in an arbitrary quantum state (not only the ground state).

As explained in Appendix \ref{Appendix_Lippman_Schwinger}, the two point correlator in the presence of an object placed at a position $z=d$, can be obtained through the Lippmann-Schwinger equation, as \Eq{FGreen_Retardada_Placa}:
\begin{align}\label{Funcion_Green_Placa_Texto}
\mathbb{G}_{\bm{k}}(x_{1}, x_{2}) & = 
u^{*}(x_{1})\left[ 1 - \mathcal{R}_{\bm{k}} e^{2\ii k_{z}d} \right]u(x_{2}),
\end{align}
where $\mathcal{R}_{\bm{k}}$ is the T-scattering matrix of the object that imposes the conditions on the field. The T-scattering matrix has all the information about the geometry and the kind of boundary conditions of the considered object.

Now we have everything to write the correction of the four-force due to the presence of a plate by studying the contribution of the additional term of the two-point correlator to the four-force in \Eq{Fuerza_1er_ordenb}.

Using the plane-wave basis defined in \Eq{Plane-Wave_basis}, we can write
\begin{eqnarray}
\mathbb{G}_{\bm{k}}(\tau, \bm{\xi}, \tau', \bm{\xi}') & = & \frac{1}{2(2\pi)^{d}\abs{\bm{k}}}
e^{+\ii k_{\mu}\left( x_\textsc{cm}^{\mu}(\tau) + \tensor{\Lambda}{_{j}^{\mu}}(\tau)\xi^{j} \right)}\\
& & \hspace{-5mm}\times\left[ 1 - \mathcal{R}_{\bm{k}} e^{2\ii k_{z}d} \right]
e^{-\ii k_{\mu}\left( x^{\mu}(\tau') + \tensor{\Lambda}{_{j}^{\mu}}(\tau'){\xi'}^{j} \right)}.\nonumber
\end{eqnarray}
After substitution of the two-point function, the spatial integrals in \eqref{Fuerza_1er_ordenb} can be easily evaluated. Indeed, we can express the integrals in \Eq{Fuerza_1er_ordenb} as Fourier transforms:
\begin{eqnarray}
\int\dd^{d}\bm{\xi}'f(\bm{\xi}')
e^{-\ii k_{\mu}\tensor{\Lambda}{_{j}^{\mu}}(\tau'){\xi'}^{j}}
& = &
\bar{f}\left(k_{\mu}\tensor{\Lambda}{_{j}^{\mu}}(\tau')\right),\\
\int\dd^{d}\bm{\xi}f(\bm{\xi})
e^{+\ii k_{\mu}\tensor{\Lambda}{_{j}^{\mu}}(\tau)\xi^{j}}
& = &
\bar{f}^{*}\left(k_{\mu}\tensor{\Lambda}{_{j}^{\mu}}(\tau)\right),
\end{eqnarray}
where $\bar{f}(\bm{k})$ is the Fourier transform of $f(\bm{\xi})$:
\begin{equation}
\bar{f}(\bm k) =  \int_{\mathbb{R}^{3}}\dd^{3}\bm \xi f(\bm \xi)e^{ - \ii \bm k\cdot \bm{\xi}}.
\end{equation}
In addition to that, we define
\begin{eqnarray}
\Upsilon(\Omega, \tau) & = & 
\bar{f}^{*}\left(k_{\mu}\tensor{\Lambda}{_{j}^{\mu}}(\tau)\right)
e^{-\ii\Omega\tau}e^{+\ii k_{\mu}x_\textsc{cm}^{\mu}(\tau)}\chi(\tau)\!\!\! \\
& & \times\int_{-\infty}^{\tau}\!\!\!\!\!\!\!\dd\tau'\chi(\tau')e^{\ii\Omega\tau'}
e^{-\ii k_{\mu}x^{\mu}(\tau')}
\bar{f}\left(k_{\mu}\tensor{\Lambda}{_{j}^{\mu}}(\tau)\right),\nonumber
\end{eqnarray}
where $\Upsilon$ depends on the properties of the detector ($\Omega$ and its smearing $f(\bm{\xi})$), on the particular trajectory chosen (through $x_\textsc{cm}^{\mu}(\tau)$ and $\tensor{\Lambda}{_{j}^{\mu}}(\tau)$), on the switching $\chi(\tau)$ and on time $\tau$. For comparison with previous literature \cite{Dalvitian}, let us consider the addition of an imaginary part to the detector gap, performing the substitution $\Omega\rightarrow \Omega +\ii\Gamma$. This can be understood in terms of a dissipation term in the detector coming from a Weiskop-Wigner decay model \cite{PhysRevA.36.2763,Weisskopf1930}, or can be understood as a convenient regulator for the integrals in momentum space. The results with the usual UDW model will be recovered in the limit \mbox{$\Gamma\rightarrow0^{+}$}, when this regulator is taken to zero, as we will do later on. Introducing this regularizator $\Gamma$, the result is
\begin{align}\label{Fuerza_1er_ordend}
\mean{\hat{F}_{\mu}} \!=& \hbar c^{2}\lambda^{2}\,\text{Re}\Bigg[
\int\frac{\dd^{d}\bm{k}}{(2\pi)^{d}}\frac{k_{\mu}}{\abs{\bm{k}}}
\left[ 1 - \mathcal{R}_{\bm{k}} e^{2\ii k_{z}d} \right]
\\
&\times\Big( a\Upsilon( - (\Omega + \ii\Gamma), \tau)  + (1-a)\Upsilon( (\Omega + \ii\Gamma), \tau) \Big)
\Bigg].\nonumber
\end{align}
Then we observe that this four-force can be split in four parts. From the correlator function, it can be split in a part that depends on the distance with the object and in another part always present in the problem, even when the detector is in free space. In addition to that, the force can be written as a weighted sum of the four-force for the ground and excited states.

\subsection{Relativistic detector in an inertial trajectory}

We particularize to  an inertial particle detector that moves parallel to a dielectric plate, keeping at all times a constant distance $d$ in the $z$ axis with the plate surface. We recall that the trajectory of the detector's centre of mass in its own reference frame is
\begin{eqnarray}
\xi^{\nu}_{\textsc{cm}}(\tau) & = & \left(\begin{array}{c}
c\tau\\
\textbf{0}
\end{array}\right).
\end{eqnarray}
In this particular case, $\tensor{\Lambda}{_{\nu}^{\mu}}(\tau) = \tensor{\Lambda}{_{\nu}^{\mu}}$ with constant $\bm{v}$ for all $\tau > 0$. The Lorentz transformation that relates the quantization (lab) frame $x^{\nu} = (ct,\bm x)$ and the detector's frame \mbox{$\xi^{\mu} = (c\tau,\bm \xi)$} is given by
\begin{eqnarray}
\left(\tensor{\Lambda}{_{\nu}^{\mu}}\right) & = & \left(\begin{array}{c|c}
\gamma & \gamma\frac{v^{i}}{c}\\
\hline
\gamma\frac{v_{j}}{c} & \delta^{i}_{j} + (\gamma - 1)\frac{v^{i}v_{j}}{ \bm{v}^{2}}
\end{array}\right),
\end{eqnarray}
where $\mu\in \{0,1,2,3\}$ and  $i, j\in\{1,2,3\}$.
The transformation can be summarized as
\begin{eqnarray}
\left(\begin{array}{c}
ct(\tau,\bm{\xi})\\
 \bm{x}(\tau,\bm{\xi})
\end{array}\right)^{\mu}
& = & \tensor{\Lambda}{_{\nu}^{\mu}}\left(\begin{array}{c}
c\tau\\
\bm{0}
\end{array}\right)^{\nu} + \tensor{\Lambda}{_{\nu}^{\mu}}\left(\begin{array}{c}
0\nonumber\\
\bm{\xi}
\end{array}\right)^{\nu}\label{Trajectory7}\\
& = & x_\textsc{cm}^{\mu}(\tau) + \tensor{\Lambda}{_{i}^{\mu}}\xi^{i}\nonumber\\
& = & \tensor{\Lambda}{_{\nu}^{\mu}}\xi^{\nu}.
\end{eqnarray}
Then, for the inertial trajectory, $\tensor{\Lambda}{_{\nu}^{\mu}}$ is independent of $\tau$, and the trajectory $x_\textsc{cm}^{\mu}(\tau) = c\tensor{\Lambda}{_{0}^{\mu}}\tau$ is linear in $\tau$. As a consequence, for this particular case, $\Upsilon(\Omega, \tau)$ is simplified into
\begin{eqnarray}\label{Def:Upsilon_Inercial}
\Upsilon(\Omega, \tau) & = & 
\Det{\bar{f}\left(k_{\mu}\tensor{\Lambda}{_{j}^{\mu}}\right)}^{2}
\beta(\Omega, \tau),
\end{eqnarray}
where we define
\begin{eqnarray}
\!\!\!\!\!\!\!\beta(\Omega, \tau)\!\! & = & \!\!
e^{-\ii\left( \Omega - k_{\mu}\tensor{\Lambda}{_{0}^{\mu}}\right)\tau}\chi(\tau)\!\!\! \int_{-\infty}^{\tau}\!\!\!\!\!\!\!\dd\tau'\chi(\tau')
e^{\ii\left( \Omega - k_{\mu}\tensor{\Lambda}{_{0}^{\mu}}\right)\tau'}\!.
\end{eqnarray}
We choose as a switching function a constant switching that has been on since a time $\tau_{0}$. We define the time interval from the moment of switching on the interaction (or prepare the state of the atom) to the moment when we evaluate the force as $\Delta \tau\coloneqq\tau-\tau_{0}$. Then we obtain
\begin{eqnarray}\label{Def_beta}
\beta(\Omega, \tau) & = & 
e^{-\ii\left( \Omega - k_{\mu}\tensor{\Lambda}{_{0}^{\mu}}\right)\tau} \int_{-\tau_{0}}^{\tau}\!\!\dd\tau'
e^{\ii\left( \Omega - k_{\mu}\tensor{\Lambda}{_{0}^{\mu}}\right)\tau'}\nonumber\\
 & = & \ii\frac{ - 1 + e^{-\ii\Delta\tau \left( \Omega - ck_{\mu}\tensor{\Lambda}{_{0}^{\mu}} \right)}}{\Omega - ck_{\mu}\tensor{\Lambda}{_{0}^{\mu}}}\nonumber\\
& = & \ii\,\alpha(\Omega)
 \left( - 1 + e^{-\ii\Delta\tau \left( \Omega - ck_{\mu}\tensor{\Lambda}{_{0}^{\mu}} \right)} \right),\label{def:alpha}
\end{eqnarray}
where $\alpha(\Omega)$ is defined implicitly in \eqref{def:alpha}. As discussed above, we consider the introduction of the Weiskop-Wigner regularizator $\Gamma$, performing the substitution \mbox{$\Omega\rightarrow \Omega +\ii\Gamma$}. Applying this substitution we obtain that the distribution (over $ck_{\mu}\tensor{\Lambda}{_{0}^{\mu}}$) $\alpha(\Omega+\ii\Gamma)=\alpha_\RR +\ii\alpha_\II$ is given by
\begin{align}
\alpha_{\RR} + \ii\alpha_{\II} & \coloneqq \frac{1}{ (\Omega + \ii\Gamma) - ck_{\mu}\tensor{\Lambda}{_{0}^{\mu}} }\\
& \hspace{-1cm}= \frac{ \Omega - ck_{\mu}\tensor{\Lambda}{_{0}^{\mu}} }{ \Gamma^{2} + \big( \Omega - ck_{\mu}\tensor{\Lambda}{_{0}^{\mu}} \big)^{2} } - \ii\,\frac{\Gamma}{ \Gamma^{2} + \big( \Omega - ck_{\mu}\tensor{\Lambda}{_{0}^{\mu}} \big)^{2} }.\nonumber
\end{align}
Note that to lift the regularization we take the limit \mbox{$\Gamma\rightarrow 0^{+}$}, which yields
\begin{align}\label{Def_alpha_g}
  &\lim_{\Gamma\rightarrow 0^{+}} (\alpha_{\RR} + \ii\alpha_{\II}) \\&\qquad = \mathcal{P}\left[1/\big( \Omega - ck_{\mu}\tensor{\Lambda}{_{0}^{\mu}} \big)\right] - \ii\pi\,\text{sgn}(\Omega)\,\delta\left[ \Omega - ck_{\mu}\tensor{\Lambda}{_{0}^{\mu}} \right],\nonumber
\end{align}
where $\mathcal{P}$ in the distribution above denotes Cauchy's principal value prescription under an integral sign and sgn is the signum function.

Using \Eq{Def:Upsilon_Inercial}, and substituting these results in \Eq{Fuerza_1er_ordend}, we obtain
\begin{align}
\!\!\mean{\hat{F}_{\mu}}
& \!= \! \hbar c^{2}\lambda^{2}\,\text{Re}\Bigg[
\!\!\int\!\frac{\dd^{d}\bm{k}}{(2\pi)^{d}}\frac{k_{\mu}}{\abs{\bm{k}}}
\left[ 1 \!- \!\mathcal{R}_{\bm{k}} e^{2\ii k_{z}d} \right]
\Det{\bar{f}\left(k_{\mu}\tensor{\Lambda}{_{j}^{\mu}}\right)}^{2}\nonumber\\
& \!\!\! \times
\left( a\beta(- (\Omega + \ii\Gamma), \tau)  + (1-a)\beta(\Omega + \ii\Gamma, \tau) \right)
\Bigg].
\end{align}
We substitute $\beta(\pm(\Omega + \ii\Gamma),\tau)$ using \Eq{Def_beta}, and obtain
\begin{align}\label{Fuerza_1er_ordene}
\!\mean{\hat{F}_{\mu}}
& \!=\!  \hbar c^{2}\lambda^{2}\,\text{Re}\Bigg[\!
\int\!\frac{\dd^{d}\bm{k}}{(2\pi)^{d}}\frac{k_{\mu}}{\abs{\bm{k}}}
\left[ 1 \!-\! \mathcal{R}_{\bm{k}} e^{2\ii k_{z}d} \right]\!
\Det{\bar{f}\left(k_{\mu}\tensor{\Lambda}{_{j}^{\mu}}\right)}^{2}\nonumber\\
&  \hspace{-1cm}\times\Bigg( a\alpha(-(\Omega + \ii\Gamma)) \left( - 1 + e^{-\ii\Delta\tau \left(- (\Omega + \ii\Gamma) - ck_{\mu}\tensor{\Lambda}{_{0}^{\mu}} \right)} \right)\\
&  \hspace{-1cm} + (1-a)\alpha(\Omega + \ii\Gamma)\left( - 1 + e^{-\ii\Delta\tau \left( (\Omega + \ii\Gamma) - ck_{\mu}\tensor{\Lambda}{_{0}^{\mu}} \right)} \right)
\Bigg)
\Bigg].\nonumber
\end{align}
This expression is general and can be particularized to specific boundary conditions. We will do so for the empty space and the conducting plate case in the following sections.

\section{Force on relativistic detectors in empty space}
Let us first compute the force on the detector in the case of a detector in free space. In the empty space case, we use the free two-point correlator as
\begin{align}\label{Green_0}
\mathbb{G}^{0}_{\bm{k}}(\tau, \bm{\xi}, \tau', \bm{\xi}') = \frac{1}{(2\pi)^{3}2\Abs{\bm k}}e^{-\ii ( c\Abs{\bm k}(t - t') - \bm{k}\cdot( \bm{x} - \bm{x}' ) )},
\end{align}
where $t=t(\tau, \bm{\xi})$ and $\bm{x} = \bm{x}(\tau, \bm{\xi})$, which in the inertial case considered takes the explicit form
\begin{align}\label{tiempo_trayectoria}
t &= \gamma \left(\tau + \frac{v_{x}}{c} \xi_{1}\right)\\
\label{posicion_trayectoria}
\bm{x}& = \left(\gamma \left(\xi_{1} +\frac{v_{x}}{c} \tau \right), \xi_{2}, \xi_{3} \right)
\end{align}
where $\xi_i$, (with $i\in \{1,2,3\}$) is the $i$-th component of the vector $\bm \xi$. With this expression for $\mathbb{G}^0_{\bm k}$ at hand, we can compute the expectation of the four-force operator. Concretely, we substitute  the trajectory \eqref{tiempo_trayectoria} and \eqref{posicion_trayectoria} and the two-point correlator \eqref{Green_0} into \eqref{Fuerza_1er_ordenb} to get the following expression
\begin{align}\label{Fuerza_1er_ordenb2}
\nonumber &\mean{\hat{F}_{\mu}(\Omega)} \!=\frac{\hbar c^{2}\lambda^{2}}{8\pi^{3}}\,\text{Re}\Bigg[
\int\dd^{d}\bm{k}\,
\frac{\abs{\bar{f}(\tilde{\bm{k}})}^{2}}{\Abs{\bm k}}k_{\mu}\chi(\tau)
\int_{-\infty}^{\tau}\!\!\!\!\!\!\dd\tau'\chi(\tau')\\
& \times
e^{\ii \tau c\tilde{k}_{0}}
e^{-\ii \tau'c\tilde{k}_{0}}
\left( a\, e^{\ii \Omega(\tau - \tau')}  + (1-a) e^{-\ii \Omega  (\tau -\tau')} \right)
\Bigg].
\end{align}
Where we have evaluated $\tilde{k}^{\mu}\coloneqq k_{\nu}\tensor{\Lambda}{_{\nu}^{\mu}}=(\tilde{k}^{0},\tilde{\bm{k}})$ in \Eq{Fuerza_1er_ordenb2} using the expression of $\Lambda$ for the Lorentz boost, assuming, without loss of generality, that the detector moves in the direction of the $x$ axis, i.e., $\bm{v}= v_{x}\bm{e}_{x}$: 
\begin{align}\label{Def_tilde_momento}
\tilde{k}_{\mu}\coloneqq k_{\nu}\Lambda^{\nu}_{\phantom{\nu}\mu} & = 
\left(\begin{array}{c}
-\abs{ \bm{k}}\\
k_{x}\\
k_{y}\\
k_{z}\\
\end{array}
\right)\left(\begin{array}{c|ccc}
\gamma & \gamma\frac{v_{x}}{c} & 0 & 0\\
\hline
\gamma\frac{v_{x}}{c} & \gamma & 0 & 0\\
0 & 0 & 1 & 0\\
0 & 0 & 0 & 1
\end{array}
\right)\nonumber\\
& = 
\left(\begin{array}{c}
- \gamma\left(\abs{ \bm{k}} - \frac{v_{x}}{c}k_{x}\right)\\
\phantom{+} \gamma \left(k_{x} - \frac{v_{x}}{c}\abs{ \bm{k}}\right) \\
k_{y}\\
k_{z}\\
\end{array}
\right).
\end{align}
\begin{align}\label{Fuerza_1er_ordenc}
\nonumber \mean{\hat{F}_{\mu}(\Omega)} \!=&-\frac{\hbar c^{2}\lambda^{2}}{8\pi^{3}}\,\text{Re}\Bigg[
\int\dd^{d}\bm{k}\,
\frac{\abs{\bar{f}(\tilde{\bm{k}})}^{2}}{\Abs{\bm k}}k_{\mu}\\
& \times\ii \left( a\,\frac{\left(1-e^{-\ii \Delta \tau  (-c \tilde{k}_{0}-\Omega )}\right)}{-c \tilde{k}_{0}-\Omega }\right.\nonumber\\
& \left. + (1-a) \frac{\left(1-e^{-\ii \Delta \tau  (-c \tilde{k}_{0}+\Omega )}\right)}{-c \tilde{k}_{0}+\Omega } \right)
\Bigg].
\end{align}
Notice, however, that as discussed before with our regularization scheme, the dependence on $\Omega$ here has to be understood as a dependence on $\Omega+\ii\Gamma$, and later on the limit $\Gamma\to0^{+}$ has to be taken. Making this explicit we get
\begin{eqnarray}\label{Fuerza_1er_ordenf}
\mean{\hat{F}_{\mu}(\Omega+\ii\Gamma)} &=&-\frac{\hbar c^{2}\lambda^{2}}{8\pi^{3}}\,\text{Re}\Bigg[
\int\dd^{d}\bm{k}\,
\frac{\abs{\bar{f}(\tilde{\bm{k}})}^{2}}{\Abs{\bm k}}k_{\mu}\\
& &\hspace{-0.5cm}\times\ii \left( a\alpha(\Omega+\ii\Gamma)\,\left(1-e^{-\ii \Delta \tau  (-c \tilde{k}_{0}-(\Omega+\ii\Gamma) )}\right)\right.\nonumber\\
& &\hspace{-1.6cm}+ \left.(1-a)\alpha(-(\Omega+\ii\Gamma)) \left(1-e^{-\ii \Delta \tau  (-c \tilde{k}_{0}+\Omega+\ii\Gamma )}\right) \right)
\Bigg],\nonumber
\end{eqnarray}

Taking the limit $\Gamma\rightarrow 0^{+}$, and particularizing for the ground state ($a=0$) yields
\begin{align}\label{4_Fuerza_Ground}
&\mean{\hat{F}_{\mu}(\Omega)}^{0}_{g}  =  \frac{\hbar c^{2}\lambda^{2}}{16\pi^{3}}\int \dd^{3}\bm{k}\,\frac{[\bar f(\tilde{\bm k})]^2}{|\bm k|}k_{\mu}\\
&\times\left[
2\alpha_{\II}\sin^{2}\left(\frac{\Delta\tau}{2}\big( \Omega - c\tilde{k}_{0} \big)\right) + 
\alpha_{\RR}\sin\left(\Delta\tau\big( \Omega - c\tilde{k}_{0} \big)\right) \right]\!,\nonumber
\end{align}
where $\alpha_\RR$ and $\alpha_\II$ have been already substituted by their limit expression \eqref{Def_alpha_g} and the superindex in $\mean{\hat{F}_{\mu}(\Omega)}^0_{g}$ denotes that we are looking at the free space case.

Also, we recall that $\tilde k_\mu$ was defined in \eqref{Def_tilde_momento} and $\tilde{\bm k}$ is its spatial part. Notice that the result for the excited state can be obtained directly from \eqref{4_Fuerza_Ground} for negative gaps, i.e.,  under the change $\Omega \rightarrow -\Omega$.

To make the calculations concrete, we are going to choose a particular form for the detector smearing $F(\bm x)$. In the case of detectors modeling atoms, the smearing function is proportional to the wavefunctions of the excited and ground state orbitals (see section II of \cite{Pozas-Kerstjens2016}). Here, we are going to consider smeared detectors of size $\sigma$ localized with a Gaussian spatial profile, i.e, 
\begin{equation}
f(\bm{\xi}) = \frac{e^{-\frac{\bm{\xi}^{2}}{\sigma^{2}}}}{\pi^{3/2} \sigma^{3}}
\Rightarrow
\bar{f}(\tilde{\bm{k}})=e^{-\frac{\sigma^{2}\tilde{\bm{k}}^{2}}{2}},
\end{equation}
where
\begin{align}
\tilde{\bm{k}}^{2} &= \gamma^{2}\left( k_{x} - \frac{v_{x}}{c}\abs{\bm{k}} \right)^{2} + k_{y}^{2} + k_{z}^{2}.
\end{align}
We recall that $\tilde k_\mu$ is defined in \eqref{Def_tilde_momento}. Applying the change of variables
\begin{align}
\Abs{\bm{k}} &= \frac{\kappa}{c\gamma \left(1-\frac{v_{x}}{c}\cos(\theta)\right)}
\end{align}
to \Eq{4_Fuerza_Ground}, we carry out the angular integrals in $\theta$ and $\varphi$ yielding
\begin{align} \label{Integral_Fuerza_Ground_Caso_Libre}
    \mean{F_{x}(\Omega)}^0_{g}=& -\gamma\frac{v_{x}}{c}\frac{\hbar\lambda^{2}}{2\pi^{2}c} \nonumber\\
    &\times\int_{0}^{\infty}\dd\kappa\, e^{-\frac{\kappa^{2}\sigma^{2}}{2 c^{2}}} \kappa^{2}\frac{\sin(\Delta\tau( \kappa + \Omega ) )}{ \kappa + \Omega }.
\end{align}

For the detector in the excited state one can quickly obtain that the expectation of the four-force is given by
\begin{align}
    \mean{\hat{F}_{\mu}(\Omega)}^{0}_{e}=\mean{\hat{F}_{\mu}(-\Omega)}^{0}_{g}.
\end{align}
For any general state of the detector (pure or mixed) given by the density matrix \eqref{Definicion_Matriz_Densidad_Atomo}, and for arbitrary boundary conditions (not only the free-space case), the expectation of the four-force operator is given by
\begin{align}\label{PabloMea}
\mean{\hat{F}_{\mu}(\Omega)}_{\hat{\rho}}= (1-a)\mean{\hat{F}_{\mu}(\Omega)}_{g} + a\mean{\hat{F}_{\mu}(\Omega)}_{e}.
\end{align}
 
From \eqref{Integral_Fuerza_Ground_Caso_Libre}, we can obtain asymptotic closed expressions for the expectation of the four-force for the limits of short ($\Delta\tau\ll \Omega^{-1}$) and long times ($\Delta\tau\gg \Omega^{-1}$). The asymptotic expressions are, for a detector in its ground state,
\begin{align}
\lim_{\Omega\Delta\tau\to 0}\mean{\hat{F}_{x}}^0_{g} & = - \gamma\frac{v_{x}}{c}\frac{\hbar c^{2}\lambda^{2}}{2\sqrt{2\pi^{3}}}\frac{\Delta\tau}{\sigma^{3}}e^{-\frac{c^{2}\Delta\tau^{2}}{2\sigma^{2}}},\\*
\lim_{\Omega\Delta\tau\to\infty}\mean{\hat{F}_{x}}^0_{g} & = \gamma\frac{v_{x}}{c}\frac{\hbar \lambda^{2}}{c\,\pi^{2}}\frac{\cos(\Delta\tau\Omega)}{\Omega\Delta\tau^{3}}.\label{Ground_Free_Force_Large_times}
\end{align}
Note that the finite-size of the detector (the width of the smearing function) plays a crucial role in the force experienced by the detector in the short time regime: The force (that initially opposes the direction of motion) experienced by a pointlike detector is divergent. In fact, the finite size of the detector is what allows the expectation of the four-force to be integrable, as one can see by inspection from \eqref{Integral_Fuerza_Ground_Caso_Libre}.

For the excited case, we can also find asymptotic expressions for the expectation value of the force in the direction of motion:
\begin{align}
\lim_{\Omega\Delta\tau\to 0}\mean{\hat{F}_{x}}^0_{e} & = - \gamma\frac{v_{x}}{c}\frac{\hbar c^{2}\lambda^{2}}{2\sqrt{2\pi^{3}}}\frac{\Delta\tau}{\sigma^{3}}e^{-\frac{c^{2}\Delta\tau^{2}}{2\sigma^{2}}},\\
\lim_{\Omega\Delta\tau\to\infty}\mean{\hat{F}_{x}}^0_{e}
& = 
- \gamma\frac{v_{x}}{c}\frac{\hbar \lambda^{2}}{2\pi c}
\left( \Omega^{2}e^{-\frac{\sigma^{2}\Omega^{2}}{2c^{2}}}
+ \frac{2}{\pi}\frac{\cos(\Delta\tau\Omega)}{\Omega\Delta\tau^{3}} \right).
\end{align}
Notice that the excited and ground state four-forces are related:
\begin{align}
\lim_{\Omega\Delta\tau\to 0}\mean{\hat{F}_{x}}^0_{e} & = \lim_{\Omega\Delta\tau\to 0}\mean{\hat{F}_{x}}^0_{g},\\
\lim_{\Omega\Delta\tau\to\infty}\mean{\hat{F}_{x}}^0_{e} & = 
- \gamma\frac{v_{x}}{c}\frac{\hbar \lambda^{2}\Omega^{2}}{2\pi c}e^{-\frac{\sigma^{2}\Omega^{2}}{2c^{2}}} - \lim_{\Omega\Delta\tau\to\infty}\mean{\hat{F}_{x}}^0_{g}.
\end{align}

\begin{figure}[H]
  \centering
  \includegraphics[width=1\linewidth]{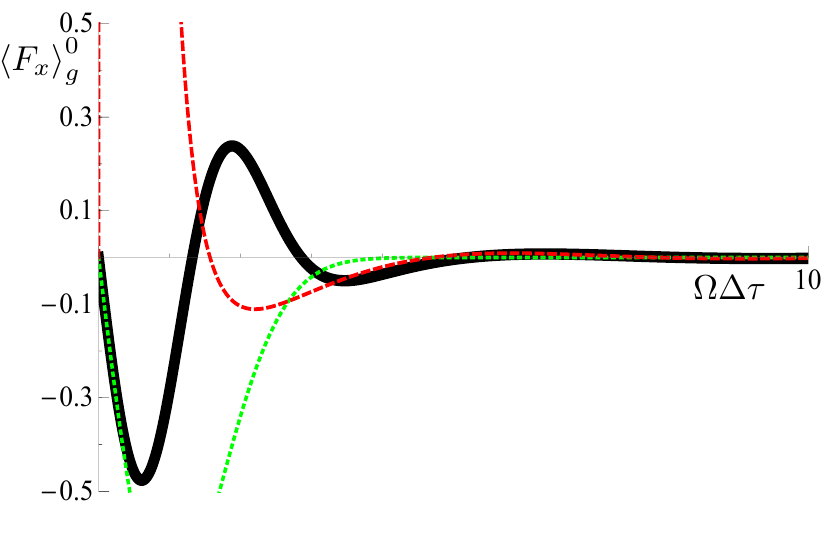}\\
  \includegraphics[width=1\linewidth]{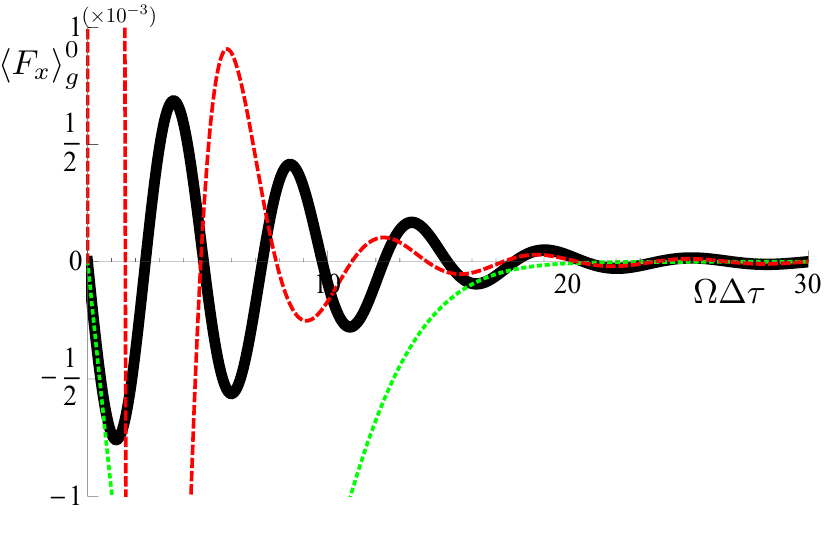}\\
\caption{ (Color online) For the detector in the ground state, friction force divided by $\frac{\hbar \lambda^{2}\Omega^{2}}{2\pi^{2}c}\gamma\frac{v_{x}}{c}$ (full thick black curve) in the direction of motion vs. $\Delta\tau$ for $\sigma\Omega/c =1$ and $\sigma\Omega/c =5$ in the upper and lower panel respectively. The short and long switching times $\Omega\Delta\tau$ limits are shown as a green dotted curve and a red dashed curve respectively. The force starts opposing the initial direction of motion, and oscillates with frequency the closer to $\Omega$ the larger $\Delta\tau\Omega$ with a decayment proportional to $\Delta\tau^{3}$.}
\label{fig:Ground_free}
\end{figure}

\begin{figure*}[t]
  \centering
\includegraphics[width=0.45\linewidth]{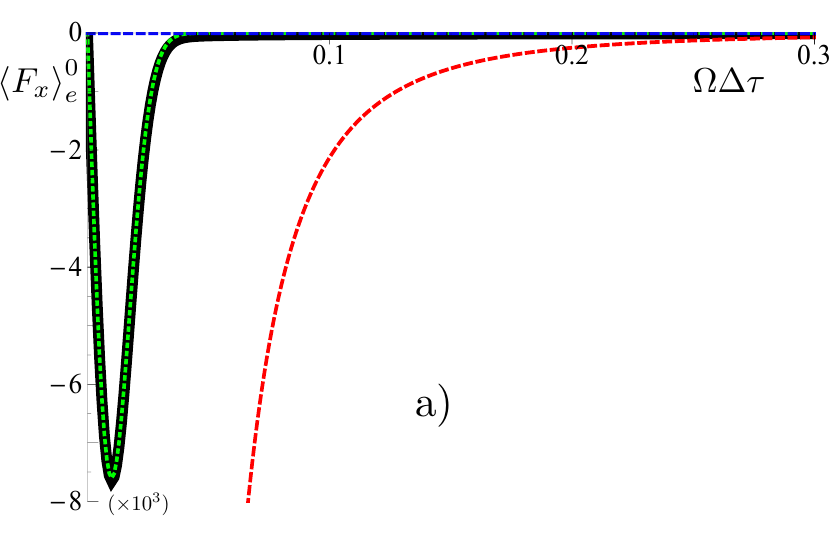}
\includegraphics[width=0.45\linewidth]{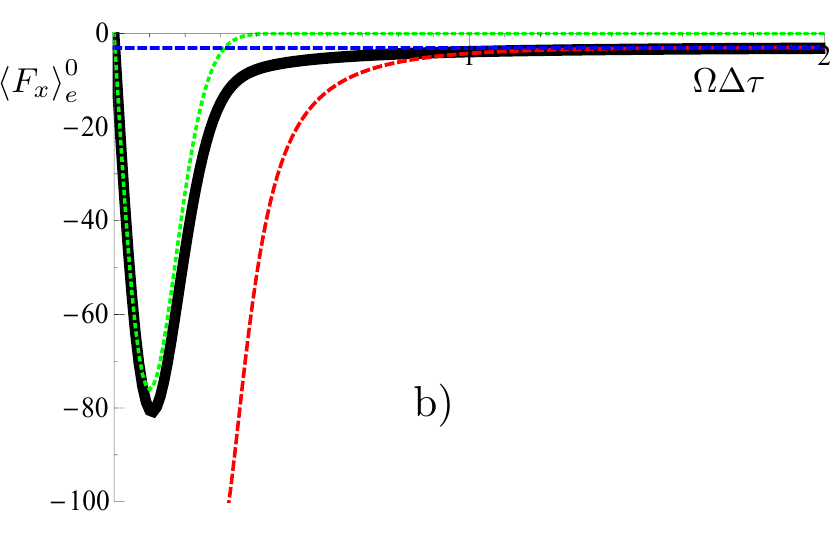}\\
\includegraphics[width=0.45\linewidth]{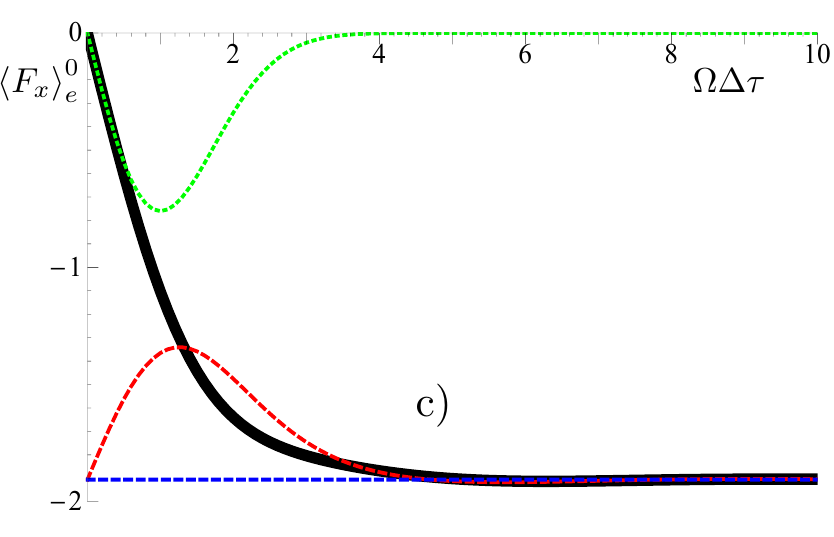}
\includegraphics[width=0.45\linewidth]{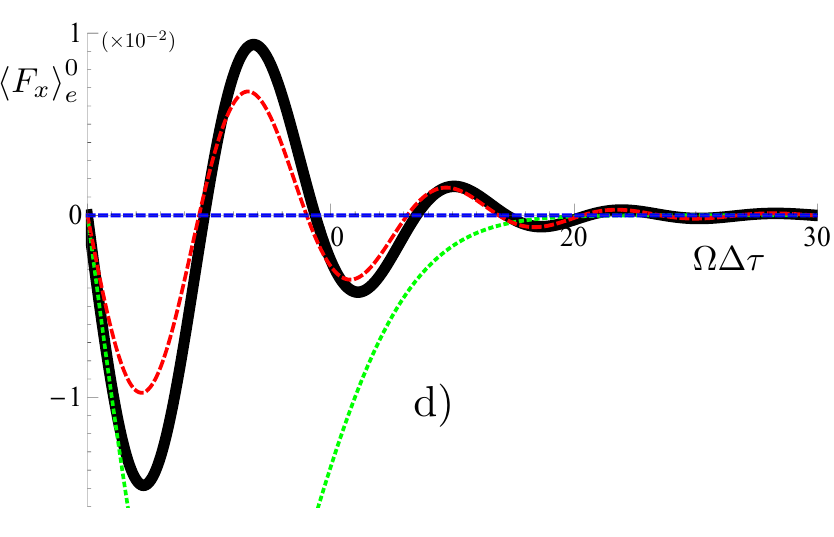}
\caption{ (Color online) For the detector in the excited state, friction force divided by $\frac{\hbar \lambda^{2}\Omega^{2}}{2\pi^{2}c}\gamma\frac{v_{x}}{c}$ (full thick black curve) in the direction of motion vs. $\Omega\Delta\tau$, for $\sigma\Omega/c \in \{10^{-2}, 10^{-1}, 1, 5\}$ for a, b, c and d panel respectively. The short and long switching times $\Omega\Delta\tau$ limits are shown as a green dotted curve and a red dashed curve respectively. For the sake of clarity, the non-oscillating part of the large switching time limit is also draw as a dashed blue curve. Depending of the value of $\Omega$, different regimes are observed: An exponential decay of the force until reaching the constant asymptotics for small $\Omega$ (observed in the cases $\sigma\Omega/c \in \{10^{-2}, 10^{-1}\}$), the disappearance of the exponential decay when the large time asymptotics surpass the maximum of the short time asymptotics for intermediate values of $\Omega$ (observed in the case $\sigma\Omega/c=1$) and the appearance of oscillations (of frequency $\Omega$) in the decay into a constant value of the force in the large $\Omega\Delta\tau$ limit (observed in the $\sigma\Omega/c=5$ case).}
\label{fig:Excited_free}
\end{figure*}

We can see the behaviour of the force experienced by the detector in Fig.~\ref{fig:Ground_free} for the ground state and in Fig.~\ref{fig:Excited_free} for the excited state.

For the ground state, there is a quantum friction that opposes the motion of the detector for short times. At longer times, the force starts oscillating between a friction force (opposing motion) and a push force (favoring motion). The frequency of oscillation is controlled by the detector's energy gap $\hbar\Omega$ (energy difference between excited and ground state). This suggests that the oscillations in the force correspond to the internal oscillations at finite times of the state of the detector between ground and excited state. These oscillations eventually decay in time as seen in \eqref{Ground_Free_Force_Large_times}. This is expected since for infinite times a detector in the ground state has a zero probability of excitation.  Consistently one should not expect any backreaction to the field in this asymptotic limit.

Finally, note from Eq. \eqref{Integral_Fuerza_Ground_Caso_Libre} that the behaviour of the force with the velocity of the particle (relative to the lab frame that sets the timescale of interaction) is proportional to $\gamma |\bm v|/c$. This tells us that the quantum friction force which opposes motion diverges as the detector approaches the speed of light.

For the excited state the phenomenology is richer. Same as in the ground state, the friction force always starts opposing motion. For most cases, the force remains a friction for all times. Only for large values of the energy gap, a detector can experience a `quantum push' force at intermediate times with the internal frequency of oscillation of the detector's internal state $\Omega$ which, might perhaps be understood as the detector emitting net momentum to the field as it decays. 

Notice that in all cases (regardless of the gap size), in the limit of infinite times, the detector always experiences a quantum friction opposing motion. Therefore we can conclude that excited detectors will always experience a net reaction force opposing motion in the very long time regime.

The excited state is particularly interesting since it displays a range of different behaviours depending on the energy difference between the excited and the ground state $\hbar\Omega$. In particular, for $\Omega \ll c\sigma$, the short time asymptote, with its exponential decay, models very well the system until the constant friction force regime is reached without the observation of any oscillation. In the regime $\Omega \gg c\sigma$, the behaviour of the force changes: at short times we still observe the initial friction force but, the exponential decay is not so prominent: the force starts oscillating with a frequency $\Omega$ until the constant friction force asymptote is reached in the long time regime. The different regimes can be seen in Fig. \ref{fig:Excited_free}.

\section{Force on detectors in the presence of a plate}\label{Seccion_Resultados_Placa}

\subsection*{General results}

In this section we are going to introduce a general boundary condition, and then particularize for an infinite plate. The formalism that we are going to introduce is valid for any kind of linear boundary conditions. This includes the typical choices of Dirichlet (e.g., perfect conductor for the electric field), Neumann or any other kind of continuity condition with the field on the other side of the plate.

In the presence of a boundary condition, the two-point correlator of the scalar field is modified by the introduction of an extra term that is derived from the Lippmann-Schwinger equation \cite{Trace_formalism_em} (see Appendix \ref{Appendix_Lippman_Schwinger}).
\begin{align}\label{PabloCaga}
\mathbb{G}_{\bm{k}} & = \mathbb{G}^{0}_{\bm{k}} + \sumint_{\bm{k}'} \mathbb{G}^{0}_{\bm{k}}\mathbb{T}_{\bm{k}\bm{k}'}\mathbb{G}^{0}_{\bm{k}'},
\end{align}
where $\mathbb{G}^{0}_{\bm k}$ is the two-point correlator for the field in free space, and $\mathbb{T}_{\bm{k}\bm{k}'}$ is the T-scattering matrix of the object that imposes the conditions on the field. Recall that the T-scattering matrix has all the information about the geometry and the kind of boundary conditions of the considered object. The symbol $\sumint$ represents the sum over the momentum variable $k'$, which in the continuum is an integral over momentum space.

In the particular case of a planar geometry (infinite plate) The T-scattering matrix is given by \cite{Trace_formalism_em,Scattering_formalism_em}
\begin{align}
\mathbb{T}_{\bm{k}\bm{k}'} & = - (2\pi)^{3}\mathcal{R}_{\bm{k}}\delta^{(3)}(\bm{k} - \bm{k}'),
\end{align}
where $\mathcal{R}_{\bm{k}}$ is the Fresnel reflection coefficient written in terms of momentum $\bm k$.

Now we have everything to write the correction of the four-force due to the presence of a plate by studying the contribution of the additional term of the two-point correlator to the four-force in \Eq{Fuerza_1er_ordenb}. We will denote this correction $\mean{\delta F_{\mu}(\Omega)}_{\hat{\rho}}$ so that
\begin{equation}
\mean{\hat{F}_{\mu}(\Omega)}_{\hat{\rho}}=\mean{\hat{F}_{\mu}(\Omega)}^0_{\hat{\rho}} + \mean{\delta F_{\mu}(\Omega)}_{\hat{\rho}}.
\end{equation}
Note that the $x$ component of the four-force will lead to the so called quantum friction\cite{Dalvitian, Buhmann, Libro_Dalvit}, and the $z$ component will lead to the dynamical extension of the Casimir force between the detector and the plate \cite{Libro_Dalvit, CasimirPolderada}.

Same as before, to evaluate the expectation of the four-force for a general state we only need to evaluate it on the ground and the excited state independently since  \eqref{PabloMea} holds.

Particularizing \eqref{PabloCaga} for a conducting plate (see \eqref{FGreen_Retardada_Placa} in Appendix \ref{Appendix_Lippman_Schwinger} for the T-matrix coefficients), and using \eqref{Fuerza_1er_ordenb}, we obtain, for the detector in the ground state

\begin{align}\label{Fuerza_ground}
\nonumber \mean{\delta F_{\mu}}_g =&  \frac{\lambda^{2}\hbar c^{2}}{8\pi^{3}}\!\int \!\dd^{3}\bm{k}\frac{k_{\mu}}{|\bm k|}e^{-\frac{\sigma^{2}\tilde{\bm{k}}^{2}}{2}}\\
&\times\left[
2A\sin^{2}\left(\frac{\Delta\tau}{2}C\right)
 + B\sin\left(\Delta\tau C\right)
\right],
\end{align}
where
\begin{align}
A = & \left( \alpha_{\II}R_{\II}V_{\II} - \alpha_{\II}R_{\RR}V_{\RR} - \alpha_{\RR}R_{\RR}V_{\II} - \alpha_{\RR}R_{\II}V_{\RR} \right),\nonumber\\
B = & \left( \alpha_{\II}R_{\RR}V_{\II} + \alpha_{\II}R_{\II}V_{\RR} + \alpha_{\RR}R_{\II}V_{\II} - \alpha_{\RR}R_{\RR}V_{\RR} \right),\nonumber\\
C = & \Omega - c \tilde{k}_{0},\nonumber\\
V = & e^{2\ii d k_{z}} = V_{\RR} + \ii V_{\II} = \cos(2d k_{z}) + \ii\sin(2d k_{z}),\nonumber\\
R = & R_{\RR} + \ii R_{\II}.
\end{align}
We see that the $y$ component of the four-force (in the lab frame) is zero. We also see that the four-force presents, for intermediate times, a transient oscillatory behaviour and that it asymptotes to a stationary value that depends on the distance to the plate and the relative velocity between the detector and the plate. 

As before, the result for the detector in the excited state is easily obtained from \Eq{Fuerza_ground} changing $\Omega$ by $-\Omega$. Also as above, for any general state of the detector (pure or mixed) and for arbitrary boundary conditions (not only the free-space case), given by the density matrix \eqref{Definicion_Matriz_Densidad_Atomo},  the expectation of the four-force operator is 
 \begin{align}\label{PabloMeaChocolate}
 \mean{\delta \hat{F}_{\mu}(\Omega)}_{\hat{\rho}}= (1-a)\mean{\delta\hat{F}_{\mu}(\Omega)}_{g} + a\mean{\delta\hat{F}_{\mu}(\Omega)}_{e},   
 \end{align}
which is analogous to the free case \eqref{PabloMea}.

Let us first consider a general case where the real and imaginary parts of the reflection coefficient are independent of the frequency. Note that this includes the physically motivated scenario of Dirichlet boundary conditions (perfect reflection) where $R_{\RR}=1$ and $R_{\II}=0$.

\subsection{Ground state}
Due to symmetry considerations in our setup, the $y$ component of the four-force (in the lab-frame) is zero $\mean{\delta F_{y}}=0$. In this section we are going to present the different results for the four-force in the different regimes studied. The full derivations can be found in Appendix \ref{Apendice_solucion_fuerzas}. 
\subsubsection{Quantum friction}
The $x$-component of the four-force is different from zero as long as the $x$-component of the relative velocity is different from zero. We consider this component of the force as a quantum friction force (because it goes opposite to the direction of motion) induced by the relative velocity with the planar plate.

\begin{figure}[H]
\centering
\includegraphics[width=1\linewidth]{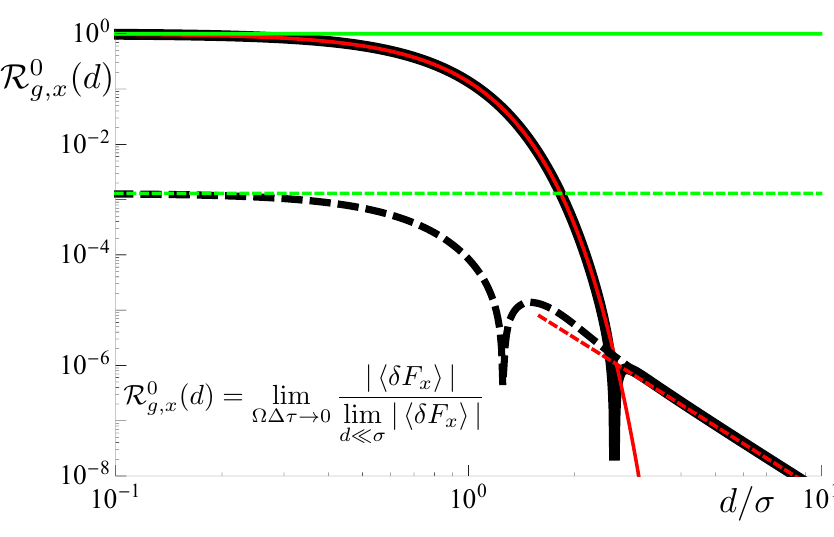}\\
\caption{ (Color online) Ratio of the magnitude of the Friction force $\mean{F_{x}}$ of the detector in the ground state $\mathcal{R}^{\Omega\Delta\tau}_{g,x}$ at $\Omega\Delta\tau \rightarrow 0$ to the value in contact $ d\ll\sigma$ for the short time limit as a function of the distance $d$ (in units of $\sigma$) at \mbox{$v_{x} =0.999 c$} (However, the figure is negligibly dependent of the value of $v_x$, even for non-relativistic speeds. This is  because we are plotting the ratio between two forces and the dependence on the velocity is always $\propto \gamma v_{x}/c$). The black thick curve is the exact numerical result obtained from \Eq{Fuerza_ground} for $\Omega\Delta\tau = 10^{-3}$. The red curve is the small distance limit shown in \Eq{Fx_Ground_plate_short_t_small_d}, the green curve is the large distance limit shown in \Eq{Fx_Ground_plate_short_t_large_d}. The dashed curves are the terms proportional to $\Delta\tau^{2}$ of the same results. The whole result is dominated by the linear term in $\Delta\tau$.
We have used $\Omega = c/\sigma$, and $R_{\RR} = R_{\II}$.}
\label{fig:Friction_Force_vs_d_short_t}
\end{figure}

\begin{figure}[H]
\centering
\includegraphics[width=1\linewidth]{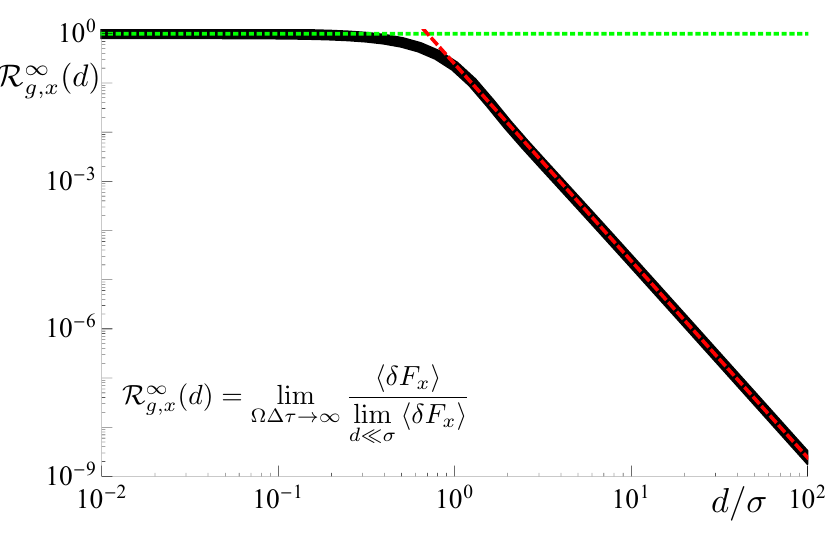}\\
\caption{ (Color online) Ratio of the Friction force $\mean{F_{x}}$ for the detector in the ground state $\mathcal{R}^{\Omega\Delta\tau}_{g,x}$ at $\Omega\Delta\tau \rightarrow \infty$ to the value in contact $ d\ll\sigma$ for the short time limit as a function of the distance $d$ (in units of $\sigma$) at \mbox{$v_{x} =0.999 c$} (However, the figure is negligibly dependent of the value of $v_x$, even for non-relativistic speeds. This is  because we are plotting the ratio between two forces and the dependence on the velocity is always $\propto \gamma v_{x}/c$). The black thick curve is the exact numerical result obtained from \Eq{Fuerza_ground} in the long switching time limit ($\Omega\Delta\tau\to\infty$). The green dotted curve is the small distance limit shown in \Eq{Fx_Ground_plate_large_t_small_d}, the red dashed curve is the large distance limit shown in \Eq{Fx_Ground_plate_large_t_large_d}. We have used $\Omega = c/\sigma$.}
\label{fig:Friction_Force_vs_d_long_t}
\end{figure}

For clarity, we summarize all the studied regimes and the formulas obtained for the final results in table \ref{TablaFrictionGround}. 
\begin{table}[H]
\begin{center}
\begin{tabular}{|c|c|c|}
\hline
\multicolumn{3}{|c|}{Ground State - Quantum Friction} \\
\hline
& $d\ll\sigma$ & $d\gg\sigma$ \\
\hline
$\Omega\Delta\tau\ll1$ & \Eq{Fx_Ground_plate_short_t_small_d} & \Eq{Fx_Ground_plate_short_t_large_d} \\
\hline
$\Omega\Delta\tau\gg1$ & \Eq{Fx_Ground_plate_large_t_small_d} & \Eq{Fx_Ground_plate_large_t_large_d} \\
\hline
\end{tabular}
\\
\caption{Different analytical limits for the $x$-component of the four-force (the quantum friction term) of the detector at the ground state for small and large switching times and distances.}
\label{TablaFrictionGround}
\end{center}
\end{table}
Notice that the dependence of the quantum friction for the planar plate case on the atomic speed is very simple and in all cases the same as in the free-space case: the friction force is proportional to $\gamma v_{x}/c$. The quantum friction for short switching times  (respectively for short and large separations from the plate) is given by
\begin{eqnarray}\label{Fx_Ground_plate_short_t_small_d}
\lim_{d\ll\sigma}\lim_{\Omega\Delta\tau\to0}\mean{\delta F_{x}} & = & - \gamma\frac{v_{x}}{c}\frac{\hbar c^{3}\Delta\tau^{2}}{\sigma^{4}}\frac{\lambda^{2}}{8\pi^{2}}R_{\II}
\left(4 \hspace{-0.5mm}+\hspace{-0.5mm} \sqrt{2\pi}\frac{\sigma\Omega}{c}\right)\nonumber\\
& & - \gamma\frac{v_{x}}{c}\frac{\hbar c^{2}\Delta\tau}{\sigma^{3}}\frac{\lambda^{2}}{(2\pi)^{3/2}}R_{\RR},
\end{eqnarray}
\begin{eqnarray}\label{Fx_Ground_plate_short_t_large_d}
\lim_{d\gg\sigma}\lim_{\Omega\Delta\tau\to0}\mean{\delta F_{x}} & = & \gamma\frac{v_{x}}{c}\frac{\hbar c^{3}\Delta\tau^{2}}{d^{4}}\frac{\lambda^{2}}{32\pi^{2}}R_{\II}\\
& & - \gamma\frac{v_{x}}{c}\frac{\hbar c^{2}\Delta\tau}{\sigma^{3}}\frac{\lambda^{2}}{(2\pi)^{3/2}}R_{\RR}e^{-\frac{2d^{2}}{\sigma^{2}}}.\nonumber
\end{eqnarray}
Conversely, in the long switching time limit, we get, for short distances from the plate,
\begin{align}\label{Fx_Ground_plate_large_t_small_d}
&\lim_{d\ll\sigma}\lim_{\Omega\Delta\tau\to\infty}\mean{\delta F_{x}} = - \frac{\hbar c}{\sigma^{2}}\frac{R_{\II}\lambda^{2}}{2\pi^{2}}\gamma\frac{v_{x}}{c}\nonumber\\
&\times\left[
1 - \sqrt{\pi}y + y^{2}\left( 2\sqrt{\pi}D(y) - e^{-y^{2}}\text{Ei}\left(y^{2}\right)\right)
\right],
\end{align}
where $y = \frac{\sigma\Omega}{\sqrt{2}c}$, $\text{Ei}(x)$ is the exponential integral function, and $D(x)$ is the Dawson integral. The expression in square brackets reduces to $1$ in the small detector limit ($\sigma\Omega\ll c$) and to 
$\frac{\sqrt{\pi}}{2y}$ in the opposite limit (when $\sigma\Omega\gg c$).
Finally, in the long switching time regime, and for long separation distances to the plate the four-force correction takes the form
\begin{eqnarray}\label{Fx_Ground_plate_large_t_large_d}
\lim_{d\gg\sigma}\lim_{\Omega\Delta\tau\to\infty}\mean{\delta F_{x}} & = & - \gamma\frac{v_{x}}{c}\frac{\hbar c^{3}}{\Omega^{2}d^{4}}\frac{\lambda^{2}}{16\pi^{2}}R_{\II}.
\end{eqnarray}
We can see the behaviour of the quantum friction force experienced by the detector in the ground state in the short time limit ($\Omega\Delta\tau\ll1$) in Fig.~\ref{fig:Friction_Force_vs_d_short_t} and in the large time limit ($\Omega\Delta\tau\gg1$) in Fig.~\ref{fig:Friction_Force_vs_d_long_t}.

\subsubsection{Casimir force}
The $z$-component of the four-force is different from zero, even at zero relative velocity between the plate and the detector. We consider this component of the force as a Casimir force (because it is parallel to the separation from the plate).

\begin{figure}[H]
\centering
\includegraphics[width=1\linewidth]{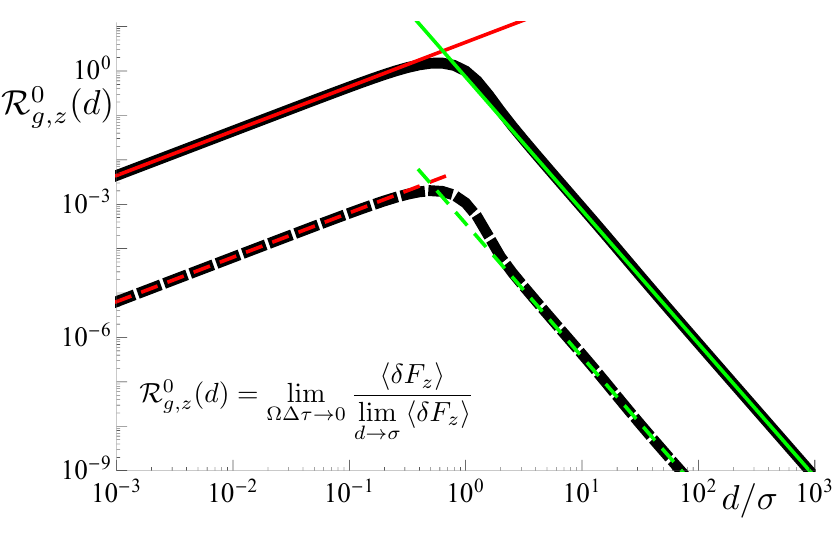}
\caption{ (Color online) Ratio of the Casimir force $\mean{F_{z}}$ for the detector in the ground state $\mathcal{R}^{\Omega\Delta\tau}_{g,z}$ at $\Omega\Delta\tau \rightarrow 0$ to the case when $ d=\sigma$ for the short time limit as a function of the distance $d$ (in units of $\sigma$) at \mbox{$v_{x} =0.999 c$} (Although in this case it is not trivial, we find that the ratio is also negligibly dependent of the value of $v_x$, even for non-relativistic speeds). The black thick curve is the exact numerical result obtained from \Eq{Fuerza_ground} for $\Omega\Delta\tau = 10^{-3}$. The red curve is the small distance limit shown in \Eq{Fz_Ground_plate_short_t_small_d}, the green curve is the large distance limit shown in \Eq{Fz_Ground_plate_short_t_large_d}. The dashed curves are the terms proportional to $\Delta\tau^{2}$ of the same results. The whole result is dominated by the linear term in $\Delta\tau$. We have used $\Omega = c/\sigma$, and $R_{\RR} = R_{\II}$.}
\label{fig:Casimir_Force_vs_d_short_t}
\end{figure}

\begin{figure}[h]
\centering
\includegraphics[width=1\linewidth]{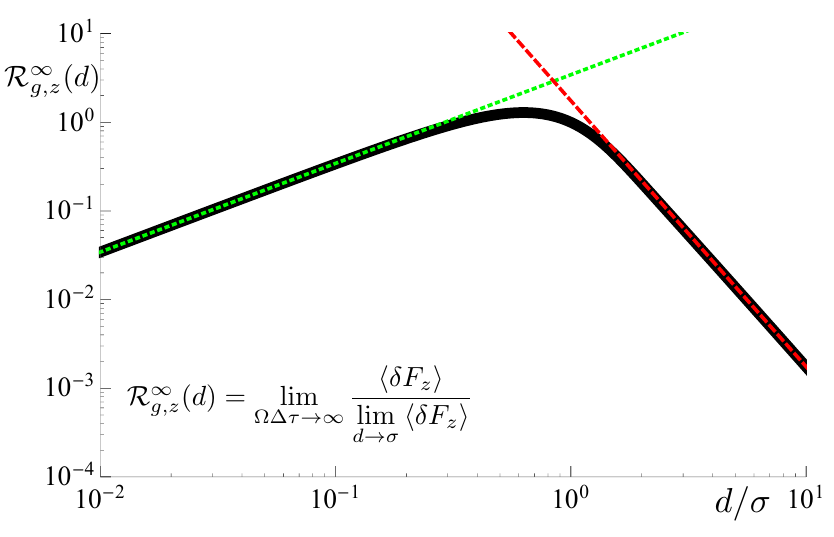}\\
\includegraphics[width=1\linewidth]{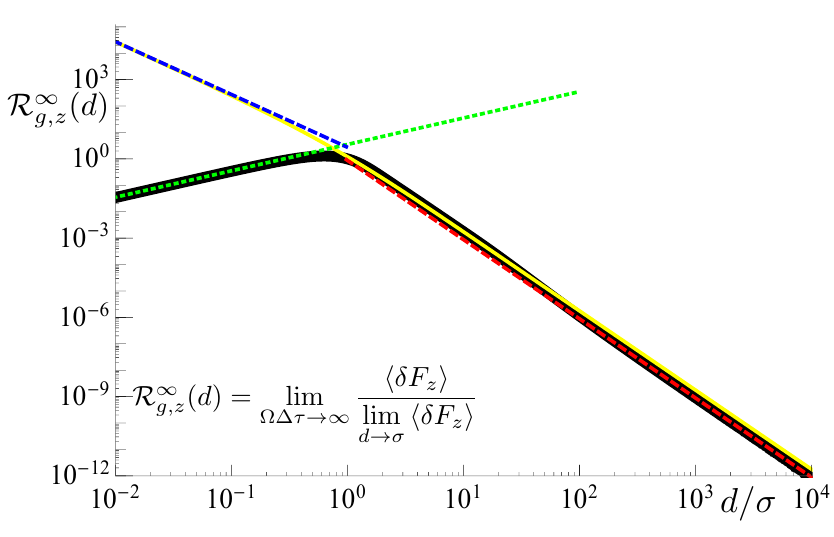}
\caption{ (Color online) Double logarithmic plot of the ratio of the Casimir force $\mean{F_{z}}$ of the detector in the ground state $\mathcal{R}^{\Omega\Delta\tau}_{g,z}$ with the case when $d=\sigma$ as a function of the distance $d$ (in units of $\sigma$) at $v_{x}=0$ (upper panel) and $v_{x} = 0.999c$ (lower panel). The black thick curve is the exact numerical result obtained from \Eq{Fuerza_ground} in the long switching time limit ($\Omega\Delta\tau\to\infty$). The green dotted curve is the small distance limit shown in \Eq{Fz_Ground_plate_large_t_small_d}, the red dashed curve is the large distance limit (in the small velocity regime) shown in \Eq{Fz_Ground_plate_large_t_large_d_small_v} (upper panel) and the large distance limit (in the high velocity regime) shown in \Eq{Fz_Ground_plate_large_t_large_d_high_v} (lower panel). Notice that in this case, the long distance regime ($d\gg\sigma$) is sensitive to the speed of the detector, unlike in all previous cases. The yellow curve is the punctual detector limit shown in \Eq{Fz_detector_puntual_small_vx}, and the blue dashed curve is the small distance limit of the punctual atom limit, showed in \Eq{Fz_detector_puntual_small_distance_limit}. We have used $\Omega = c/\sigma$.}
\label{fig:Casimir_Force_vs_d}
\end{figure}

Same as in the previous section, we summarize all the studied regimes and the formulas obtained for the final results in table \ref{TablaCasimirGround}.
\begin{table}[H]
\begin{center}
\begin{tabular}{|c|c|c|}
\hline
\multicolumn{3}{|c|}{Ground State - Casimir Force} \\
\hline
& $d\ll\sigma$ & $d\gg\sigma$ \\
\hline
$\Omega\Delta\tau\ll1$ & \Eq{Fz_Ground_plate_short_t_small_d} & \Eq{Fz_Ground_plate_short_t_large_d} \\
\hline
$\Omega\Delta\tau\gg1$ & \Eq{Fz_Ground_plate_large_t_small_d} & \Eq{Fz_Ground_plate_large_t_large_d_high_v} and \Eq{Fz_Ground_plate_large_t_large_d_small_v} \\
\hline
\end{tabular}
\\
\caption{Different analytical limits for the $z$-component of the four-force (the Casimir force term) of the detector at the ground state for small and large switching times and distances.}
\label{TablaCasimirGround}
\end{center}
\end{table}
The Casimir force for short switching times  (respectively for short and large separations from the plate) is given by
\begin{eqnarray}\label{Fz_Ground_plate_short_t_small_d}
\lim_{d\ll\sigma}\lim_{\Omega\Delta\tau\to0}\mean{\delta F_{z}} & = & - \frac{\hbar c^{3} d\Delta\tau^{2}}{\sigma^{5}}\frac{R_{\RR}\lambda^{2}}{12\pi^{2}}
\left( 3\sqrt{2\pi} + 4\frac{\sigma\Omega}{c} \right)\nonumber\\
& & + \frac{\hbar c^{2} d\Delta\tau}{\sigma^{4}}\frac{2\lambda^{2}}{3\pi^{2}}R_{\II},\\
\label{Fz_Ground_plate_short_t_large_d}
\lim_{d\gg\sigma}\lim_{\Omega\Delta\tau\to0}\mean{\delta F_{z}} & = & - \frac{\hbar c^{2}\Omega}{d^{3}}\frac{7\lambda^{2}}{128\pi^{2}}\Delta\tau^{2}R_{\RR}\nonumber\\
& & + \frac{\hbar c^{2}\Delta\tau}{d^{3}}\frac{7\lambda^{2}}{64\pi^{2}} R_{\II},
\end{eqnarray}
and in the long switching time limit, we get
\begin{align}\label{Fz_Ground_plate_large_t_small_d}
& \lim_{d\ll \sigma}\lim_{\Omega\Delta\tau\to\infty}\mean{\delta F_{z}} = - \frac{\hbar c d}{\sigma^{3}}\frac{\sqrt{2}R_{\RR}\lambda^{2}}{3\pi^{2}}\\
&\times\left[
\frac{\sqrt{\pi}}{2} - y + \sqrt{\pi}y^{2}
 + y^{3}e^{-y^{2}}\left( \text{Ei}\left(y^{2}\right) - \pi\text{erfi}(y) \right)
 \right],\nonumber
\end{align}
where $y = \frac{\sigma\Omega}{\sqrt{2}c}$, $\text{Ei}(x)$ is the exponential integral function, and $\text{erfi}(y)\coloneqq-\ii\, \text{erf}(\ii y)$ is the imaginary error function.
In the large separation distance regime, we have,
\begin{eqnarray}\label{Fz_Ground_plate_large_t_large_d_small_v}
\lim_{d\gg \sigma}\lim_{v_{x}\ll c}\lim_{\Omega\Delta\tau\to\infty}\mean{\delta F_{z}} & = & - \frac{\hbar c^{2}}{\Omega d^{3}}\frac{R_{\RR}\lambda^{2}}{8\pi^{2}},
\end{eqnarray}
\begin{eqnarray}\label{Fz_Ground_plate_large_t_large_d_high_v}
\lim_{d\gg \sigma}\lim_{v_{x}\to c}\lim_{\Omega\Delta\tau\to\infty}\mean{\delta F_{z}} & = & - \frac{\hbar c^{2}}{\Omega d^{3}}\frac{R_{\RR}\lambda^{2}}{16\pi^{2}}.
\end{eqnarray}
It is also possible to obtain an analytical result for the small detector size limit in the small velocity limit. If we make $v_{x} = 0$ and $\sigma = 0$ in \Eq{Fuerza_ground}, we get, after an analytical regularization
\begin{align}\label{Fz_detector_puntual_small_vx}
&\lim_{\sigma \to 0^{+}}\lim_{v_{x}\ll c}\lim_{\Omega\Delta\tau\to\infty}\mean{\delta F_{z}} = - \frac{\hbar c}{d^{2}}\frac{R_{\RR}\lambda^{2}}{16\pi^{2}}\\
& \times\hspace{-1mm}\Big[ \text{SI}(x) (x\sin(x) + \cos(x) )
- 2\text{Ci}(x) (x \cos(x) - \sin(x))\Big],\nonumber
\end{align}
with $x=\frac{2 d\Omega}{c}$, $\text{SI}(x)\coloneqq(\pi - 2\text{Si}(x))$, $\text{Si}(x)$ is the sine integral function and $\text{Ci}(x)$ the cosine integral function. The large distance limit of \Eq{Fz_detector_puntual_small_vx} is
\begin{eqnarray}\label{Fz_detector_puntual_large_distance_limit}
\lim_{d \gg c/\Omega}\lim_{\sigma \to 0^{+}}\lim_{v_{x}\ll c}\lim_{\Omega\Delta\tau\to\infty}\mean{\delta F_{z}} & = & - \frac{\hbar c^{2}}{\Omega d^{3}}\frac{R_{\RR}\lambda^{2}}{8\pi^{2}},
\end{eqnarray}
and the short distance limit is
\begin{eqnarray}\label{Fz_detector_puntual_small_distance_limit}
\lim_{d \ll c/\Omega}\lim_{\sigma \to 0^{+}}\lim_{v_{x}\ll c}\lim_{\Omega\Delta\tau\to\infty}\mean{\delta F_{z}} & = & - \frac{\hbar c}{d^{2}}\frac{R_{\RR}\lambda^{2}}{16\pi}.
\end{eqnarray}
The importance of the finite size of the detector is clear here, where we see that the inclusion of the finite size modifies the behaviour of the force in the short distance limit, from the divergence shown in \Eq{Fz_detector_puntual_small_vx} to a linear behaviour without spurious divergences, as seen in \Eq{Fz_Ground_plate_large_t_small_d}.
We can see the behaviour of the Casimir force experienced by the detector in ground state in the short time limit ($\Omega\Delta\tau\ll1$) in Fig.~\ref{fig:Casimir_Force_vs_d_short_t} and in the large time limit ($\Omega\Delta\tau\gg1$) in Fig.~\ref{fig:Casimir_Force_vs_d}.

\subsection{Excited state}
Again, due to symmetry considerations in our setup, the $y$ component of the four-force (in the lab-frame) is zero $\mean{\delta F_{y}}=0$. When we consider the excited state we see that we have a new contribution to the four-force: The results for the excited state can be naturally divided into two contributions arising from the decomposition \eqref{Def_alpha_g}: the first one arises from the integral of a principal part (that we will call $\mean{\delta F_{\mu}}_{\mathcal{P}}$ and was already present in the ground state case) and the second from the integral of a Dirac delta distribution (that we will call $\mean{\delta F_{\mu}}_{\delta}$), that is new in the excited case. Therefore, we can write
\begin{eqnarray}
\mean{\delta F_{\mu}} & = & \mean{\delta F_{\mu}}_{\mathcal{P}} + \mean{\delta F_{\mu}}_{\delta}.
\end{eqnarray}
\subsubsection{Quantum friction}
Same as for the ground state, the $x$-component of the four-force is different from zero as long as the relative velocity is the $x$ direction is different from zero. We also call this component `quantum friction' same as in the case of the ground state. However, notice that since the initial state of the detector is the excited state, it can cede energy and momentum to the field through spontaneous emission and it is sometimes possible in this case to get a positive force in this component (in the direction of motion instead of opposite to it). 

For clarity, we summarize all the studied regimes and the formulas obtained for the final results in tables \ref{TablaFrictionExcited_PV} and \ref{TablaFrictionExcited_delta}.
\begin{table}[H]
\begin{center}
\begin{tabular}{|c|c|c|}
\hline
\multicolumn{3}{|c|}{Excited State - Quantum Friction} \\
\hline
Principal value contribution & $d\ll\sigma$ & $d\gg\sigma$ \\
\hline
$\Omega\Delta\tau\ll1$ & \Eq{Fx_Excited_plate_short_t_small_d} & \Eq{Fx_Excited_plate_short_t_large_d} \\
\hline
$\Omega\Delta\tau\gg1$ & \Eq{Fx_P_Excited_plate_large_t_small_d} & \Eq{Fx_P_Excited_plate_large_t_large_d} \\
\hline
\end{tabular}
\\
\caption{Different analytical limits for the $x$-component of the Principal value contribution of the four-force (the quantum friction term) of the detector at the excited state for small and large switching times and distances.}
\label{TablaFrictionExcited_PV}
\end{center}
\end{table}

\begin{figure}[ht]
\centering
\includegraphics[width=1\linewidth]{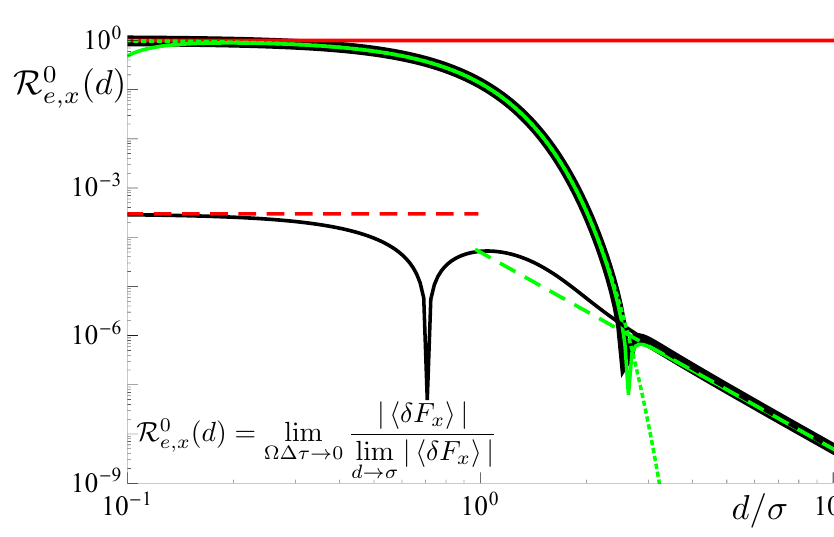}
\caption{  (Color online) Ratio of the magnitude of the Friction force $\mean{F_{x}}$ of the detector in the excited state $\mathcal{R}^{\Omega\Delta\tau}_{e,x}$ at $\Omega\Delta\tau \rightarrow 0$ to the value in contact $ d\ll\sigma$ for the short time limit as a function of the distance $d$ (in units of $\sigma$) at \mbox{$v_{x} =0.999 c$} (However, the figure is negligibly dependent of the value of $v_x$, even for non-relativistic speeds. This is  because we are plotting the ratio between two forces and the dependence on the velocity is always $\propto \gamma v_{x}/c$). The black thick curve is the exact numerical result obtained from \Eq{Fuerza_ground} for $\Omega\Delta\tau = 10^{-3}$. The red curve is the small distance limit shown in \Eq{Fx_Excited_plate_short_t_small_d}, the green curve is the large distance limit shown in \Eq{Fx_Excited_plate_short_t_large_d}. The dashed curves are the terms proportional to $\Delta\tau^{2}$ of the same results. The whole result is dominated by the linear term in $\Delta\tau$, and the dotted curves are the terms proportional to $\Delta\tau$ of the same results.
We have used $\Omega = c/\sigma$, and $R_{\RR} = R_{\II}$.}
\label{fig:Fx_Excited_vs_d_short_t}
\end{figure}

\begin{figure}[ht]
\centering
\includegraphics[width=1\linewidth]{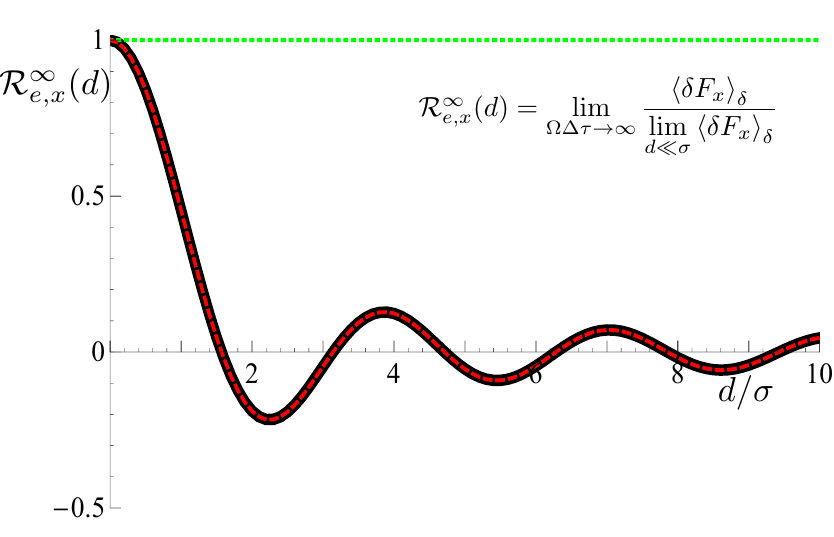}
\caption{(Color online) Ratio of the delta-contribution to the Friction force $\mean{F_{x}}_{\delta}$ for the detector in the excited state $\mathcal{R}^{\Omega\Delta\tau}_{e,x}$ at $\Omega\Delta\tau \rightarrow \infty$ to the value in contact $ d\ll\sigma$ for the short time limit as a function of the distance $d$ (in units of $\sigma$) at \mbox{$v_{x} =0.999 c$} (However, the figure is negligibly dependent of the value of $v_x$, even for non-relativistic speeds. This is  because we are plotting the ratio between two forces and the dependence on the velocity is always $\propto \gamma v_{x}/c$). The black thick curve is the exact numerical result obtained from \Eq{Fuerza_ground} in the long switching time limit ($\Omega\Delta\tau\to\infty$). The green dotted curve is the small distance limit shown in \Eq{Fx_d_Excited_plate_large_t_small_d}, the red dashed curve is the large distance limit shown in \Eq{Fx_d_Excited_plate_large_t_large_d}. We have used $\Omega = c/\sigma$.}
\label{fig:Fx_delta_Excited_vs_d}
\end{figure}

\begin{figure}[ht]
\centering
\includegraphics[width=1\linewidth]{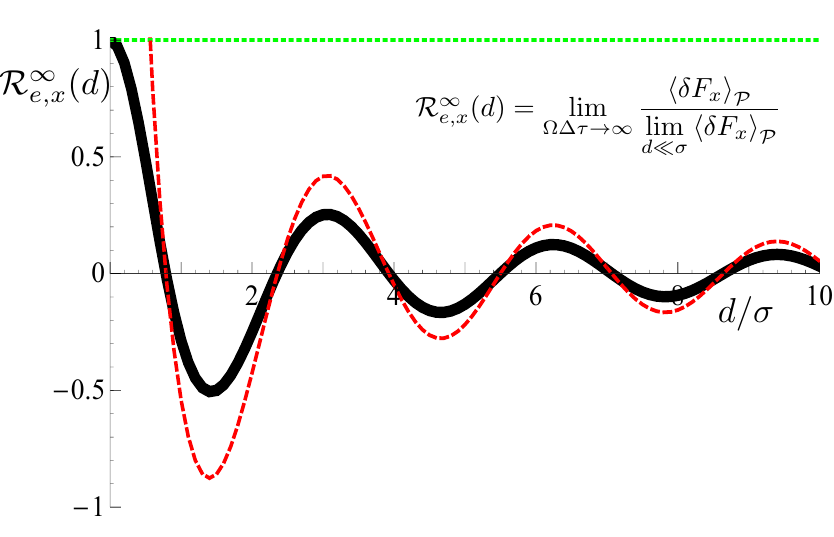}
\caption{(Color online) Ratio of the principal-value-contribution to the Friction force $\mean{F_{x}}_{\mathcal{P}}$ for the detector in the excited state $\mathcal{R}^{\Omega\Delta\tau}_{e,x}$ at $\Omega\Delta\tau \rightarrow \infty$ to the value in contact $ d\ll\sigma$ for the short time limit as a function of the distance $d$ (in units of $\sigma$) at \mbox{$v_{x} =0.999 c$} (However, the figure is negligibly dependent of the value of $v_x$, even for non-relativistic speeds. This is  because we are plotting the ratio between two forces and the dependence on the velocity is always $\propto \gamma v_{x}/c$). The black thick curve is the exact numerical result obtained from \Eq{Fuerza_ground} in the long switching time limit ($\Omega\Delta\tau\to\infty$). The green dotted curve is the small distance limit shown in \Eq{Fx_P_Excited_plate_large_t_small_d}, the red dashed curve is the large distance limit shown in \Eq{Fx_P_Excited_plate_large_t_large_d}. We have used $\Omega = c/\sigma$.}
\label{fig:Fx_P_Excited_vs_d}
\end{figure}

\begin{table}[H]
\begin{center}
\begin{tabular}{|c|c|c|}
\hline
\multicolumn{3}{|c|}{Excited State - Quantum Friction} \\
\hline
$\delta$-contribution & $d\ll\sigma$ & $d\gg\sigma$ \\
\hline
$\Omega\Delta\tau\ll1$ & 0 & 0 \\
\hline
$\Omega\Delta\tau\gg1$ & \Eq{Fx_d_Excited_plate_large_t_small_d} & \Eq{Fx_d_Excited_plate_large_t_large_d} \\
\hline
\end{tabular}
\\
\caption{Different analytical limits for the $x$-component of the four-force (the quantum friction term) of the Dirac delta contribution of the detector at the excited state for small and large switching times and distances.}
\label{TablaFrictionExcited_delta}
\end{center}
\end{table}
Again, we remark that the dependence of the quantum friction for the planar plate case on the atomic speed is very simple and in all cases the same as in the free-space case even for an excited atom: the friction force is proportional to $\gamma v_{x}/c$.  The quantum friction for short switching times  (respectively for short and large separations from the plate) is given by
\begin{eqnarray}\label{Fx_Excited_plate_short_t_small_d}
\lim_{d\ll\sigma}\lim_{\Omega\Delta\tau\to0}\mean{\delta F_{x}} & = & - \gamma\frac{v_{x}}{c}\frac{\hbar c^{2}}{\sigma^{3}}\frac{\lambda^{2}}{2\pi^{2}}\\
& & \hspace{-1cm}\times \left[ R_{\RR}\Delta\tau\sqrt{\frac{\pi}{2}} + R_{\II}\frac{c\Delta\tau^{2}}{\sigma}\left( 1 - \sqrt{\frac{\pi}{2}}\frac{\Omega\sigma}{2c} \right) \right],\nonumber
\end{eqnarray}
\begin{eqnarray}\label{Fx_Excited_plate_short_t_large_d}
\lim_{d\gg\sigma}\lim_{\Omega\Delta\tau\to0}\mean{\delta F_{x}} & = & - \gamma\frac{v_{x}}{c}\frac{\hbar c^{2}}{\sigma^{3}} \frac{\lambda^{2}}{4\pi^{2}}\\
& & \times\left[ R_{\RR}\Delta\tau\sqrt{2\pi}e^{-\frac{2d^{2}}{\sigma^{2}}} - R_{\II}\frac{c\sigma^{3}}{4d^{4}}\frac{\Delta\tau^{2}}{2} \right].\nonumber
\end{eqnarray}
For large switching times we get
\begin{eqnarray}\label{Fx_d_Excited_plate_large_t_small_d}
\lim_{d\ll\sigma}\lim_{\Delta\tau\Omega\to\infty}\mean{\delta F_{x}}_{\delta} & = & - \gamma\frac{v_{x}}{c}\frac{\hbar\Omega^{2}}{c}\frac{\lambda^{2}}{2\pi}R_{\RR}
e^{-\frac{\sigma^{2}\Omega^{2}}{2c^{2}}},
\end{eqnarray}
\begin{eqnarray}\label{Fx_P_Excited_plate_large_t_small_d}
\lim_{d\ll\sigma}\lim_{\Delta\tau\Omega\to\infty}\mean{\delta F_{x}}_{\mathcal{P}} & = & \gamma\frac{v_{x}}{c}\frac{\hbar\Omega^{2}}{c}\frac{\lambda^{2}}{2\pi}R_{\II}\\
& & \hspace{-1cm}\times G_{4,5}^{3,2}\left(\frac{\sigma ^2 \Omega ^2}{2 c^2}\left\vert
\begin{array}{c}
 -1,-\frac{1}{2},-\frac{5}{4},-\frac{3}{4} \\
 -1,-\frac{1}{2},0,-\frac{5}{4},-\frac{3}{4} \\
\end{array}
\right.\right),\nonumber
\end{eqnarray}
where $G$ is the Meijer G function. Note that this force can be either a friction or an acceleration depending on the size of the detector and on the excitation energy. Note that the expressions can easily be simplified for the pointlike limit: when $\sigma\Omega \ll c$, the Meijer G function tends to $\frac{ - c^{2}}{\pi\sigma^{2}\Omega^{2}}$, while when $\sigma\Omega \gg c$, the Meijer G function tends to $\frac{c^{3}}{\sqrt{2\pi}\sigma^{3}\Omega^{3}}$.
\begin{equation}\label{Fx_P_Excited_plate_large_t_large_d}
\hspace{-2mm}\lim_{d\gg\sigma}\lim_{\Delta\tau\Omega\to\infty}\mean{\delta F_{x}}_{\mathcal{P}} = - \gamma\frac{v_{x}}{c}\frac{\hbar\Omega}{d}\frac{\lambda^{2}}{4\pi}R_{\II}
\cos\left(\frac{2d\Omega}{c}\right),
\end{equation}
\begin{eqnarray}\label{Fx_d_Excited_plate_large_t_large_d}
\lim_{d\gg\sigma}\lim_{\Delta\tau\Omega\to\infty}\mean{\delta F_{x}}_{\delta} & = & - \gamma\frac{v_{x}}{c}\frac{\hbar\Omega}{d}\frac{\lambda^{2}}{4\pi}R_{\RR}
e^{-\frac{\sigma^{2}\Omega^{2}}{2c^{2}}}\nonumber\\
& & \times\sin\left(\frac{2d\Omega}{c}\right).
\end{eqnarray}
We can see the behaviour of the quantum friction force experienced by the detector in the excited state in the short time limit ($\Omega\Delta\tau\ll1$) in Fig.~\ref{fig:Fx_Excited_vs_d_short_t} and in the large time limit ($\Omega\Delta\tau\gg1$) in Figs.~\ref{fig:Fx_delta_Excited_vs_d} and \ref{fig:Fx_P_Excited_vs_d}.


\subsubsection{Casimir force}
\begin{figure}[h]
\centering
\includegraphics[width=1\linewidth]{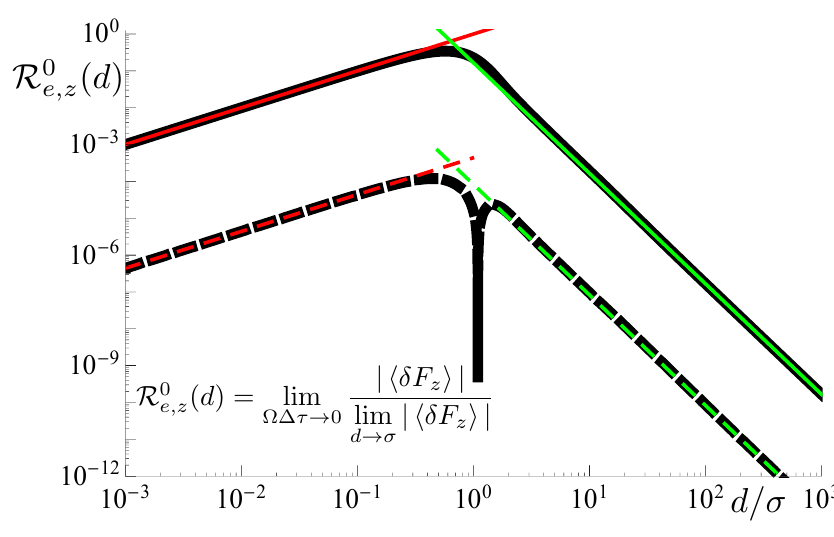}
\caption{ (Color online) Ratio of the magnitude of the Casimir force $\mean{F_{z}}$ for the detector in the excited state $\mathcal{R}^{\Omega\Delta\tau}_{e,z}$ at $\Omega\Delta\tau \rightarrow 0$ to the case when $d=\sigma$ for the short time limit as a function of the distance $d$ (in units of $\sigma$) at \mbox{$v_{x} =0.999 c$} (Although in this case it is not trivial, we find that the ratio is also negligibly dependent of the value of $v_x$, even for non-relativistic speeds). The black thick curve is the exact numerical result obtained from \Eq{Fuerza_ground} for $\Omega\Delta\tau = 10^{-3}$. The red curve is the small distance limit shown in \Eq{Fz_Excited_plate_short_t_small_d}, the green curve is the large distance limit shown in \Eq{Fz_Excited_plate_short_t_large_d}. The dashed curves are the terms proprtional to $\Delta\tau^{2}$ of the same results. The whole result is dominated by the linear term in $\Delta\tau$.
We have used $\Omega = c/\sigma$, and $R_{\RR} = R_{\II}$.}
\label{fig:Casimir_Force_Excited_vs_d_short_t}
\end{figure}

\begin{figure}[h]
\centering
\includegraphics[width=1\linewidth]{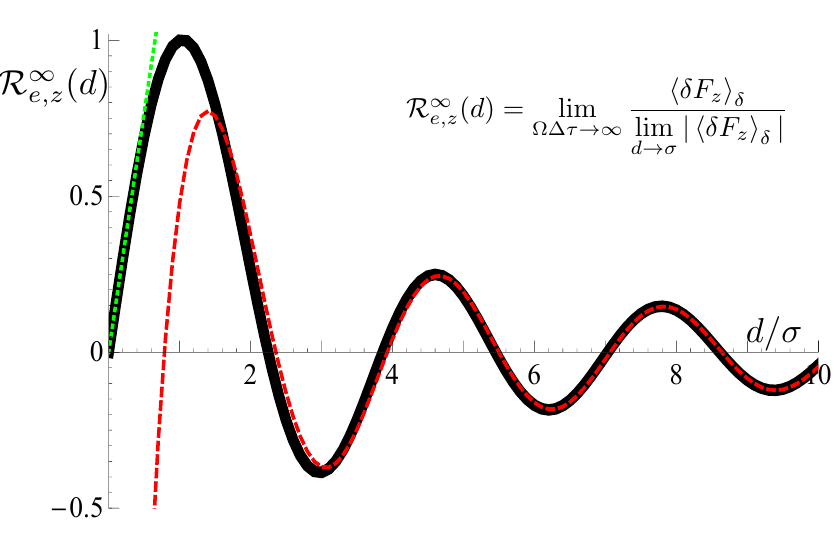}
\caption{ (Color online) Ratio of the delta-contribution to the Casimir force $\mean{F_{z}}_{\delta}$ of the detector in the excited state $\mathcal{R}^{\Omega\Delta\tau}_{e,z}$ with \Eq{Fz_d_Excited_plate_large_t_small_d} at $d=\sigma$ as a function of the distance $d$ (in units of $\sigma$) (We find that the ratio is also negligibly dependent of the value of $v_x$, even for non-relativistic speeds). The black thick curve is the exact numerical result obtained from \Eq{Fuerza_ground} in the long switching time limit ($\Omega\Delta\tau\to\infty$). The green dotted curve is the small distance limit shown in \Eq{Fz_d_Excited_plate_large_t_small_d}, the red dashed curve is the large distance limit shown in \Eq{Fz_d_Excited_plate_large_t_large_d}. We have used $\Omega = c/\sigma$.}
\label{fig:Fz_delta_Excited_vs_d_large_t}
\end{figure}

\begin{figure}[h]
\centering
\includegraphics[width=1\linewidth]{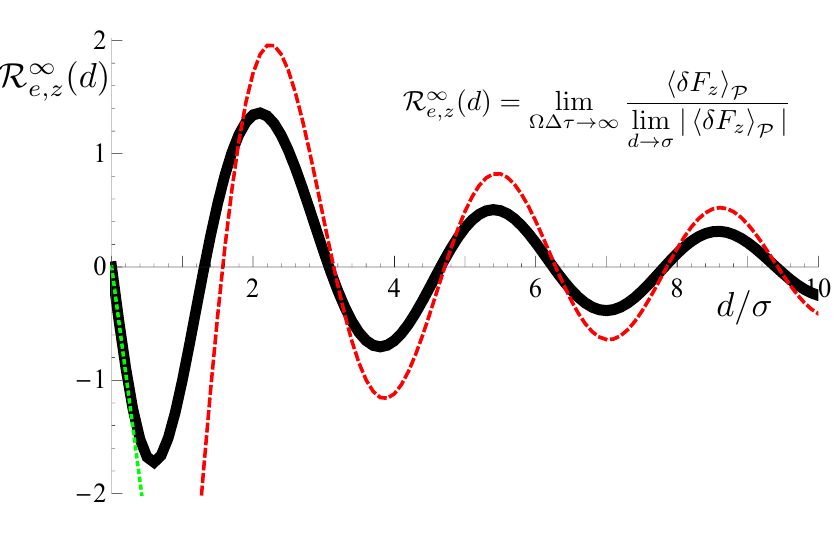}
\caption{(Color online) Ratio of the principal-value-contribution to the Casimir force $\mean{F_{z}}_{\mathcal{P}}$ of the detector in the excited state $\mathcal{R}^{\Omega\Delta\tau}_{e,z}$ with \Eq{Fz_P_Excited_plate_large_t_small_d} at $d=\sigma$ as a function of the distance $d$ (in units of $\sigma$) (We find that the ratio is also negligibly dependent of the value of $v_x$, even for non-relativistic speeds). The black thick curve is the exact numerical result obtained from \Eq{Fuerza_ground} in the long switching time limit ($\Omega\Delta\tau\to\infty$). The green dotted curve is the small distance limit shown in \Eq{Fz_P_Excited_plate_large_t_small_d}, the red dashed curve is the large distance limit shown in \Eq{Fz_P_Excited_plate_large_t_large_d}. We have used $\Omega = c/\sigma$.}
\label{fig:Fz_P_Excited_vs_d_large_t}
\end{figure}

The $z$-component of the four-force is different from zero, even at zero relative velocity between detector and plate. For clarity, we summarize all the studied regimes and the formulas obtained for the final results in tables \ref{TablaCasimirExcited_PV} and \ref{TablaCasimirExcited_delta}.
\begin{table}[H]
\begin{center}
\begin{tabular}{|c|c|c|}
\hline
\multicolumn{3}{|c|}{Excited State - Casimir Force} \\
\hline
Principal value contribution & $d\ll\sigma$ & $d\gg\sigma$ \\
\hline
$\Omega\Delta\tau\ll1$ & \Eq{Fz_Excited_plate_short_t_small_d} & \Eq{Fz_Excited_plate_short_t_large_d} \\
\hline
$\Omega\Delta\tau\gg1$ & \Eq{Fz_P_Excited_plate_large_t_small_d} & \Eq{Fz_P_Excited_plate_large_t_large_d} \\
\hline
\end{tabular}
\\
\caption{Different analytical limits for the $z$-component of the four-force (the Casimir force term) of the Principal value contribution of the detector at the excited state for small and large switching times and distances.}
\label{TablaCasimirExcited_PV}
\end{center}
\end{table}
\begin{table}[H]
\begin{center}
\begin{tabular}{|c|c|c|}
\hline
\multicolumn{3}{|c|}{Excited State - Casimir Force} \\
\hline
$\delta$-contribution & $d\ll\sigma$ & $d\gg\sigma$ \\
\hline
$\Omega\Delta\tau\ll1$ & 0 & 0 \\
\hline
$\Omega\Delta\tau\gg1$ & \Eq{Fz_d_Excited_plate_large_t_small_d} & \Eq{Fz_d_Excited_plate_large_t_large_d} \\
\hline
\end{tabular}
\\
\caption{Different analytical limits for the $z$-component of the four-force (the Casimir force term) of the Dirac delta contribution of the detector at the excited state for small and large switching times and distances.}
\label{TablaCasimirExcited_delta}
\end{center}
\end{table}
The Casimir force for short switching times  (respectively for short and large separations from the plate) is given by
\begin{align}\label{Fz_Excited_plate_short_t_small_d}
&\lim_{d\ll\sigma}\lim_{\Delta\tau\Omega\to 0}\mean{\delta F_{z}}  =  \frac{\hbar c^{2}d}{\sigma^{4}}\frac{\lambda^{2}}{3\pi^{2}}\\
&  \qquad\qquad\times\left[ 2R_{\II}\Delta\tau - R_{\RR}\frac{c\Delta\tau^{2}}{\sigma}\left( \frac{3}{4}\sqrt{2\pi} - \frac{\sigma\Omega}{c}\right)\right],\nonumber
\end{align}
\begin{align}\label{Fz_Excited_plate_short_t_large_d}
\lim_{d\gg\sigma}\lim_{\Delta\tau\Omega\to 0}\mean{\delta F_{z}}  = \hspace{-1mm}
- \frac{\hbar c^{2}}{d^{3}}\frac{7\lambda^{2}}{64\pi^{2}}
\left[ R_{\II}\Delta\tau + R_{\RR}\Omega\frac{\Delta\tau^{2}}{2}\right]\hspace{-1mm}.
\end{align}
In the regime of long switching times the force correction takes the form
\begin{eqnarray}\label{Fz_d_Excited_plate_large_t_small_d}
\lim_{d\ll\sigma}\lim_{\Delta\tau\Omega\to\infty}\mean{\delta F_{z}}_{\delta} & = & \frac{\hbar\Omega^{3}d}{c}\frac{\lambda^{2}}{3\pi}R_{\II}
e^{-\frac{\sigma^{2}\Omega^{2}}{2c^{2}}},
\end{eqnarray}
\begin{eqnarray}\label{Fz_P_Excited_plate_large_t_small_d}
\lim_{d\ll\sigma}\lim_{\Omega\Delta\tau\to\infty}\mean{\delta F_{z}}_{\mathcal{P}} & = &
\frac{\hbar \Omega^{3}d}{c^{2}}\frac{\lambda^{2}}{3\pi}R_{\RR}\\
& & \hspace{-1cm}\times G_{4,5}^{3,2}\left(\frac{\sigma ^2 \Omega ^2}{2 c^2}\left\vert
\begin{array}{c}
 -\frac{3}{2},-1,-\frac{7}{4},-\frac{5}{4} \\
 -\frac{3}{2},-1,0,-\frac{7}{4},-\frac{5}{4} \\
\end{array}
\right.
\right),\nonumber
\end{eqnarray}
where $G$ is the Meijer G function. Again, note that this force can be either attractive or repulsive depending on the size of the detector and on the excitation energy. Once again, note that the expressions can easily be simplified for the pointlike limit: when $\sigma\Omega \ll c$, the Meijer G function tends to $\frac{ - c^{3}}{\sqrt{2\pi}\sigma^{3}\Omega^{3}}$, while when $\sigma\Omega \gg c$, the Meijer G function tends to $\frac{2c^{4}}{\pi\sigma^{4}\Omega^{4}}$.
\begin{equation}\label{Fz_d_Excited_plate_large_t_large_d}
\hspace{-3mm}\lim_{d\gg\sigma}\lim_{\Delta\tau\Omega\to\infty}\mean{\delta F_{z}}_{\delta}\hspace{-1mm}  =  \hspace{-1mm}- \frac{\hbar\Omega}{d}\frac{R_{\II}\lambda^{2}}{4\pi}
e^{-\frac{\sigma^{2}\Omega^{2}}{2c^{2}}}
\cos\left(\frac{2d\Omega}{c}\right)\hspace{-1mm},
\end{equation}
\begin{equation}\label{Fz_P_Excited_plate_large_t_large_d}
\hspace{-2mm}\lim_{d \gg \sigma}\lim_{\Delta\tau\Omega\to\infty}\mean{\delta F_{z}}_{\mathcal{P}} = - \frac{\hbar\Omega}{d}\frac{R_{\RR}\lambda^{2}}{4\pi}
\sin\left(\frac{2d\Omega}{c}\right).
\end{equation}
We can see the behaviour of the Casimir force experienced by the detector in the excited state in the short time limit ($\Omega\Delta\tau\ll1$) in Fig.~\ref{fig:Casimir_Force_Excited_vs_d_short_t} and in the large time limit ($\Omega\Delta\tau\gg1$) in Figs.~\ref{fig:Fz_delta_Excited_vs_d_large_t} and \ref{fig:Fz_P_Excited_vs_d_large_t}.


\section{Conclusions}
In this article, we have developed a covariant formalism of the dynamical interaction between an arbitrary moving Unruh-DeWitt detector and an scalar quantum field in the presence of macroscopic objects even in relativistic regimes.

In particular, we have studied, at leading order in time dependent perturbation theory, the different components of the four-force operator over a particle detector (e.g., an atom) as a function of time starting from an initial time where the interaction was switched on. We have developed the general formalism to compute Casimir  forces and quantum friction dynamically, for arbitrary linear boundary conditions and arbitrary relativistic motion of the particle detector.

Furthermore, we have evaluated in full detail the expectation value of the quantum friction and Casimir forces in two particular regimes: free space, and in the presence of a parallel infinite conducting plate for relativistic constant velocity trajectories. Moreover, we have considered arbitrary initial states of the particle detector, which is treated as a fully quantum system, showing that quantum coherence does not play a relevant role in the leading order of the Casimir and quantum friction forces.

We have proved that the four-force of the detector is the weighted sum of two terms: the four-force of the detector in the ground state and the four-force of the detector in the excited state. This is true even for arbitrary superpositions of excited and ground states. 


We have also considered a spatial smearing for the detector, instead of the usual pointlike nature of the atom used in most past literature on Casimir and quantum friction. Not only this generalizes the point-like case and constitutes a more realistic model for atoms (see, e.g, \cite{Pozas-Kerstjens2016,Nuestro1}), but also the non-pointlike character of the detector avoids the presence of spurious divergent forces that were present in previous studies. Specifically, we show that in the limit of very short distance to a plate, the quantum friction tends to a constant force and the Casimir force peaks at a given distance and then goes to zero as the limit of zero distance is taken. This is in stark contrast with the divergent result that we would obtain for pointlike detectors. In addition to that, we do not find any round-trip time \cite{Armata2016} where the force diverges and change its behaviour. Indeed, as suggested in \cite{Armata2016}, the smearing solves this problem. 

As for the quantitative results, we have studied in detail the covariant expression for the force at the short and long time interaction limits in the presence of a plate. In that case, we have shown that the the quantum friction is proportional to $\gamma\frac{v_{x}}{c}$ and that, perhaps surprisingly, the Casimir force is almost independent of the relative velocity between detector and plate, except for very relativistic velocities and large distances. We have also studied the four-force for a detector in free space, when no additional object is present. Due to the form of the Lippmann-Schwinger equation, the free-space contribution to the four-force is always present and that has to be summed up to the terms that appear due to the presence of  external objects.

The formalism developed here for general relativistic trajectories is easily generalizable to the electromagnetic field (with the techniques in \cite{Pozas-Kerstjens2016,Lopp2018,Nuestro1}), more realistic models of macroscopic objects (See \cite{Scattering_formalism_em,Trace_formalism_em,Scattering_formalism_Dynamical_em2013}) and more realistic detectors, such as multilevel atoms.

\section{Acknowledgments}
E. M-M was partially funded by  the National Sciences and Engineering Research Council of Canada through the Discovery  programme and the Ontario Early Researcher award. 
The research of P.R.-L. leading to these results has received funding from the People Programme (Marie Curie Actions) of the European Union's Seventh Framework Programme (FP7/2007-2013) under REA grant agreement nº 302005 and the Spanish Grant FIS2014-52486-R. P.R.-L. also acknowledges partial support from TerMic (Grant No. FIS2014-52486-R, Spanish Government), CONTRACT (Grant No. FIS2017-83709-R, Spanish Government) and from Juan de la Cierva - Incorporacion program (Ref: I JCI-2015-25315, Spanish Government).

\begin{widetext}

\appendix

\section{Explicit calculation of the trace of \Eq{Definicion_mathcalF}}\label{Apendice_Calculo_Traza}
In this appendix, we are going to show explicitly the evaluation of the trace of \Eq{Definicion_mathcalF} when the state of the field is the vacuum $\rho_{0,\phi}=\ket{0}\!\bra{0}$ to obtain \Eq{Traza_MathcalF}. Applying the trace to \Eq{Definicion_mathcalF} with the field in the vacuum state, we have
\begin{align}\label{Traza_MathcalF_Appendix}
& \text{tr}\Big[ \hat{\mathcal{F}}_{\mu}(\tau,\bm{\xi},\tau',\bm{\xi}')\Big]\nonumber\\
& = \text{tr}\Bigg[ \int\dd^{d}\bm{k}\,k_{\mu}\int \dd^{d} \bm{k}'
\Big(
   \hat{a}_{\bm{k}}^{\dagger}u_{\bm{k}}(x^{\mu}(\tau,\bm{\xi}))
 - \hat{a}_{\bm{k}}u_{\bm{k}}^{*}(x^{\mu}(\tau,\bm{\xi}))
\Big)\Big(
  \hat{a}_{\bm{k}'}^{\dagger}u_{\bm{k}'}(x^{\mu}(\tau',\bm{\xi}'))
+ \hat{a}_{\bm{k}'}u_{ \bm{k}'}^{*}(x^{\mu}(\tau',\bm{\xi}'))
\Big)\ket{0}\!\bra{0}\Bigg]\nonumber\\
& = \text{tr}\Bigg[ \int\dd^{d}\bm{k}\,k_{\mu}\int \dd^{d} \bm{k}'
\Big(
   \hat{a}_{\bm{k}}^{\dagger}u_{\bm{k}}(x^{\mu}(\tau,\bm{\xi}))
 - \hat{a}_{\bm{k}}u_{\bm{k}}^{*}(x^{\mu}(\tau,\bm{\xi}))
\Big)
u_{\bm{k}'}(x^{\mu}(\tau',\bm{\xi}'))
\ket{1_{\bm{k}'}}\!\bra{0}\Bigg]\nonumber\\
& = \text{tr}\Bigg[ \int\dd^{d}\bm{k}\,k_{\mu}\int \dd^{d} \bm{k}'
\Big(
   \hat{a}_{\bm{k}}^{\dagger}\ket{1_{\bm{k}'}}\!\bra{0}u_{\bm{k}}(x^{\mu}(\tau,\bm{\xi}))
 - \hat{a}_{\bm{k}}\ket{1_{\bm{k}'}}\!\bra{0}u_{\bm{k}}^{*}(x^{\mu}(\tau,\bm{\xi}))
\Big)
u_{\bm{k}'}(x^{\mu}(\tau',\bm{\xi}'))
\Bigg]\nonumber\\
& = \text{tr}\Bigg[ \int\dd^{d}\bm{k}\,k_{\mu}\int \dd^{d} \bm{k}'
\Big(
   \ket{1_{\bm{k}}1_{\bm{k}'}}\!\bra{0}u_{\bm{k}}(x^{\mu}(\tau,\bm{\xi}))
 - \delta( \bm{k} - \bm{k}' )\ket{0}\!\bra{0}u_{\bm{k}}^{*}(x^{\mu}(\tau,\bm{\xi}))
\Big)
u_{\bm{k}'}(x^{\mu}(\tau',\bm{\xi}'))
\Bigg]\nonumber\\
& = \int\dd^{d}\bm{k}\,k_{\mu}\int \dd^{d} \bm{k}'
\Big(
   \text{tr}\Big[\ket{1_{\bm{k}}1_{\bm{k}'}}\!\bra{0}\Big]u_{\bm{k}}(x^{\mu}(\tau,\bm{\xi}))
 - \delta( \bm{k} - \bm{k}' )\text{tr}\Big[\ket{0}\!\bra{0}\Big]u_{\bm{k}}^{*}(x^{\mu}(\tau,\bm{\xi}))
\Big)
u_{\bm{k}'}(x^{\mu}(\tau',\bm{\xi}'))\nonumber\\
& = - \int\dd^{d}\bm{k}\,k_{\mu}\int \dd^{d} \bm{k}'
\delta( \bm{k} - \bm{k}' )u_{\bm{k}}^{*}(x^{\mu}(\tau,\bm{\xi}))u_{\bm{k}'}(x^{\mu}(\tau',\bm{\xi}'))\nonumber\\
& = - \int\dd^{d}\bm{k}\,k_{\mu}u_{\bm{k}}^{*}(x^{\mu}(\tau,\bm{\xi}))u_{\bm{k}}(x^{\mu}(\tau',\bm{\xi}')),
\end{align}
where we have used $\hat{a}_{\bm{k}}^{\dagger}\ket{0} = \ket{1_{\bm{k}}}$, $\hat{a}_{\bm{k}}\ket{1_{\bm{k}'}} = \delta^{(d)}( \bm{k} - \bm{k}' )\ket{0}$, $\text{tr}\Big[\ket{1_{\bm{k}}1_{\bm{k}'}}\!\bra{0}\Big] = 0$, $\text{tr}\Big[\ket{0}\!\bra{0}\Big] = 1$ and the Dirac delta to carry out the integral over $\bm{k}'$. This is the result shown in \Eq{Traza_MathcalF} that we wanted to prove.

\section{Derivation of the results of section \ref{Seccion_Resultados_Placa}}\label{Apendice_solucion_fuerzas}
In this appendix we are going to show the procedure to obtain the different analytical results obtained in this paper for the spatial components of the four-force. We are going to show the derivation of Eqs.~\eqref{Fz_Ground_plate_large_t_small_d}, \eqref{Fz_Ground_plate_large_t_large_d_small_v} and  \eqref{Fz_Ground_plate_large_t_large_d_high_v}. The rest of results are obtained with the same procedure.

We start from Eqs.~\eqref{Fuerza_ground} and \eqref{PabloMeaChocolate}, we apply a Taylor series in $\Delta\tau$ in order to obtain the results in the short time regime, and we apply the substitution $\sin(\Delta\tau C)\to 0$ and $[\sin(\Delta\tau C/2)]^{2}\to 1/2$ in the long switching time regime. I.e., we keep the contribution of the small frequency of the integrand.

Particularizing for the ground state of the detector, the $z$ component of the force in the lab frame, i.e., $\mu=z$ in \Eq{Fuerza_ground}, and in the long time limit $\Omega\Delta\tau\gg 1$, the Casimir force is
\begin{eqnarray}\label{Fz_Ground_plate_large_t0}
\lim_{\Omega\Delta\tau\to\infty}\mean{\delta F_{z}} & = & - \hbar c^{2}\lambda^{2}\int\frac{\dd^{3}\bm{k}}{(2\pi)^{3}}e^{-\frac{\sigma^{2}}{2}\gamma^{2}\left( k - \frac{v_{x}}{c}k_{x} \right)^{2}}\frac{k_{z}}{k}\frac{ R_{\II}\cos(2dk_{z}) + R_{\RR}\sin(2dk_{z}) }{c\gamma\left( k - \frac{v_{x}}{c}k_{x}\right) + \Omega }.
\end{eqnarray}
Let us first consider a general case where the real and imaginary parts of the reflection coefficient are independent of the frequency. Note that this includes the physically motivated scenario of Dirichlet boundary conditions (perfect reflection) where $R_{\RR}=1$ and $R_{\II}=0$.

We choose to express the integral in spherical coordinates where the $x$ axis is taken in the direction of the detector's velocity. Explicitly,
\begin{align}
k_{x} & = k\cos(\theta),\nonumber\\
k_{y} & = k\sin(\theta)\sin(\varphi),\nonumber\\
k_{z} & = k\sin(\theta)\cos(\varphi).
\end{align}
Performing the integral over $\varphi$ first yields 
\begin{eqnarray}\label{Fz_Ground_plate_large_t}
\lim_{\Omega\Delta\tau\to\infty}\mean{\delta F_{z}} & = & - \int_{0}^{\infty}\dd k\int_{0}^{\pi}\dd\theta
\frac{\hbar c^{2} \lambda^{2}}{4\pi^{2}}e^{-\frac{\sigma^{2}k^{2}\gamma^{2}}{2}\left( 1 - \frac{v_{x}}{c}\cos(\theta) \right)^{2}}k^{2}\sin^{2}(\theta) 
\frac{R_{\RR}J_{1}\left( 2d k\sin(\theta)\right)}{c k\gamma  \left( 1 - \frac{v_{x}}{c}\cos(\theta) \right) + \Omega },
\end{eqnarray}
where the dependence with $R_{\II}$ cancels out. In the study of the $x$-component, the integrand is proportional to $\sin(\theta)\cos(\theta)J_{0}(2dk\sin(\theta))$ instead to $\sin^{2}(\theta)J_{1}(2dk\sin(\theta))$. Applying the change of variables
\begin{align}
k &= \frac{s}{c \gamma\left( 1 - \frac{v_{x}}{c}\cos(\theta) \right)},
\end{align}
we get
\begin{eqnarray}\label{Fz_Ground_plate_large_tA2}
\lim_{\Omega\Delta\tau\to\infty}\mean{\delta F_{z}} & = & - \frac{\hbar}{c\gamma^{3}}\frac{\lambda^{2}}{4\pi^{2}}
\int_{0}^{\infty}\dd s\int_{0}^{\pi}\dd\theta\frac{s^{2}e^{-\frac{\sigma^{2}s^{2}}{2c^{2}}}}{s + \Omega}
R_{\RR}\frac{ \sin(\theta)^{2}}{\left( 1 - \frac{v_{x}}{c}\cos(\theta)\right)^{3} }J_{1}\left( \frac{2ds}{c\gamma}\frac{\sin(\theta)}{1 - \frac{v_{x}}{c}\cos(\theta)} \right).
\end{eqnarray}

\paragraph{Small distance limit}
To obtain the short distance limit, a Taylor expansion in $d$ of \Eq{Fz_Ground_plate_large_tA2} leads to
\begin{eqnarray}\label{Fz_Ground_plate_large_t2_small_d}
\lim_{d\ll \sigma}\lim_{\Omega\Delta\tau\to\infty}\mean{\delta F_{z}} & = & - \frac{\hbar d }{c^{2}\gamma^{4}}\frac{\lambda^{2}}{4\pi^{2}}
\int_{0}^{\infty}\dd s\int_{0}^{\pi}\dd\theta\frac{s^{3}e^{-\frac{\sigma^{2}s^{2}}{2c^{2}}}}{s + \Omega}
R_{\RR}\frac{ \sin(\theta)^{3}}{\left( 1 - \frac{v_{x}}{c}\cos(\theta)\right)^{4} }.
\end{eqnarray}
Carrying out the integral over $\theta$, we get the velocity independent result 
\begin{eqnarray}\label{Fz_Ground_plate_large_t2_small_d2}
\hspace{-5mm}\lim_{d\ll \sigma}\lim_{\Omega\Delta\tau\to\infty}\mean{\delta F_{z}} & = & - \frac{\hbar}{c^{2}} d\frac{\lambda^{2}}{3\pi^{2}}R_{\RR}
\int_{0}^{\infty}\hspace{-2mm}\dd s\frac{s^{3}e^{-\frac{\sigma^{2}s^{2}}{2c^{2}}}}{s + \Omega}.
\end{eqnarray}
This integral in $t$ admits a closed form, yielding the final result
\begin{eqnarray}\label{Fz_Ground_plate_large_t2_small_d3}
\lim_{d\ll \sigma}\lim_{\Omega\Delta\tau\to\infty}\mean{\delta F_{z}} & = & - \frac{\hbar c d}{\sigma^{3}}\frac{R_{\RR}\lambda^{2}}{3\pi^{2}}\sqrt{2}\left[
\frac{\sqrt{\pi}}{2} - y + \sqrt{\pi}y^{2}
 + y^{3}e^{-y^{2}}\left( \text{Ei}\left(y^{2}\right) - \pi\text{erfi}(y) \right)
\right],
\end{eqnarray}
where $y = \frac{\sigma\Omega}{\sqrt{2}c}$, $\text{Ei}(x)$ is the exponential integral function, and $\text{erfi}(y)\coloneqq-\ii\, \text{erf}(\ii y)$ is the imaginary error function. This is the result shown in \Eq{Fz_Ground_plate_large_t_small_d}.
\paragraph{Large distance limit}
Using \eqref{Limite_Integralz_grandes_v} from the appendix \ref{Apendice_integrales_en_theta}, and that, at large distances, the Gaussian profile can be approximated by a Dirac delta, we simplify \Eq{Fz_Ground_plate_large_tA2} into
\begin{eqnarray}\label{Fz_Ground_plate_large_t2_large}
\lim_{d\gg \sigma}\lim_{v_{x}\to c}\lim_{\Omega\Delta\tau\to\infty}\mean{\delta F_{z}} & = & \frac{\hbar}{d}\frac{\lambda^{2}}{4\pi^{2}}R_{\RR}
\int_{0}^{\infty}\dd s\frac{s}{s + \Omega}\cos\left(\frac{2ds}{c}\right).
\end{eqnarray}
This is a divergent integral, but can be solved by an analytical continuation of a convergent integral as
\begin{eqnarray}\label{Fz_Ground_plate_large_t2_large2}
\lim_{d\gg \sigma}\lim_{v_{x}\to c}\lim_{\Omega\Delta\tau\to\infty}\mean{\delta F_{z}} & = & \frac{\hbar }{d}\frac{\lambda^{2}}{4\pi^{2}}R_{\RR}\frac{c}{2}\partial_{d}
\int_{0}^{\infty}\dd s\frac{\sin\left(\frac{2ds}{c}\right)}{s + \Omega}.
\end{eqnarray}
After carrying out this integral, the large distance limit is obtained as
\begin{eqnarray}\label{Fz_Ground_plate_large_t2_large_d}
\lim_{d\gg \sigma}\lim_{v_{x}\to c}\lim_{\Omega\Delta\tau\to\infty}\mean{\delta F_{z}} & = & - \frac{\hbar c^{2}}{\Omega d^{3}}\frac{R_{\RR}\lambda^{2}}{16\pi^{2}}.
\end{eqnarray}
This is the result shown in \Eq{Fz_Ground_plate_large_t_large_d_high_v}.
\paragraph{Small velocity limit}
In this particular limit, we apply a Taylor expansion to \Eq{Fz_Ground_plate_large_tA2} in $v_{x}$ around $v_{x}=0$. We will show in appendix \ref{Apendice_integrales_en_theta} that, in the small velocity limit, the dominant contribution is independent of $v_{x}$, and equal to
\begin{eqnarray}\label{Fz_Ground_plate_large_t_small_vx}
\lim_{v_{x}\ll c}\lim_{\Omega\Delta\tau\to\infty}\mean{\delta F_{z}} & = & - \frac{\hbar}{c}\frac{\lambda^{2}}{4\pi^{2}}
\int_{0}^{\infty}\dd s\frac{s^{2}e^{-\frac{\sigma^{2}s^{2}}{2c^{2}}}}{s + \Omega}
R_{\RR}
\left[ \frac{ \sin\left(\frac{2ds}{c}\right) - 2\frac{ds}{c}\cos\left(\frac{2ds}{c}\right)}{2\left(\frac{ds}{c}\right)^{2}} \right].
\end{eqnarray}
Since the short distance limit obtained in \Eq{Fz_Ground_plate_large_t2_small_d3} is valid for all velocities, we do not need to repeat the calculation here. In contrast, the high distance limit obtained in \Eq{Fz_Ground_plate_large_t2_large_d} is valid for high velocities, therefore the result at small velocities will not be the same and needs to be computed. In the large distance limit, the Gaussian profile can be approached by a Dirac delta, then we obtain
\begin{eqnarray}\label{Fz_Ground_plate_large_t_small_vx_high_d}
\lim_{d\gg\sigma}\lim_{v_{x}\ll c}\lim_{\Omega\Delta\tau\to\infty}\mean{\delta F_{z}} & = & - \frac{\hbar c}{d^{2}}\frac{\lambda^{2}}{8\pi^{2}}R_{\RR}
\int_{0}^{\infty}\dd s
\frac{ \sin\left(\frac{2ds}{c}\right) - 2\frac{ds}{c}\cos\left(\frac{2ds}{c}\right)}{s + \Omega}.
\end{eqnarray}
This is a divergent integral, but can be regularized by an analytical continuation of a convergent integral as
\begin{eqnarray}\label{Fz_Ground_plate_large_t_small_vx_high_d2}
\lim_{d\gg\sigma}\lim_{v_{x}\ll c}\lim_{\Omega\Delta\tau\to\infty}\mean{\delta F_{z}} 
& = & - \frac{\hbar c}{d^{2}}\frac{\lambda^{2}}{8\pi^{2}}R_{\RR}
\left[ 1 - d\partial_{d}\right]
\int_{0}^{\infty}\dd s
\frac{ \sin\left(\frac{2ds}{c}\right)}{s + \Omega}.
\end{eqnarray}
After carrying out this integral, the large distance limit is obtained as
\begin{eqnarray}
\lim_{d\gg\sigma}\lim_{v_{x}\ll c}\lim_{\Omega\Delta\tau\to\infty}\mean{\delta F_{z}} 
& = & - \frac{\hbar c^{2}}{\Omega d^{3}}\frac{\lambda^{2}}{8\pi^{2}}R_{\RR}.
\end{eqnarray}
Note that we have obtained an analytical result for the small detector case as
\begin{eqnarray}\label{Fz_Ground_plate_large_t_small_vx_punctual_atom}
\lim_{\sigma\to 0}\lim_{v_{x}\ll c}\lim_{\Omega\Delta\tau\to\infty}\mean{\delta F_{z}} & = & - \frac{\hbar c}{d^{2}}\frac{\lambda^{2}}{8\pi^{2}}R_{\RR}\left[
\text{Ci}(x) \left(\sin(x) - x\cos(x) \right) + \left(\frac{\pi }{2}-\text{Si}(x)\right) \left( x\sin(x) + \cos(x) \right)
\right],
\end{eqnarray}
with $x=\frac{2d\Omega}{c}$. The large distance limit is, therefore
\begin{eqnarray}\label{Fz_Ground_plate_large_t_small_vx_high_d3}
\lim_{d\gg\sigma}\lim_{v_{x}\ll c}\lim_{\Omega\Delta\tau\to\infty}\mean{\delta F_{z}} & = & - \frac{\hbar c^{2}}{\Omega d^{3}}\frac{\lambda^{2}}{8\pi^{2}}R_{\RR},
\end{eqnarray}
and the small distance limit of the pointlike detector is
\begin{eqnarray}\label{Fz_Ground_plate_large_t_small_vx_punctual_atom_small_d}
\lim_{d\ll c\Omega}\lim_{\sigma\to 0}\lim_{v_{x}\ll c}\lim_{\Omega\Delta\tau\to\infty}\mean{\delta F_{z}} & = & - \frac{\hbar c}{d^{2}}\frac{\lambda^{2}}{16\pi}R_{\RR}.
\end{eqnarray}

\section{Different limits of the angular integral in \texorpdfstring{$\theta$}{theta}}\label{Apendice_integrales_en_theta}
In this appendix we are going to obtain the different asymptotic results of
\begin{eqnarray}
I_{0} = \int_{0}^{\pi}\dd\theta\frac{ \sin(\theta)\cos(\theta)}{\left( 1 - \frac{v_{x}}{c}\cos(\theta)\right)^{3} }J_{0}\left( \frac{2dt}{\gamma}\frac{\sin(\theta)}{1 - \frac{v_{x}}{c}\cos(\theta)} \right),
\end{eqnarray}
\begin{eqnarray}
I_{1} = \int_{0}^{\pi}\dd\theta\frac{ \sin^{2}(\theta)}{\left( 1 - \frac{v_{x}}{c}\cos(\theta)\right)^{3} }J_{1}\left( \frac{2dt}{\gamma}\frac{\sin(\theta)}{1 - \frac{v_{x}}{c}\cos(\theta)} \right),
\end{eqnarray}
used in section \ref{Seccion_Resultados_Placa}.
\subsection{Limit of small velocities}
Up to the linear term in the small velocities limit, we have
\begin{eqnarray}
\lim_{v_{x}\ll c}I_{0} & = & \int_{0}^{\pi}\dd\theta\sin(\theta)\cos(\theta)\left[ 
J_{0}\left( 2dt\sin(\theta)\right)
 - \frac{v_{x}}{c}\left( 3 J_{0}(2dt\sin(\theta)) - 2dt\sin(\theta)J_{1}(2dt\sin(\theta))\right)
 + \dots\right],
\end{eqnarray}
\begin{eqnarray}
\lim_{v_{x}\ll c}I_{1} & = & \int_{0}^{\pi}\dd\theta\sin^{2}(\theta)\left[ 
J_{1}\left( 2dt\sin(\theta)\right)
 - 2\frac{v_{x}}{c} \cos(\theta)\left( d t\sin(\theta)J_{2}(2dt\sin(\theta)) - 2 J_{1}( 2dt\sin(\theta) ) \right) + \dots \right].
\end{eqnarray}
Then it is possible to carry out the integrals in $\theta$, and we obtain, up to linear order in velocities
\begin{eqnarray}
\lim_{v_{x}\ll c}I_{0} & = & \frac{v_{x}}{c}\frac{\sin(2dt)}{dt},
\end{eqnarray}
\begin{eqnarray}
\lim_{v_{x}\ll c}I_{1} & = & -\frac{\cos(2dt)}{dt} + \frac{\sin(2dt)}{2d^{2}t^{2}}.
\end{eqnarray}
Those results will be useful in the small velocities regime, in particular, in the large distance limit.

\subsection{Limit of small distances}
In the small distances regime, we have, up to linear order in $d$
\begin{eqnarray}
\lim_{d\ll \sigma}I_{0} & = & \int_{0}^{\pi}\dd\theta\frac{ \sin(\theta)\cos(\theta)}{\left( 1 - \frac{v_{x}}{c}\cos(\theta)\right)^{3} }\left[ 1 + \mathcal{O}\left(d^{2}\right) \right],
\end{eqnarray}
\begin{eqnarray}
\lim_{d\ll \sigma}I_{1} & = & \int_{0}^{\pi}\dd\theta\frac{ \sin^{3}(\theta)}{\left( 1 - \frac{v_{x}}{c}\cos(\theta)\right)^{4} }\left[ \frac{d t}{\gamma} + \mathcal{O}\left(d^{2}\right) \right].
\end{eqnarray}
Then it is possible to carry out the integrals in $\theta$, and we obtain, up to linear order in distances
\begin{eqnarray}
\lim_{d\ll \sigma}I_{0} = 2\frac{v_{x}}{c}\gamma^{4},
\end{eqnarray}
\begin{eqnarray}
\lim_{d\ll \sigma}I_{1} = \frac{4}{3}\gamma^{3}d t.
\end{eqnarray}
Note that those results are valid for all velocities.

\subsection{Limit of large distances at large velocities}
In this subsection, we are going to study the limit of large distances for large velocities of $I_{0}$ and $I_{1}$, i.e. the following results
\begin{eqnarray}\label{Limite_Integralx_grandes_v}
C_{0} = \lim_{v_{x}\to c}\lim_{d\gg\sigma}\int_{0}^{\pi}\dd\theta\frac{ \sin(\theta)\cos(\theta)}{\left( 1 - \frac{v_{x}}{c}\cos(\theta)\right)^{3} }J_{0}\left( \frac{2dt}{\gamma}\frac{\sin(\theta)}{1 - \frac{v_{x}}{c}\cos(\theta)} \right)
& = & \gamma^{4}\frac{v_{x}}{c}\frac{\sin(2dt)}{dt},
\end{eqnarray}
\begin{eqnarray}\label{Limite_Integralz_grandes_v}
C_{1} = \lim_{v_{x}\to c}\lim_{d\gg\sigma}\int_{0}^{\pi}\dd\theta\frac{ \sin^{2}(\theta)}{\left( 1 - \frac{v_{x}}{c}\cos(\theta)\right)^{3} }J_{1}\left( \frac{2dt}{\gamma}\frac{\sin(\theta)}{1 - \frac{v_{x}}{c}\cos(\theta)} \right)
& = & - \gamma^{3}\frac{\cos(2dt)}{dt}.
\end{eqnarray}
$C_{0}$ and $C_{1}$ are used in the calculation of the large distance regime of the $x$ (quantum friction) and $z$ (Casimir force) components of the four-force respectively.
First of all, we use the asymptotic limit of Bessel functions for large argument $\lim_{ad\gg1}J_{0}(ad)\to\sqrt{\frac{2}{\pi a d}}\sin(ad + \pi/4)$ and $\lim_{ad\gg1}J_{1}(ad)\to-\sqrt{\frac{2}{\pi a d}}\cos(ad + \pi/4)$. Then we have
\begin{eqnarray}
C_{0} & = & \sqrt{\frac{\gamma}{\pi d t}}
\lim_{v_{x}\to c}\lim_{d\gg\sigma}\int_{0}^{\pi}\dd\theta
\frac{ \sqrt{\sin(\theta)}\cos(\theta)}{\left( 1 - \frac{v_{x}}{c}\cos(\theta)\right)^{5/2} }\sin\left(\frac{\pi}{4} +  \frac{2dt}{\gamma}\frac{\sin(\theta)}{1 - \frac{v_{x}}{c}\cos(\theta)} \right),
\end{eqnarray}
\begin{eqnarray}
C_{1} & = & - \sqrt{\frac{\gamma}{\pi d t}}
\lim_{v_{x}\to c}\lim_{d\gg\sigma}\int_{0}^{\pi}\dd\theta\frac{ \sin^{3/2}(\theta)}{\left( 1 - \frac{v_{x}}{c}\cos(\theta)\right)^{5/2} }\cos\left( \frac{\pi}{4} + \frac{2dt}{\gamma}\frac{\sin(\theta)}{1 - \frac{v_{x}}{c}\cos(\theta)} \right).
\end{eqnarray}
After that, we separate the region of integration into two parts, the first one from $\theta=0$ to $\theta=\pi/2$, and the second one from $\theta=\pi/2$ to $\theta=\pi$.
Then, we apply the change of variable $\sin(\theta) = S_{\pm}$ to the two integrals, taking into account that $\cos(\theta)$ transforms into $+\sqrt{ 1 - S_{+}^{2} }$ in the integral that runs from $\theta=0$ to $\theta=\pi/2$ and into $-\sqrt{ 1 - S_{-}^{2} }$ in the integral that runs from $\theta=\pi/2$ to $\theta=\pi$ (then we have $\dd\theta = \frac{\dd S_{\pm}}{\pm\sqrt{ 1 - S_{\pm}^{2} }}$ for each integral) in order to keep the correct criterion of signs of $\cos(\theta)$ in the first and second quadrants. Then we have
\begin{eqnarray}
C_{0} & = & \sqrt{\frac{\gamma}{\pi d t}}
\lim_{v_{x}\to c}\lim_{d\gg\sigma}\int_{0}^{1}\dd S_{+}
\frac{\sqrt{S_{+}}}{\left( 1 - \frac{v_{x}}{c}\sqrt{ 1 - S_{+}^{2}}\right)^{5/2} }\sin\left(\frac{\pi}{4} +  \frac{2dt}{\gamma}\frac{S_{+}}{1 - \frac{v_{x}}{c}\sqrt{ 1 - S_{+}^{2}}} \right)\nonumber\\
& & - \sqrt{\frac{\gamma}{\pi d t}}
\lim_{v_{x}\to c}\lim_{d\gg\sigma}\int_{0}^{1}\dd S_{-}
\frac{ \sqrt{S_{-}}}{\left( 1 + \frac{v_{x}}{c}\sqrt{ 1 - S_{-}^{2}}\right)^{5/2} }\sin\left(\frac{\pi}{4} +  \frac{2dt}{\gamma}\frac{S_{-}}{1 + \frac{v_{x}}{c}\sqrt{ 1 - S_{-}^{2}}} \right),
\end{eqnarray}
\begin{eqnarray}
C_{1} & = & - \sqrt{\frac{\gamma}{\pi d t}}
\lim_{v_{x}\to c}\lim_{d\gg\sigma}\int_{0}^{1}\frac{\dd S_{+}}{+\sqrt{ 1 - S_{+}^{2} }}\frac{ S_{+}^{3/2}}{\left( 1 - \frac{v_{x}}{c}\sqrt{ 1 - S_{+}^{2} }\right)^{5/2} }\cos\left( \frac{\pi}{4} + \frac{2dt}{\gamma}\frac{S_{+}}{1 - \frac{v_{x}}{c}\sqrt{ 1 - S_{+}^{2} }} \right)\nonumber\\
& & + \sqrt{\frac{\gamma}{\pi d t}}
\lim_{v_{x}\to c}\lim_{d\gg\sigma}\int_{0}^{1}\frac{\dd S_{-}}{-\sqrt{ 1 - S_{-}^{2} }}\frac{ S_{-}^{3/2}}{\left( 1 + \frac{v_{x}}{c}\sqrt{ 1 - S_{-}^{2} }\right)^{5/2} }\cos\left( \frac{\pi}{4} + \frac{2dt}{\gamma}\frac{S_{-}}{1 + \frac{v_{x}}{c}\sqrt{ 1 - S_{-}^{2} }} \right).
\end{eqnarray}
Note the subtle change of signs in the second integral due to the sign of $\cos(\theta)$ in the second quadrant.

We apply another change of variables
\begin{align}
\alpha_{\pm} &= \frac{S_{\pm}}{1 \mp \frac{v_{x}}{c}\sqrt{ 1 - S_{\pm}^{2} }},
\end{align}
obtaining that
\begin{eqnarray}
C_{0} & = & \sqrt{\frac{\gamma}{\pi d t}}
\lim_{v_{x}\to c}\lim_{d\gg\sigma}\int_{0}^{1}\dd S_{+}
\frac{\alpha_{+}^{5/2}}{S_{+}^{2}}
\sin\left(\frac{\pi}{4} +  \frac{2dt}{\gamma}\alpha_{+} \right)\nonumber\\
& & - \sqrt{\frac{\gamma}{\pi d t}}
\lim_{v_{x}\to c}\lim_{d\gg\sigma}\int_{0}^{1}\dd S_{-}
\frac{\alpha_{-}^{5/2}}{S_{-}^{2}}\sin\left(\frac{\pi}{4} +  \frac{2dt}{\gamma}\alpha_{-} \right),
\end{eqnarray}
\begin{eqnarray}
C_{1} & = & - \sqrt{\frac{\gamma}{\pi d t}}
\lim_{v_{x}\to c}\lim_{d\gg\sigma}\int_{0}^{1}\dd S_{+}
\frac{\alpha_{+}^{5/2}}{S_{+}}
\cos\left(\frac{\pi}{4} + \frac{2dt}{\gamma}\alpha_{+} \right)
\frac{v_{x}}{c}\frac{\alpha_{+}}{\alpha_{+} - S_{+}}\nonumber\\
& & + \sqrt{\frac{\gamma}{\pi d t}}
\lim_{v_{x}\to c}\lim_{d\gg\sigma}\int_{0}^{1}\dd S_{-}
\frac{\alpha_{-}^{5/2}}{S_{-}}
\cos\left(\frac{\pi}{4} + \frac{2dt}{\gamma}\alpha_{-} \right)
\frac{v_{x}}{c}\frac{\alpha_{-}}{\alpha_{-} - S_{-}}.
\end{eqnarray}
The Jacobian of the transformation is
\begin{align}
\dd\alpha_{\pm} &= \frac{\dd}{\dd S_{\pm}}\left(\frac{S_{\pm}}{1 \mp \frac{v_{x}}{c}\sqrt{ 1 - S_{\pm}^{2} }}\right)\dd S_{\pm}\nonumber\\
&= \dd S_{\pm}\left[ \mp\frac{S_{\pm}^{2}\frac{v_{x}}{c}}{\sqrt{1 - S_{\pm}^{2}}\left( 1 \mp \frac{v_{x}}{c}\sqrt{1 - S_{\pm}^{2}}\right)^{2}} + \frac{1}{1 \mp \frac{v_{x}}{c}\sqrt{1 - S_{\pm}^{2}}}\right]\nonumber\\
&= \frac{\dd S_{\pm}}{J_{\pm}}.
\end{align}
Note that, if we invert this change of variable, we get
\begin{align}
S_{+} &= \alpha_{+}\frac{1 - \left(\frac{v_{x}}{c}\right)\sqrt{ 1 + \alpha_{+}^{2}\left( \left(\frac{v_{x}}{c}\right)^{2} - 1 \right) }}{\alpha_{+}^{2}\left(\frac{v_{x}}{c}\right)^{2} + 1}
\hspace{1cm}\forall\alpha_{+}\in(0,1) \parallel \forall S_{+}\in\left(0,\frac{1}{\gamma}\right),
\end{align}
\begin{align}
S_{+} &= \alpha_{+}\frac{1 + \left(\frac{v_{x}}{c}\right)\sqrt{ 1 + \alpha_{+}^{2}\left( \left(\frac{v_{x}}{c}\right)^{2} - 1 \right) }}{\alpha_{+}^{2}\left(\frac{v_{x}}{c}\right)^{2} + 1}
\hspace{1cm}\forall\alpha_{+}\in(1,\gamma) \parallel \forall S_{+}\in\left(\frac{1}{\gamma},1\right),
\end{align}
\begin{align}
S_{-} &= \alpha_{-}\frac{1 + \left(\frac{v_{x}}{c}\right)\sqrt{ 1 + \alpha_{-}^{2}\left( \left(\frac{v_{x}}{c}\right)^{2} - 1 \right) }}{\alpha_{-}^{2}\left(\frac{v_{x}}{c}\right)^{2} + 1}
\hspace{1cm}\forall\alpha_{-}\in(0,1) \parallel \forall S_{-}\in(0,1).
\end{align}
Then we observe that the change of variables is different, in the first integral, for the interval $S_{+}\in(0,\frac{1}{\gamma})$ than for the other interval $S_{+}\in(\frac{1}{\gamma},1)$, therefore we have to divide this integral in two parts, taking into account that the transformation from $S_{+}$ into $\alpha_{+}$ is subtly different in each integral. Then, we have
\begin{eqnarray}
C_{0} & = & \sqrt{\frac{\gamma}{\pi d t}}
\lim_{v_{x}\to c}\lim_{d\gg\sigma}\int_{0}^{\gamma}\dd\alpha_{+}J_{+}
\frac{\alpha_{+}^{5/2}}{S_{+}^{2}}
\sin\left(\frac{\pi}{4} +  \frac{2dt}{\gamma}\alpha_{+} \right)\nonumber\\
& & - \sqrt{\frac{\gamma}{\pi d t}}
\lim_{v_{x}\to c}\lim_{d\gg\sigma}\int_{1}^{\gamma}\dd\alpha_{+}J_{+}
\frac{\alpha_{+}^{5/2}}{S_{+}^{2}}
\sin\left(\frac{\pi}{4} +  \frac{2dt}{\gamma}\alpha_{+} \right)\nonumber\\
& & - \sqrt{\frac{\gamma}{\pi d t}}
\lim_{v_{x}\to c}\lim_{d\gg\sigma}\int_{0}^{1}\dd\alpha_{-}J_{-}
\frac{\alpha_{-}^{5/2}}{S_{-}^{2}}\sin\left(\frac{\pi}{4} +  \frac{2dt}{\gamma}\alpha_{-} \right),
\end{eqnarray}
\begin{eqnarray}
C_{1} & = & - \sqrt{\frac{\gamma}{\pi d t}}\frac{v_{x}}{c}
\lim_{v_{x}\to c}\lim_{d\gg\sigma}\int_{0}^{\gamma}\dd\alpha_{+}J_{+}
\frac{\alpha_{+}^{5/2}}{S_{+}}
\cos\left(\frac{\pi}{4} + \frac{2dt}{\gamma}\alpha_{+} \right)
\frac{\alpha_{+}}{\alpha_{+} - S_{+}}\nonumber\\
& & + \sqrt{\frac{\gamma}{\pi d t}}\frac{v_{x}}{c}
\lim_{v_{x}\to c}\lim_{d\gg\sigma}\int_{1}^{\gamma}\dd\alpha_{+}J_{+}
\frac{\alpha_{+}^{5/2}}{S_{+}}
\cos\left(\frac{\pi}{4} + \frac{2dt}{\gamma}\alpha_{+} \right)
\frac{\alpha_{+}}{\alpha_{+} - S_{+}}\nonumber\\
& & + \sqrt{\frac{\gamma}{\pi d t}}\frac{v_{x}}{c}
\lim_{v_{x}\to c}\lim_{d\gg\sigma}\int_{0}^{1}\dd\alpha_{-}J_{-}
\frac{\alpha_{-}^{5/2}}{S_{-}}
\cos\left(\frac{\pi}{4} + \frac{2dt}{\gamma}\alpha_{-} \right)
\frac{\alpha_{-}}{\alpha_{-} - S_{-}}.
\end{eqnarray}
In the next step, we separate the first integral in two parts, the first one for $\alpha_{+}\in(0,1)$, and the second one for $\alpha_{+}\in(1,\gamma)$. Next, we write $S_{\pm}$ explicitly as functions of $\alpha_{\pm}$, $S_{\pm} = S_{\pm}(\alpha_{\pm})$ taking into account in which dominion of $\alpha_{\pm}$ we are, and we drop the sub-indices $\pm$ because, at the end of the day, $\alpha_{\mp}$ are dummy dimensionless integration variables. Then we have two integrals, one for the interval $\alpha_{+}\in(0,1)$ and another one for the interval $\alpha_{+}\in(1,\gamma)$.

In the next step, we apply a high velocity limit to the integrand of the first integral $v_{x}\to c$, and an asymptotic expansion of $\alpha$ around $\alpha = \gamma$ to the integrand of the second integral, obtaining
\begin{eqnarray}
C_{0} & = & \sqrt{\frac{\gamma}{\pi d t}}
\lim_{d\gg\sigma}\int_{0}^{1}\dd\alpha 
\sin\left(\frac{\pi}{4} + \frac{2dt}{\gamma}\alpha \right)
\sqrt{\alpha}\left[ \frac{1}{1 - \frac{v_{x}}{c}} + \mathcal{O}\left[\alpha^{0}\right] \right]\nonumber\\
& & +
\sqrt{\frac{\gamma}{\pi d t}}
\lim_{d\gg\sigma}\int_{1}^{\gamma}\dd\alpha 
\sin\left(\frac{\pi}{4} + \frac{2dt}{\gamma}\alpha \right)
\sqrt{\alpha}\left[ \frac{v_{x}}{c}\frac{\sqrt{2}\gamma^{5/2}}{\sqrt{\gamma - \alpha}}  + \mathcal{O}\left[(\gamma - \alpha)^{1/2}\right]\right],
\end{eqnarray}
\begin{eqnarray}
C_{1} & = & \frac{v_{x}}{c}\sqrt{\frac{\gamma}{\pi d t}}
\lim_{d\gg\sigma}\int_{0}^{1}\dd\alpha 
\cos\left(\frac{\pi}{4} + \frac{2dt}{\gamma}\alpha \right)
\alpha^{3/2}\left[ - 1 + \mathcal{O}\left[\alpha^{2}\right] \right]\nonumber\\
& & -
\sqrt{\frac{\gamma}{\pi d t}}
\lim_{d\gg\sigma}\int_{1}^{\gamma}\dd\alpha 
\cos\left(\frac{\pi}{4} + \frac{2dt}{\gamma}\alpha \right)
\alpha^{3/2}\left[ \frac{\sqrt{2\gamma}}{\sqrt{\gamma - \alpha}}  + \mathcal{O}\left[(\gamma - \alpha)^{1/2}\right]\right].
\end{eqnarray}
The first integral can be carried out analytically, in the large distance limit we obtain
\begin{eqnarray}
\sqrt{\frac{\gamma}{\pi d t}}
\lim_{d\gg\sigma}\int_{0}^{1}\dd\alpha 
\sin\left(\frac{\pi}{4} + \frac{2dt}{\gamma}\alpha \right)
\left[ \frac{\sqrt{\alpha}}{1 - \frac{v_{x}}{c}} \right] 
& = &
\frac{1}{1 - \frac{v_{x}}{c}}\sqrt{\frac{\gamma}{2\pi d t}}
\frac{\gamma}{2dt}\left[\sin \left(\frac{2 d t}{\gamma }\right)-\cos \left(\frac{2 d t}{\gamma }\right)\right],
\end{eqnarray}
\begin{eqnarray}
 - \frac{v_{x}}{c}\sqrt{\frac{\gamma}{\pi d t}}
\lim_{d\gg\sigma}\int_{0}^{1}\dd\alpha 
\cos\left(\frac{\pi}{4} + \frac{2dt}{\gamma}\alpha \right)
\alpha^{3/2}
& = & 
\frac{v_{x}}{c}\sqrt{\frac{\gamma}{2\pi d t}}
\frac{\gamma^{2}}{2d^{2}t^{2}}\left[\cos\left(\frac{2 d t}{\gamma }\right)-\sin\left(\frac{2 d t}{\gamma }\right)\right].
\end{eqnarray}
The second one can be carried out as well in the high velocity limit: first we approximate the lower limit of the integral from $1$ to $0$. After applying the change of variable $\alpha = \gamma\beta$, then, the limits of integration change to $\beta\in(1/\gamma,1)$ and, in the high velocity limit, we have $1/\gamma\to0$. When we apply the high velocity limit and the large distance limit, we get the following result
\begin{eqnarray}
\sqrt{\frac{\gamma}{\pi d t}}
\lim_{v\to c}\lim_{d\gg\sigma}\int_{1}^{\gamma}\dd\alpha 
\sin\left(\frac{\pi}{4} + \frac{2dt}{\gamma}\alpha \right)
\left[ \frac{v_{x}}{c}\frac{\sqrt{2\alpha}\gamma^{5/2}}{\sqrt{\gamma - \alpha}} \right]
& = & \gamma^{4}\sqrt{\frac{2}{\pi d t}}
\frac{v_{x}}{c}\lim_{v\to c}\lim_{d\gg\sigma}\int_{1/\gamma}^{1}\dd\beta
\sqrt{\beta}\frac{\sin\left(\frac{\pi}{4} + 2dt\beta \right)}{\sqrt{1 - \beta}}\nonumber\\
& = & \gamma^{4}\sqrt{\frac{2}{\pi d t}}
\frac{v_{x}}{c}\lim_{d\gg\sigma}\int_{0}^{1}\dd\beta
\sqrt{\beta}\frac{\sin\left(\frac{\pi}{4} + 2dt\beta \right)}{\sqrt{1 - \beta}}\nonumber\\
& = & \gamma^{4}\sqrt{\frac{2}{\pi d t}}
\frac{v_{x}}{c}
\sqrt{\frac{\pi}{2dt}}\left[ \sin(2dt) + \frac{\cos(2dt)}{8dt}
\right],
\end{eqnarray}
\begin{eqnarray}
- \sqrt{\frac{\gamma}{\pi d t}}
\lim_{v\to c}\lim_{d\gg\sigma}\int_{1}^{\gamma}\dd\alpha 
\cos\left(\frac{\pi}{4} + \frac{2dt}{\gamma}\alpha \right)
\alpha^{3/2}\left[ \frac{\sqrt{2\gamma}}{\sqrt{\gamma - \alpha}}\right]
& = & - \gamma^{3}\sqrt{\frac{2}{\pi d t}}
\lim_{v\to c}\lim_{d\gg\sigma}\int_{1/\gamma}^{1}\dd\beta
\cos\left(\frac{\pi}{4} + 2dt\beta \right)
\frac{\beta^{3/2}}{\sqrt{1 - \beta}}\nonumber\\
& = & - \gamma^{3}\sqrt{\frac{2}{\pi d t}}
\lim_{d\gg\sigma}\int_{0}^{1}\dd\beta
\cos\left(\frac{\pi}{4} + 2dt\beta \right)
\frac{\beta^{3/2}}{\sqrt{1 - \beta}}\nonumber\\
& = & - \gamma^{3}\sqrt{\frac{1}{2\pi d t}}
\sqrt{\frac{2\pi}{d t}}\left[ \cos(2dt) - \frac{3}{8d t}\sin(2dt)\right].
\end{eqnarray}
Then, the dominant contribution at large distances is
\begin{eqnarray}
C_{0} & = & \gamma^{4}\frac{v_{x}}{c}\frac{\sin(2dt)}{dt},
\end{eqnarray}
\begin{eqnarray}
C_{1} & = & - \gamma^{3}\frac{\cos(2dt)}{d t}.
\end{eqnarray}
which are the results we wanted to demonstrate.

\section{The Lippman-Swchinger Equation applied to the two point correlator in the presence of a macroscopic object.}\label{Appendix_Lippman_Schwinger}
In this appendix we are going to introduce a general boundary condition, and then particularize for an infinite plate. The formalism that we are going to introduce is valid for any kind of linear boundary conditions. This includes the typical choices of Dirichlet (e.g., perfect conductor for the electric field), Neumann or any other kind of continuity condition with the field on the other side of the plate.

In the presence of a boundary condition, the two-point correlator of the scalar field is modified by the introduction of an extra term that is derived from the Lippmann-Schwinger equation \cite{Trace_formalism_em} 
\begin{align}\label{PabloCagaSchwinger}
\mathbb{G}_{\bm{k}} & = \mathbb{G}^{0}_{\bm{k}} + \sumint_{\bm{k}'} \mathbb{G}^{0}_{\bm{k}}\mathbb{T}_{\bm{k}\bm{k}'}\mathbb{G}^{0}_{\bm{k}'},
\end{align}
where $\mathbb{G}^{0}_{\bm k}$ is the two-point correlator for the field in free space, and $\mathbb{T}_{\bm{k}\bm{k}'}$ is the T-scattering matrix of the object that imposes the conditions on the field. The T-scattering matrix has all the information about the geometry and the kind of boundary conditions of the considered object. The symbol $\sumint$ represents the sum over the momentum variable $k'$, which in the continuum is an integral over the momentum space, but depending on the multipole basis used, it can be a continuous or discrete variable \cite{Scattering_formalism_em}. The general expression for the free two-point correlator function in the multipolar basis is
\begin{eqnarray}
\mathbb{G}_{0}( \bm{r}_{0} -  \bm{r}_{1}) = \int_{\bm{k}}\left[
\phi_{\bm{k}}^{out}(\bm{r}_{0})\bar{\phi}_{\bm{k}}^{reg}(\bm{r}_{1})\Theta( z_{0} - z_{1} )
 + \phi_{\bm{k}}^{reg}(\bm{r}_{0})\bar{\phi}_{\bm{k}}^{out}(\bm{r}_{1})\Theta( z_{1} - z_{0} )
\right],
\end{eqnarray}
where $\phi_{\bm{k}}^{reg/out}$ are the incoming/outgoing multipoles to/from the origin of coordinates. The outgoing multipoles are regular at infinity, while the incoming multipoles are regular at the origin of coordinates (for all complex frequencies $\omega$ with positive real and imaginary part), those multipoles will be related with $u_{\bm{k}}$ and $u_{\bm{k}}^{*}$ later. $\Theta(x)$ is the Heaviside theta function, and we use the the definition $\int_{\bm{k}} = \int\dd^{3}\bm{k}$. We are going to obtain the general two-point correlator in the presence of one object. Later, we will particularize the result for Cartesian multipoles and the object to an infinite plate. Finally, we are going to use this expression of the Lippmann-Schwinger equation to obtain the method of images explicitly and the expression of the two point correlator used in the text in \Eq{Funcion_Green_Placa_Texto}, which is a central result, widely used in all the calculations of the paper.

In general, the $\mathbb{T}$ matrix operator that defines the interaction of the object with the scalar field is not defined in the same cordinate system as the multipoles of the free two-point correlator (for example, the $\mathbb{T}$ matrix of a plate is calculated from a coordinate system centered in one point of its surface. That of a sphere is calculated in in a frame set on its centre. For a cylinder the origin is taken to be at a point of its axis, etc), while, in our paper, the free two-point function is defined in the quantization frame and pulled back to the smeared trajectory of the detector. Therefore, we have to apply a change of coordinates from the multipole basis where the two-point function is defined (centred in $\bm x=0$) to the multipole basis where the $\mathbb{T}$ matrix is defined, and it is done by the use of the Translation matrices $\mathcal{X}$ \cite{Scattering_formalism_em,Trace_formalism_em}.
\begin{eqnarray}\label{Translation_Matrices_Def1}
\bar{\phi}_{\bm{k}}^{reg}( \bm{r}_{\alpha}) & = &\int_{\bm{v}}
\mathcal{X}^{\dagger}_{\bm{k}, \bm{v}}( \bm{X}_{\alpha\beta})
\bar{\phi}_{ \bm{v}}^{reg}( \bm{r}_{\beta}),\\
\phi_{\bm{q}}^{reg}( \bm{r}_{\alpha}) & = &\int_{\bm{w}}
\phi_{ \bm{w}}^{reg}( \bm{r}_{\beta})
\mathcal{X}_{ \bm{w}, \bm{q}}( \bm{X}_{\alpha\beta}),
\end{eqnarray}
where $\bm{X}_{\alpha\beta} = \bm{r}_{\alpha} - \bm{r}_{\beta}$ describes the relative position of the two origin of coordinates. $\mathcal{X}$ and $\mathcal{X}^{\dagger}$ correspond to $\mathbb{V}$ and $\mathbb{W}$ in \cite{Scattering_formalism_em} respectively. Note that $\bm{r}_{\alpha}$ is the same point of the space as $\bm{r}_{\beta}$, but represented in the translated reference system (see Fig.~1 of \cite{Scattering_formalism_em}). We are going to write all the expressions in terms of the $\mathbb{T}$ operator defined in its own coordinate system \cite{Scattering_formalism_em,Trace_formalism_em}
\begin{eqnarray}
\mathcal{T}_{\bm{k},\bm{u}} = - \int\dd^{3}\bm{r}_{2}\int\dd^{3} \bm{r}_{3}\bar{\phi}_{\bm{k}}^{reg}( \bm{r}_{2})\mathbb{T}( \bm{r}_{2}, \bm{r}_{3})\phi_{\bm{u}}^{reg}( \bm{r}_{3}),
\end{eqnarray}
therefore, we have to translate the mutipoles of the free two point correlators to the coordinate system where the $\mathbb{T}$ operator is defined. Assuming $\bm{r}_{4,z} > \bm{r}_{0,z} > \bm{r}_{1,z}$, where $\bm{r}_{4}$ is the point where the object is placed, we apply a direct substitution in the second term of the r.h.s. of the Lippmann-Schwinger equation obtaining
\begin{eqnarray}
\mathbb{G}_{0}\mathbb{T}\mathbb{G}_{0}
& = & \int\dd^{3}\bm{r}_{2}\int\dd^{3}\bm{r}_{3}\mathbb{G}_{0}(\bm{r}_{0},\bm{r}_{2})\mathbb{T}(\bm{r}_{2},\bm{r}_{3})\mathbb{G}_{0}(\bm{r}_{3},\bm{r}_{1})\nonumber\\
& = & \int\dd^{3}\bm{r}_{2}\int\dd^{3}\bm{r}_{3}
\int_{\bm{k}}{_{0}}\phi_{\bm{k}}^{out}(\bm{r}_{0}){_{0}}\bar{\phi}_{\bm{k}}^{reg}(\bm{r}_{2})
{_{44}}\mathbb{T}(\bm{r}_{2},\bm{r}_{3})
\int_{\bm{q}}{_{1}}\phi_{\bm{q}}^{reg}(\bm{r}_{3}){_{1}}\bar{\phi}_{\bm{q}}^{out}(\bm{r}_{1}).
\end{eqnarray}
Each two point correlator is defined in one system of reference centered in one point. The sub-index $i$ in ${_{i}}\phi_{\bm{k}}^{reg/out}(\bm{r})$ indicates that the point where the multipole basis is centered is $\bm{r}_{i}$, therefore, ${_{i}}\phi_{\bm{k}}^{reg/out}(\bm{r}_{i}) = \phi_{\bm{k}}^{reg/out}(\bm{0})$. Note that we assume that the object is placed in $\bm{r}_{4}$, and that $\bm{r}_{2}$ and $\bm{r}_{3}$ are integration variables. We have to change the system of coordinates where the scalar multipoles are defined from the frame of the source to the frame of the object (of the $\mathbb{T}$ operator). Symbolically,
\begin{eqnarray}\label{Translation_Matrices_Def2}
{_{0}}\bar{\phi}_{\bm{k}}^{reg}(\bm{r}_{2}) & = &\int_{\bm{v}}
\mathcal{X}^{\dagger}_{\bm{k},\bm{v}}(\bm{X}_{40})
{_{4}}\bar{\phi}_{\bm{v}}^{reg}(\bm{r}_{2}),\\
{_{1}}\phi_{\bm{q}}^{reg}(\bm{r}_{3}) & = &\int_{\bm{w}}
{_{4}}\phi_{\bm{w}}^{reg}(\bm{r}_{3})
\mathcal{X}_{\bm{w},\bm{q}}( \bm{X}_{41}),
\end{eqnarray}
where $\bm{X}_{40}$ and $\bm{X}_{41}$ are the relative distance between the two coordinate systems considered here. Note that this definition is equivalent to the one shown in \Eq{Translation_Matrices_Def1}. In our particular case, $\bm{X}_{40} = (x,y,d)$. In what follows, we will write the second term of the r.h.s of the Lippmann-Schwinger equation in the basis centered in $\bm{r}_{0}$
\begin{eqnarray}
\mathbb{G}_{0}\mathbb{T}\mathbb{G}_{0}
& = & \int\dd^{3}\bm{r}_{2}\int\dd^{3}\bm{r}_{3}
\int_{\bm{k}}{_{0}}\phi_{\bm{k}}^{out}(\bm{r}_{0}){_{0}}\bar{\phi}_{\bm{k}}^{reg}(\bm{r}_{2})
{_{44}}\mathbb{T}(\bm{r}_{2},\bm{r}_{3})
\int_{\bm{q}}{_{1}}\phi_{\bm{q}}^{reg}(\bm{\tilde{r}}'_{P}){_{1}}\bar{\phi}_{\bm{q}}^{out}(\bm{r}_{1})\nonumber\\
& = & \int\dd^{3}\bm{r}_{2}\int\dd^{3}\bm{r}_{3}
\int_{\bm{k}}{_{0}}\phi_{\bm{k}}^{out}(\bm{r}_{0})\int_{\bm{v}}
\mathcal{X}^{\dagger}_{\bm{k},\bm{v}}(\bm{X}_{40})
{_{4}}\bar{\phi}_{\bm{v}}^{reg}(\bm{r}_{2})
{_{44}}\mathbb{T}(\bm{r}_{2},\bm{r}_{3})
\int_{\bm{q}}\int_{\bm{w}}
{_{4}}\phi_{\bm{w}}^{reg}(\bm{r}_{3})
\mathcal{X}_{\bm{w},\bm{q}}(\bm{X}_{41}){_{1}}\bar{\phi}_{\bm{q}}^{out}(\bm{r}_{1})\nonumber\\
& = & \int_{\bm{k}}\int_{\bm{v}}\int_{\bm{w}}\int_{\bm{q}}
{_{0}}\phi_{\bm{k}}^{out}(\bm{r}_{0})
\mathcal{X}^{\dagger}_{\bm{k},\bm{v}}(\bm{X}_{40})
\left[ 
\int\dd^{3}\bm{r}_{2}\int\dd^{3}\bm{r}_{3}
{_{4}}\bar{\phi}_{\bm{v}}^{reg}(\bm{r}_{2})
{_{44}}\mathbb{T}(\bm{r}_{2},\bm{r}_{3})
{_{4}}\phi_{\bm{w}}^{reg}(\bm{r}_{3})
\right]
\mathcal{X}_{\bm{w},\bm{q}}(\bm{X}_{41}){_{1}}\bar{\phi}_{\bm{q}}^{out}(\bm{r}_{1})\nonumber\\
& = & - \int_{\bm{k}}\int_{\bm{v}}\int_{\bm{w}}\int_{\bm{q}}
{_{0}}\phi_{\bm{k}}^{out}(\bm{r}_{0})
\mathcal{X}^{\dagger}_{\bm{k},\bm{v}}(\bm{X}_{40})
\mathcal{T}_{\bm{v},\bm{w}}
\mathcal{X}_{\bm{w},\bm{q}}(\bm{X}_{41})
{_{1}}\bar{\phi}_{\bm{q}}^{out}(\bm{r}_{1}).
\end{eqnarray}
The two-point correlator in the presence of one object is
\begin{eqnarray}
\mathbb{G}_{1} = \mathbb{G}_{0} + \mathbb{G}_{0}\mathbb{T}_{1}\mathbb{G}_{0}
& = & \int_{\bm{k}}
{_{0}}\phi_{\bm{k}}^{out}(\bm{r}_{0}){_{0}}\bar{\phi}_{\bm{k}}^{reg}(\bm{r}_{1})
- \int_{\bm{k}}\int_{\bm{v}}\int_{\bm{w}}\int_{\bm{q}}
{_{0}}\phi_{\bm{k}}^{out}(\bm{r}_{0})
\mathcal{X}^{\dagger}_{\bm{k},\bm{v}}(\bm{X}_{40})
\mathcal{T}_{\bm{v},\bm{w}}
\mathcal{X}_{\bm{w},\bm{q}}(\bm{X}_{41})
{_{1}}\bar{\phi}_{\bm{q}}^{out}(\bm{r}_{1})\nonumber\\
& = & \int_{\bm{k}}
{_{0}}\phi_{\bm{k}}^{out}(\bm{r}_{0})
\left[
{_{0}}\bar{\phi}_{\bm{k}}^{reg}(\bm{r}_{1})
- \int_{\bm{v}}\int_{\bm{w}}\int_{\bm{q}}
\mathcal{X}^{\dagger}_{\bm{k},\bm{v}}(\bm{X}_{40})
\mathcal{T}_{\bm{v},\bm{w}}
\mathcal{X}_{\bm{w},\bm{q}}(\bm{X}_{41})
{_{1}}\bar{\phi}_{\bm{q}}^{out}(\bm{r}_{1})
\right].
\end{eqnarray}
The formula shown here is general, and valid for any multipolar basis chosen, only changing the integration variable $\bm{k}$ by the corresponding set of continuous and discrete integration and summation indices of the particular multipolar basis we choose. The two point correlator is the sum of a free correlator, that can be understood as a wave travelling from $\bm{r}_{1}$ to $\bm{r}_{0}$ plus another one that is another wave travelling from $\bm{r}_{1}$ to $\bm{r}_{4}$, where the wave finds the object and is scattered by $\mathcal{T}_{\bm{k},\bm{q}}$, and travels from $\bm{r}_{4}$ to $\bm{r}_{0}$. This expression is general for arbitrary shaped objects. Note as well that the $\mathcal{T}_{\bm{v},\bm{w}}$ operator defined here corresponds to the scalar analog of $\mathcal{F}^{ee}_{\beta,\alpha}(\omega)$ defined in \cite{Scattering_formalism_em}. If we particularize to the multipolar cartesian basis, the translation matrices are defined as \cite{Scattering_formalism_em}
\begin{eqnarray}
\mathcal{X}_{\bm{k},\bm{v}}(\bm{X}_{\alpha\beta}) & = & e^{ - \ii\bm{k}\cdot\bm{X}_{\alpha\beta}}\delta(\bm{k} - \bm{v}),\\
\mathcal{X}^{\dagger}_{\bm{k},\bm{v}}(\bm{X}_{\alpha\beta}) & = & e^{\ii\bm{k}\cdot\bm{X}_{\alpha\beta}}\delta(\bm{k} - \bm{v}),
\end{eqnarray}
where we have used $\omega_{\bm{k}}^{2} = \bm{k}_{\parallel}^{2} + k_{z}^{2}$ to obtain
\begin{eqnarray}
\delta(\omega_{\bm{k}} - \omega_{\bm{q}}) & = & \frac{\abs{\omega}}{\abs{k_{z}}}\delta(k_{z} - q_{z}).
\end{eqnarray}
Then, in the cartesian multipolar basis we have
\begin{eqnarray}
\mathbb{G}_{1}(\bm{r}_{0},\bm{r}_{1})
& = & \int_{\bm{k}}
{_{0}}\phi_{\bm{k}}^{out}(\bm{r}_{0})
\left[
{_{0}}\bar{\phi}_{\bm{k}}^{reg}(\bm{r}_{1})
- \int_{\bm{v}}\int_{\bm{w}}\int_{\bm{q}}
\mathcal{X}^{\dagger}_{\bm{k},\bm{v}}(\bm{X}_{40})
\mathcal{T}_{\bm{v},\bm{w}}
\mathcal{X}_{\bm{w},\bm{q}}(\bm{X}_{41})
{_{1}}\bar{\phi}_{\bm{q}}^{out}(\bm{r}_{1})
\right]\nonumber\\
& = & \int_{\bm{k}}
{_{0}}\phi_{\bm{k}}^{out}(\bm{r}_{0})
\left[
{_{0}}\bar{\phi}_{\bm{k}}^{reg}(\bm{r}_{1})
- \int_{\bm{v}}\int_{\bm{w}}\int_{\bm{q}}
e^{\ii\bm{k}\cdot\bm{X}_{40}}\delta(\bm{k} - \bm{v})
\mathcal{T}_{\bm{v},\bm{w}}
e^{ - \ii\bm{w}\cdot\bm{X}_{41}}\delta(\bm{w} - \bm{q})
{_{1}}\bar{\phi}_{\bm{q}}^{out}(\bm{r}_{1})
\right]\nonumber\\
& = & \int_{\bm{k}}
{_{0}}\phi_{\bm{k}}^{out}(\bm{r}_{0})
\left[
{_{0}}\bar{\phi}_{\bm{k}}^{reg}(\bm{r}_{1})
- \int_{\bm{q}}
e^{\ii\bm{k}\cdot\bm{X}_{40}}
\mathcal{T}_{\bm{k},\bm{q}}
e^{-\ii\bm{q}\cdot\bm{X}_{41}}
{_{1}}\bar{\phi}_{\bm{q}}^{out}(\bm{r}_{1})
\right].
\end{eqnarray}
The $\mathbb{T}$ matrix of a plate is diagonal in this particular coordinate basis \cite{Scattering_formalism_em}.
\begin{eqnarray}
\mathcal{T}_{\bm{k},\bm{q}} & = & \mathcal{R}_{\bm{k}}\delta(\bm{k} - \bm{q}),
\end{eqnarray}
where $\mathcal{R}_{\bm{k}}$ is the Fresnel reflection coefficient for the particular boundary conditions of the plate. In this case, the two-point correlator in the presence of a plate is
\begin{eqnarray}
\mathbb{G}_{1}(\bm{r}_{0},\bm{r}_{1})
& = & \int_{\bm{k}}
{_{0}}\phi_{\bm{k}}^{out}(\bm{r}_{0})
\left[
{_{0}}\bar{\phi}_{\bm{k}}^{reg}(\bm{r}_{1})
- \int_{\bm{q}}
e^{\ii\bm{k}\cdot\bm{X}_{40}}
\mathcal{T}_{\bm{k},\bm{q}}
e^{-\ii\bm{q}\cdot\bm{X}_{41}}
{_{1}}\bar{\phi}_{\bm{q}}^{out}(\bm{r}_{1})
\right]\nonumber\\
& = & \int_{\bm{k}}
{_{0}}\phi_{\bm{k}}^{out}(\bm{r}_{0})
\left[
{_{0}}\bar{\phi}_{\bm{k}}^{reg}(\bm{r}_{1})
- \int_{\bm{q}}
e^{\ii\bm{k}\cdot\bm{X}_{40}}
\mathcal{R}_{\bm{k}}\delta(\bm{k} - \bm{q})
e^{-\ii\bm{q}\cdot\bm{X}_{41}}
{_{1}}\bar{\phi}_{\bm{q}}^{out}(\bm{r}_{1})
\right]\nonumber\\
& = & \int_{\bm{k}}
{_{0}}\phi_{\bm{k}}^{out}(\bm{r}_{0})
\left[
{_{0}}\bar{\phi}_{\bm{k}}^{reg}(\bm{r}_{1})
- e^{\ii\bm{k}\cdot\bm{X}_{40}}
\mathcal{R}_{\bm{k}}
e^{-\ii\bm{k}\cdot\bm{X}_{41}}
{_{1}}\bar{\phi}_{\bm{k}}^{out}(\bm{r}_{1})
\right].
\end{eqnarray}
where
\begin{eqnarray}
\phi_{\bm{k}}^{reg/out}(\bm{r})
& = \frac{1}{\sqrt{2\abs{\bm{k}}}}e^{\mp\ii(\abs{\bm{k}}t - \ii \bm{k}\cdot\bm{r})}
& = \frac{1}{\sqrt{2\abs{\bm{k}}}}e^{\mp\ii k_{\mu}r^{\mu}},\\
\bar{\phi}_{\bm{k}}^{reg/out}(\bm{r})
& = \frac{1}{\sqrt{2\abs{\bm{k}}}}e^{\pm\ii(\abs{\bm{k}}t - \ii \bm{k}\cdot\bm{r})}
& = \frac{1}{\sqrt{2\abs{\bm{k}}}}e^{\pm\ii k_{\mu}r^{\mu}}.
\end{eqnarray}
Therefore we have $\bar{\phi}_{\bm{k}}^{out}(\bm{r}) = \bar{\phi}_{\bm{k}}^{reg}(\bm{r})$.
\subsection{Method of images}
Until now, we have assumed that $\bm{r}_{4,z} > \bm{r}_{0,z} > \bm{r}_{1,z}$. If we want to obtain the expression of the method of images from this result, we will have $\bar{\bm{r}}_{1,z} > \bm{r}_{4,z} > \bm{r}_{0,z}$ (where $\bar{\bm{r}}_{1}$ is the image of $\bm{r}_{1}$). First, we have to exchange $\bar{\phi}_{\bm{k}}^{out}(\bm{r})$ by $\bar{\phi}_{\bm{k}}^{reg}(\bm{r})$, but the translation matrices have to be modified as well. If we want to do this transformation properly, keeping that $\bar{\phi}_{\bm{k}}^{out}(\bm{r})$ is regular at infinity and $\bar{\phi}_{\bm{k}}^{reg}(\bm{r})$ is regular at the origin, we must be careful and remember that, for complex $\omega$ in the first quadrant of the complex plane, the $k_{z}$ component also belongs to the first quadrant of the complex plane as well. It means that its imaginary part produces a (positive) real value that must be kept invariant under the change of $\bm{r}_{1}$ by $\bar{\bm{r}}_{1}$. In particular, we have
\begin{eqnarray}
-\Imag{k_{z}}\abs{X_{41,z}} & = & - \Imag{k_{z}}\abs{\bar{X}_{41,z}}.
\end{eqnarray}
We eliminate the absolute values using that $\bar{r}_{1,z}  > r_{4,z} > r_{0,z} > r_{1,z}$
\begin{eqnarray}
-\Imag{k_{z}}(r_{4,z} - r_{1,z}) &= & - \Imag{k_{z}}(\bar{r}_{1,z} - r_{4,z}),
\end{eqnarray}
obtaining that $\bar{r}_{1,z}$ is
\begin{eqnarray}
\bar{r}_{1,z} & = - r_{1,z} + 2r_{4,z} & = r_{1,z} + 2(r_{4,z} - r_{1,z}) = r_{1,z} + 2X_{41,z}.
\end{eqnarray}
The $x$ and $y$ components are kept invariant, therefore, we have that
\begin{eqnarray}
\bar{\bm{r}}_{1} & = & \bm{r}_{1} + 2\left( \bm{X}_{41}\cdot\hat{z}, \right)\hat{z}
\end{eqnarray}
and using $\bar{\phi}_{\bm{k}}^{out}(\bm{r}) = \bar{\phi}_{\bm{k}}^{reg}(\bm{r})$, the two points correlator is written as
\begin{eqnarray}
\mathbb{G}_{1}(\bm{r}_{0},\bm{r}_{1})
& = & \int_{\bm{k}}
{_{0}}\phi_{\bm{k}}^{out}(\bm{r}_{0})
\left[
{_{0}}\bar{\phi}_{\bm{k}}^{reg}(\bm{r}_{1})
- e^{\ii\bm{k}\cdot\bm{X}_{40}}
\mathcal{R}_{\bm{k}}
e^{-\ii\bm{k}\cdot\bm{X}_{41}}
{_{1}}\bar{\phi}_{\bm{k}}^{out}(\bm{r}_{1})
\right]\nonumber\\
& = & \int_{\bm{k}}
{_{0}}\phi_{\bm{k}}^{out}(\bm{r}_{0})
\left[
{_{0}}\bar{\phi}_{\bm{k}}^{reg}(\bm{r}_{1})
- e^{\ii\bm{k}\cdot\bm{X}_{40}}
\mathcal{R}_{\bm{k}}
e^{-\ii\bm{k}\cdot\bar{\bm{X}}_{41}}
{_{1}}\bar{\phi}_{\bm{k}}^{reg}(\bm{r}_{1})
\right].
\end{eqnarray}
We use that the $\mathcal{R}_{\bm{k}}$ operator and $\mathcal{X}$ operator commutes for the cartesian basis, and $e^{-\ii\bm{k}\cdot\bar{\bm{X}}_{41}} = e^{\ii\bm{k}\cdot\bar{\bm{X}}_{14}}$, to obtain
\begin{eqnarray}
\mathbb{G}_{1}(\bm{r}_{0},\bm{r}_{1})
& = & \int_{\bm{k}}
{_{0}}\phi_{\bm{k}}^{out}(\bm{r}_{0})
\left[
{_{0}}\bar{\phi}_{\bm{k}}^{reg}(\bm{r}_{1})
- \mathcal{R}_{\bm{k}}
e^{\ii\bm{k}\cdot\bm{X}_{40}}
e^{\ii\bm{k}\cdot\bar{\bm{X}}_{14}}
{_{1}}\bar{\phi}_{\bm{k}}^{reg}(\bm{r}_{1})
\right]\nonumber\\
& = & \int_{\bm{k}}
{_{0}}\phi_{\bm{k}}^{out}(\bm{r}_{0})
\left[
{_{0}}\bar{\phi}_{\bm{k}}^{reg}(\bm{r}_{1})
- \mathcal{R}_{\bm{k}}
e^{\ii\bm{k}\cdot\bar{\bm{X}}_{10}}
{_{1}}\bar{\phi}_{\bm{k}}^{reg}(\bm{r}_{1})
\right]\nonumber\\
& = & \int_{\bm{k}}
{_{0}}\phi_{\bm{k}}^{out}(\bm{r}_{0})
\left[
{_{0}}\bar{\phi}_{\bm{k}}^{reg}(\bm{r}_{1})
- \mathcal{R}_{\bm{k}}
e^{2\ii\bm{k}\cdot\left( \bm{X}_{41}\cdot\hat{z} \right)\hat{z}}
e^{\ii\bm{k}\cdot\bm{X}_{10}}
{_{1}}\bar{\phi}_{\bm{k}}^{reg}(\bm{r}_{1})
\right]\nonumber\\
& = & \int_{\bm{k}}
{_{0}}\phi_{\bm{k}}^{out}(\bm{r}_{0})
\left[
{_{0}}\bar{\phi}_{\bm{k}}^{reg}(\bm{r}_{1})
- \mathcal{R}_{\bm{k}}
e^{2\ii\bm{k}\cdot\left( \bm{X}_{41}\cdot\hat{z} \right)\hat{z}}
{_{0}}\bar{\phi}_{\bm{k}}^{reg}(\bm{r}_{1})
\right]\nonumber\\
& = & \int_{\bm{k}}
{_{0}}\phi_{\bm{k}}^{out}(\bm{r}_{0})
\left[
1
- \mathcal{R}_{\bm{k}}e^{2\ii\bm{k}\cdot\left( \bm{X}_{41}\cdot\hat{z} \right)\hat{z}}
\right]{_{0}}\bar{\phi}_{\bm{k}}^{reg}(\bm{r}_{1}).
\end{eqnarray}
Here we have combined two translation matrices into one (or separate one into two) because, in general,
\begin{eqnarray}
\int_{\bm{q}}\mathcal{X}_{\bm{k}\bm{q}}(\bm{X}_{\alpha\beta})\mathcal{X}_{\bm{q}\bm{v}}(\bm{X}_{\beta\gamma}) = \mathcal{X}_{\bm{k}\bm{v}}(\bm{X}_{\alpha\gamma}).
\end{eqnarray}
Finally, we use the definition given in \Eq{Translation_Matrices_Def1} instead the equivalent one given in \Eq{Translation_Matrices_Def2} for $2\left( \bm{X}_{41}\cdot\hat{z} \right)\hat{z}$, we obtain
\begin{eqnarray}
\mathbb{G}_{1}(\bm{r}_{0},\bm{r}_{1})
& = & \int_{\bm{k}}
{_{0}}\phi_{\bm{k}}^{out}(\bm{r}_{0})
\left[
{_{0}}\bar{\phi}_{\bm{k}}^{reg}(\bm{r}_{1})
- \mathcal{R}_{\bm{k}}
{_{0}}\bar{\phi}_{\bm{k}}^{reg}(\bm{r}_{1} + 2\left( \bm{X}_{41}\cdot\hat{z} \right)\hat{z})
\right]\nonumber\\
& = & \int_{\bm{k}}
{_{0}}\phi_{\bm{k}}^{out}(\bm{r}_{0})
\left[
{_{0}}\bar{\phi}_{\bm{k}}^{reg}(\bm{r}_{1})
- \mathcal{R}_{\bm{k}}
{_{0}}\bar{\phi}_{\bm{k}}^{reg}(\bar{\bm{r}}_{1})
\right].
\end{eqnarray}
This last expression is the method of images, where we represent the effect of a planar boundary condition as the presence of a reflection of an image-charge weighted by $\mathcal{R}_{\bm{k}}$. In our particular case, a direct substitution gives $\phi_{\bm{k}}^{reg/out}(\bm{r}) = u_{\bm{k}}^{*}(\bm{r})$, $\bar{\phi}_{\bm{k}}^{reg/out}(\bm{r}) = u_{\bm{k}}(\bm{r})$, and $\bm{X}_{21}\cdot\hat{z} = d$, then
\begin{eqnarray}\label{FGreen_Retardada_Placa}
\mathbb{G}_{1}(\bm{r}_{0},\bm{r}_{1})
& = & \int_{\bm{k}}
u_{\bm{k}}^{*}(\bm{r}_{0})
\left[
1
- \mathcal{R}_{\bm{k}}
e^{2\ii k_{z}d}
\right]u_{\bm{k}}(\bm{r}_{1}),
\end{eqnarray}
which is the two point correlatior used in \Eq{Funcion_Green_Placa_Texto}.


\end{widetext}

\bibliography{bibliography}

\end{document}